\documentclass[prd,floatfix,preprint,tightenlines,amsmath,amssymb,showpacs,superscriptaddress,nofootinbib,twocolumn]{revtex4-1}
\usepackage{amsmath}
\usepackage{graphicx}
\usepackage{epstopdf}
\usepackage{dcolumn}
\usepackage{color}
\usepackage{bm}
\usepackage{overpic}
\usepackage{rotating}
\usepackage{times}
\usepackage{multirow}
\usepackage{float}
\usepackage{mathrsfs}
\usepackage{booktabs}
\usepackage{array,booktabs}
\usepackage{lineno} %% single column
\usepackage{epstopdf}
\newcommand{\ee}{e^+e^-}
\newcommand{\go}{\rightarrow}
\newcommand{\DDpipi}{\pi^+\pi^-D^+D^-}
\newcommand{\Dpipi}{D^+\pi_d^+\pi_d^-}
\newcommand{\Kpipi}{K^-\pi^+\pi^+}

\newcommand{\DststD}{D_1(2420)^+D^-}
\newcommand{\pipipsipp}{\pi^+\pi^-\psi(3770)}
\newcommand{\PHSP}{\pi^+\pi^-D^+D^-(\rm{PHSP})}
\newcommand{\LcLc}{\Lambda_c^+\bar{\Lambda}_c^-}
\newcommand{\pp}{\pi^+\pi^-}
\newcommand{\pipi}{\pi_d^+\pi_d^-}
\newcommand{\DDpi}{D^0D^-\pi^+}
\def\Journal#1#2#3#4{{#1} {\bf #2}, #3 (#4)}

\def\EPJC{Eur. Phys. J. C}

\begin{document}
   % \linenumbers
%	\normalsize
%	\parskip=5pt plus 1pt minus 1pt
%	\hyphenpenalty=10000
%	\tolerance=1000
	
	\lefthyphenmin=2
	\righthyphenmin=2
	\tolerance=1000
	\uchyph=0
	
	\normalsize
	\parskip=5pt plus 1pt minus 1pt
	
\title{\boldmath Measurement of $\ee\go \DDpipi$ cross sections
at center-of-mass energies from 4.190 to 4.946~GeV}

\author{
	\begin{small}
		\begin{center}
			M.~Ablikim$^{1}$, M.~N.~Achasov$^{11,b}$, P.~Adlarson$^{69}$, M.~Albrecht$^{4}$, R.~Aliberti$^{30}$, A.~Amoroso$^{68A,68C}$, M.~R.~An$^{34}$, Q.~An$^{65,52}$, X.~H.~Bai$^{60}$, Y.~Bai$^{51}$
% ...
Yi2020~Wang$^{55}$, Z.~Wang$^{1,52}$, Z.~Y.~Wang$^{1,57}$, Ziyi~Wang$^{57}$, D.~H.~Wei$^{12}$, F.~Weidner$^{62}$, S.~P.~Wen$^{1}$, D.~J.~White$^{61}$, U.~Wiedner$^{4}$, G.~Wilkinson$^{63}$
% ...
			\\
			\vspace{0.2cm}
			(BESIII Collaboration)\\
			\vspace{0.2cm} {\it
				$^{1}$ Institute of High Energy Physics, Beijing 100049, People's Republic of China\\
% ...
$^{4}$ Bochum Ruhr-University, D-44780 Bochum, Germany\\
% ...
$^{11}$ G.I. Budker Institute of Nuclear Physics SB RAS (BINP), Novosibirsk 630090, Russia\\
				$^{12}$ Guangxi Normal University, Guilin 541004, People's Republic of China\\
% ...
$^{30}$ Johannes Gutenberg University of Mainz, Johann-Joachim-Becher-Weg 45, D-55099 Mainz, Germany\\
% ...
$^{34}$ Liaoning Normal University, Dalian 116029, People's Republic of China\\
% ...
$^{51}$ Southeast University, Nanjing 211100, People's Republic of China\\
				$^{52}$ State Key Laboratory of Particle Detection and Electronics, Beijing 100049, Hefei 230026, People's Republic of China\\
% ...
$^{55}$ Tsinghua University, Beijing 100084, People's Republic of China\\
% ...
$^{57}$ University of Chinese Academy of Sciences, Beijing 100049, People's Republic of China\\
% ...
$^{60}$ University of Jinan, Jinan 250022, People's Republic of China\\
				$^{61}$ University of Manchester, Oxford Road, Manchester, M13 9PL, United Kingdom\\
				$^{62}$ University of Muenster, Wilhelm-Klemm-Str. 9, 48149 Muenster, Germany\\
				$^{63}$ University of Oxford, Keble Rd, Oxford, UK OX13RH\\
% ...
$^{65}$ University of Science and Technology of China, Hefei 230026, People's Republic of China\\
% ...
$^{68}$ University of Turin and INFN, (A)University of Turin, I-10125, Turin, Italy; (B)University of Eastern Piedmont, I-15121, Alessandria, Italy; (C)INFN, I-10125, Turin, Italy\\
				$^{69}$ Uppsala University, Box 516, SE-75120 Uppsala, Sweden\\
% ...
$^{b}$ Also at the Novosibirsk State University, Novosibirsk, 630090, Russia\\
% ...
		}
		\vspace{0.4cm}
			\end{center}
	\end{small}
}

% \date{\today}
	
\begin{abstract}

Using data samples collected with the BESIII detector operating at
the BEPCII storage ring, we measure the cross sections of the
$\ee\go \DDpipi$ process at center-of-mass energies
from 4.190 to 4.946~GeV with a partial reconstruction method.
Two resonance structures are seen
and the resonance parameters are determined from a fit to the
cross section line shape. The first resonance we observe has
a mass of (4373.1 $\pm$ 4.0 $\pm$ 2.2)~MeV/$c^2$ and
a width of (146.5 $\pm$ 7.4 $\pm$ 1.3)~MeV, in agreement with those
of the $Y(4390)$ state; the other resonance has
a mass of (4706 $\pm$ 11 $\pm$ 4)~MeV/$c^2$,
a width of (45 $\pm$ 28 $\pm$ 9)~MeV, and a statistical
significance of $4.1$ standard deviations ($\sigma$). This is the first evidence
for a vector state at this mass value.
The spin-$3$ $D$-wave charmonium state $X(3842)$ is searched for
through the $\ee\go \pp X(3842)\go \DDpipi$ process, and evidence
with a significance of $4.2\sigma$ is found
in the data samples with center-of-mass energies from 4.600 to
4.700~GeV.
		
\end{abstract}
	
% \pacs{13.25.Gv, 12.38.Qk, 14.20.Gk, 14.40.Cs}
	
\maketitle

\section{\boldmath Introduction}

The charmonium states with masses below the open charm threshold
and a few vector states above the open charm threshold are well-established~\cite{pdg}, and they agree well with theoretical
calculations based on QCD~\cite{Brambilla:2004jw,review4,Brambilla:2014jmp}
and QCD-inspired potential models~\cite{eichten,godfrey,barnes}.
The vector charmonia $\psi(4040)$, $\psi(4160)$, and $\psi(4415)$
were assigned as the $3^3S_1$, $2^3D_1$, and $4^3S_1$ states,
respectively, since only these three structures were observed
in the total $\ee$ annihilation cross section~\cite{bes2_psis}.

However, a few more vector states, the $Y$ states, were discovered
by the BaBar and Belle $B$-factory experiments~\cite{PBFB}. These include
the $Y(4260)$~\cite{babar_y4260}, the
$Y(4360)$~\cite{babar_y4360,belle_y4660}, and the
$Y(4660)$~\cite{belle_y4660}. They are produced via the
initial state radiation (ISR) process in $\ee$ annihilation and,
thus, are vector states with quantum numbers $J^{PC}=1^{--}$, the same
as the excited $\psi$ states listed above.
These states were observed in hidden-charm final states
in contrast to the excited $\psi$ states peaking in the
inclusive hadronic cross section~\cite{bes2_psis,uglov-kmatrix}. The final states in the latter are dominated by
open-charm meson pairs.

In potential models, five vector charmonium states with
masses between 4.0 and 4.7~GeV/$c^2$ are expected, namely the
$\psi(3^3S_1)$, $\psi(2^3D_1)$, $\psi(4^3S_1)$, $\psi(3^3D_1)$, and
$\psi(5^3S_1)$. The first three are often identified as
the $\psi(4040)$, $\psi(4160)$, and $\psi(4415)$ states, respectively.
The masses of the as yet undiscovered
$\psi(3D)$ and $\psi(5S)$ are expected to be higher than
4.4~GeV/$c^2$. However, six vector states have
been identified in the mass region between 4.0 and 4.7~GeV/$c^2$, as listed above. This makes the $Y(4260)$,
the $Y(4360)$, and perhaps the $Y(4660)$ states good candidates for new
types of exotic particles, and has stimulated theoretical work regarding
their interpretation.  They have been variously considered as candidates for tetraquark states, molecular states,
hybrid states, and
hadro-charmonia~\cite{review1,review2,review3,review4}.

With masses above the open-charm thresholds, both $Y$ and excited
$\psi$ states should couple to open-charm final states, and many
studies have been performed to measure the cross sections of
two-body final states with a pair of charmed mesons~\cite{DD, 2018DD, 2020DsstDs1, 2021DsstDsj} and
three-body final states with a pair of charmed mesons and
a light meson~\cite{bes3_ddstarpi}. Although four-body
final states with a pair of charmed mesons and a pair of
light mesons~\cite{2019HY,2020ZY} have also been studied,
and the production of intermediate two-body
($D_1(2420)\bar{D}+c.c.$) and three-body ($\pp\psi(3770)$)
states have been observed, the total cross section of
the four-body final states has not been reported. In such final states, new
exotic particles and new decay modes of known $Y$ and excited $\psi$
states can be searched for.

In this paper, we report the first measurement of the cross sections
of the $\ee\go \DDpipi$ process with the data samples taken at 37
center-of-mass energies ($\sqrt{s}$) from 4.190 to 4.946~GeV,
the study of the decays of the excited $\psi$ and $Y$ states into
this final state,
and the observation of a new resonant structure
in the cross section line shape.

Two of the $D$-wave spin-triplet states $\psi(1^3D_1)$ ($\psi(3770)$)
and $\psi(1^3D_2)$ ($\psi_2(3823)$) have been observed in the $\ee$
annihilation process $\ee\to \pp\psi(1D)$~\cite{2019HY,2020ZY,2015X3823}.
As their spin partner, the $\psi(1^3D_3)$ ($X(3842)$) observed by
LHCb~\cite{2019X3842} can also be produced in a similar process and
can be searched for in the $\DDpipi$ final state, since the $X(3842)$
decays to $D\bar{D}$.

\section{\boldmath Detector and data samples}
\label{sec:data_sets}
	
The BESIII detector~\cite{Ablikim:2009aa} records $e^+e^-$ collisions
provided by the BEPCII storage ring~\cite{Yu:IPAC2016-TUYA01}.
The cylindrical core of the BESIII detector covers 93\% of the full solid
angle and consists of a helium-based multilayer drift chamber~(MDC),
a plastic scintillator time-of-flight system~(TOF),
and a CsI(Tl) electromagnetic calorimeter~(EMC),
which are all enclosed in a superconducting solenoidal magnet
providing a 1.0~T magnetic field. The solenoid is supported by
an octagonal flux-return yoke with resistive plate counter muon
identification modules interleaved with steel.
The charged-particle momentum resolution at $1~{\rm GeV}/c$ is $0.5\%$,
and the $dE/dx$ resolution is $6\%$ for electrons from Bhabha scattering.
The EMC measures photon energies with a resolution of $2.5\%$ ($5\%$)
at $1$~GeV in the barrel (end cap) region. The time resolution in
the TOF barrel region is 68~ps, while that in the end cap region
is 110~ps. The end cap TOF system was upgraded in 2015 using
multi-gap resistive plate chamber technology, providing
a time resolution of 60~ps~\cite{etof}.
	
In this analysis, the experimental data samples used are listed
in Table~\ref{tab:CS}. The center-of-mass energy is measured using dimuon
events with a precision of 0.8~MeV for data samples with $\sqrt{s}$
smaller than 4.610 GeV~\cite{2016mumu, 2021mumu} and
using $\Lambda_{c}^{+}\bar{\Lambda}_{c}^{-}$ events with a
precision of 0.6~MeV for data samples with $\sqrt{s}$ larger than or equal to 4.610 GeV \cite{2022LcLc}.
The integrated luminosity is determined by analyzing large angle
Bhabha scattering events with an uncertainty of 1.0\%~\cite{2019SWM, 2022lum, 2022LcLc}.
The integrated luminosity of the total data sample is $17.4~\rm{fb}^{-1}$.

\begin{table*}[htp]
	\centering
	\caption{Yields and cross sections results for the $\ee\go\DDpipi$ process at different center-of-mass energies. Here, $\sigma$ is the cross section of the $e^{+}e^{-}\rightarrow\pi^{+}\pi^{-}D^{+}D^{-}$ process, where the first uncertainties are statistical and the second systematic; $\mathscr{L}$, $S$, and $\sigma_{\rm ul}$ are the integrated luminosity, statistical significance, and upper limit of the cross section at 90\% confidence level, respectively. $N_{\rm signal}$ and $N_{\rm sideband}$ are the number of $\ee\go\DDpipi$ events from fits to $RM(\Dpipi)$ distributions in $M(\Kpipi)$ signal and sideband regions, respectively.}
	\label{tab:CS}
	\resizebox{\textwidth}{95mm}{
	\begin{tabular}{
      c
			c
			c
			r@{$\ \pm\ $}l
			r@{$\ \pm\ $}l
			r@{$\ \pm\ $}c@{$\ \pm\ $}l
			c
			c
		}
	\toprule
	\hline\hline
    $\sqrt{s}$ nominal value (GeV) & $\sqrt{s}$ (MeV) & $\mathscr{L}$ ($\rm{pb}^{-1}$) & \multicolumn{2}{c}{$N_{\rm{signal}}$} & \multicolumn{2}{c}{$N_{\rm{sideband}}$} & \multicolumn{3}{c}{$\sigma\ (\rm{pb})$} & $S$ & $\sigma_{\rm{ul}}$ (\rm{pb})\\
	\hline
	\midrule
   4.190  & $4188.59\pm0.15\pm0.68$ & $\ \ $570.0  &  $-$8  &  10  & $-$17  &  11  & 0.1  & 0.9  & 0.0  & - & 1.0  \\
	 4.200  & $4199.15\pm0.05\pm0.34$ & $\ \ $526.0   &  $-$5  &  11  & $-$15  &  12  & 0.2  & 1.0  & 0.0  & - & 1.2  \\
	 4.210  & $4207.73\pm0.14\pm0.61$& $\ \ $572.1  &  15  &  13  &  19  &  14  & 0.3  & 1.0  & 0.1  & $\ \ \ 1.2 \sigma$ & 2.6  \\
	 4.220  & $4217.13\pm0.14\pm0.67$ & $\ \ $569.2   &  17  &  12  &  14  &  13  & 0.7  & 0.9  & 0.1  & $\ \ \ 1.5 \sigma$ & 2.6  \\
	 4.230  & $4225.54\pm0.05\pm0.65$ & 1100.9  & 119  &  25  &  12  &  20  & 3.4  & 0.8  & 0.3  & $\ \ \ 5.9 \sigma$ & -     \\
	 4.237  & $4235.77\pm0.04\pm0.30$ & $\ \ $530.3   &  25  &  14  & $-$29  &  13  & 2.6  & 1.0  & 0.2  & $\ \ \ 1.9 \sigma$ & 3.5  \\
	 4.245  & $4241.66\pm0.12\pm0.73$ & $\ \ \ \ $55.9   &  5   &  6   &  $-$3  &  4   & 4.0  & 3.7  & 0.3  & $\ \ \ 0.9 \sigma$ & 9.0  \\
	 4.246  & $4243.97\pm0.04\pm0.30$ & $\ \ $538.1   & 101  &  19  &  1   &  15  & 6.1  & 1.3  & 0.7  & $\ \ \ 6.6 \sigma$ & -     \\
	 4.260  & $4258.00\pm0.06\pm0.60$ & $\ \ $825.7   & 159  &  26  &  17  &  22  & 5.6  & 1.1  & 0.5  & $\ \ \ 7.5 \sigma$ & -     \\
	 4.270  & $4266.81\pm0.04\pm0.32$  & $\ \ $531.1   &  61  &  18  & $-$27  &  17  & 4.3  & 1.2  & 0.4  & $\ \ \ 3.6 \sigma$ & 6.7  \\
	 4.280  & $4277.78\pm0.11\pm0.52$  & $\ \ $175.7   &  25  &  12  &  2   &  11  & 4.2  & 2.4  & 0.4  & $\ \ \ 2.2 \sigma$ & 9.0  \\
	 4.290  & $4288.43\pm0.06\pm0.34$ & $\ \ $502.4   & 140  &  23  &  4   &  20  & 8.6  & 1.6  & 0.7  & $\ \ \ 7.1 \sigma$ & -     \\
	 4.310  & $4307.89\pm0.17\pm0.63$ & $\ \ \ \ $45.1   &  25  &  8   &  $-$4  &  7   & 17.1 & 5.5  & 1.5  & $\ \ \ 3.4 \sigma$ & 30 \\
	 4.315  & $4312.68\pm0.06\pm0.35$ & $\ \ $501.2   & 263  &  29  &  9   &  23  & 15.4 & 1.9  & 1.3  & $\ \ \ \ 11\sigma$ & -     \\
	 4.340  & $4337.93\pm0.06\pm0.35$ & $\ \ $505.0   & 666  &  42  &  20  &  27  & 36.9 & 2.5  & 3.1  & $\ \ \ \ \ 21\sigma$ & -     \\
	 4.360  & $4358.26\pm0.05\pm0.62$ & $\ \ $544.0   & 1038 &  53  &  8   &  34  & 48.2 & 2.6  & 4.1  & $\ \ \ \ \ 26\sigma$ & -     \\
	 4.380  & $4377.88\pm0.06\pm0.35$ & $\ \ $522.7   & 1184 &  67  & $-$35  &  37  & 61.6 & 3.6  & 5.2  & $\ \ \ \ \ 25\sigma$ & -     \\
	 4.390  & $4387.40\pm0.17\pm0.65$  & $\ \ \ \ $55.6   & 111  &  18  &  19  &  13  & 46.2 & 8.8  & 3.9  & $\ \ \ 7.4 \sigma$ & -     \\
	 4.400  & $4396.83\pm0.06\pm0.36$ & $\ \ $507.8   & 1217 &  62  &  61  &  43  & 61.9 & 3.5  & 5.2  & $\ \ \ \ \ 24\sigma$ & -     \\
	 4.420  & $4415.94\pm0.04\pm0.62$ & 1090.7  & 3144 & 112  & 216  &  71  & 67.7 & 2.6  & 5.8  & $\ \ \ \ \ 37\sigma$ & -     \\
	 4.440  & $4437.59\pm0.06\pm0.35$ & $\ \ $569.9   & 1588 &  85  & 140  &  59  & 65.1 & 3.9  & 5.9  & $\ \ \ \ \ 23\sigma$ & -     \\
	 4.470  & $4467.06\pm0.11\pm0.73$  & $\ \ $111.1   & 192  &  35  &  36  &  25  & 36.0 & 7.8  & 3.9  & $\ \ \ 6.8 \sigma$ & -     \\
	 4.530  & $4527.14\pm0.11\pm0.72$  & $\ \ $112.1   & 141  &  34  &  17  &  28  & 30.4 & 8.6  & 3.1  & $\ \ \ 4.1 \sigma$ & 41     \\
	 4.575  & $4574.50\pm0.18\pm0.70$  &  $\ \ \ \ $48.9   &  39  &  18  &  12  &  19  & 15.5 & 9.9  & 1.4  & $\ \ \ 2.2 \sigma$ & 38 \\
	 4.600  & $4599.53\pm0.07\pm0.74$ & $\ \ $586.9   & 811  &  74  & $-$16  &  69  & 31.2 & 3.1  & 2.8  & $\ \ \ \ \ 12\sigma$ & -     \\
	 4.612  & $4611.86\pm0.12\pm0.32$ & $\ \ $103.8  & 139  &  31  &  42  &  29  & 27.3 & 8.1  & 2.3  & $\ \ \ 4.9 \sigma$ & 40 \\
	 4.620  & $4628.00\pm0.06\pm0.32$  & $\ \ $521.5  & 758  &  90  &  30  &  72  & 33.7 & 4.4  & 2.9  & $\ \ \ \ \ 11\sigma$ & -     \\
	 4.640  & $4640.91\pm0.06\pm0.38$ & $\ \ $552.4  & 725  &  85  & $-$65  &  71  & 32.2 & 3.9  & 2.8  & $\ \ \ \ \ 10\sigma$ & -     \\
	 4.660  & $4661.24\pm0.06\pm0.29$ & $\ \ $529.6  & 814  &  93  & $-$51  &  73  & 38.1 & 4.6  & 3.4  & $\ \ \ \ \ 11\sigma$ & -     \\
	 4.680  & $4681.92\pm0.08\pm0.29$ & 1669.3  & 2427 & 156  & $-$12  & 128  & 33.7 & 2.3  & 2.8  & $\ \ \ \ \ 19\sigma$ & -     \\
	 4.700  & $4698.82\pm0.10\pm0.39$ & $\ \ $536.5  & 1020 &  85  & $-$58  &  76  & 45.7 & 4.1  & 4.0  & $\ \ \ \ \ 13\sigma$ & -     \\
	 4.740  & $4739.70\pm0.20\pm0.30$  & $\ \ $164.3  & 330  &  45  &  47  &  41  & 39.8 & 6.5  & 3.4  & $\ \ \ 8.2 \sigma$ & -     \\
	 4.750  & $4750.05\pm0.12\pm0.29$ & $\ \ $367.2  & 781  &  71  &  71  &  59  & 43.2 & 4.5  & 3.8  & $\ \ \ \ \ 13\sigma$ & -     \\
	 4.780  & $4780.54\pm0.12\pm0.33$ & $\ \ $512.8  & 1042 &  94  & 217  &  78  & 39.6 & 4.3  & 3.3  & $\ \ \ \ \ 14\sigma$ & -     \\
	 4.840  & $4843.07\pm0.20\pm0.31$ & $\ \ $527.3  & 1050 & 100  &  10  &  81  & 43.4 & 4.5  & 3.7  & $\ \ \ \ \ 13\sigma$ & -     \\
	 4.914  & $4918.02\pm0.34\pm0.35$ & $\ \ $208.1  & 471  &  67  &  40  &  58  & 48.6 & 7.9  & 4.2  & $\ \ \ 8.2 \sigma$ & -     \\
	 4.946  & $4950.93\pm0.36\pm0.44$ & $\ \ $160.4  & 247  &  51  &  80  &  51  & 29.4 & 8.2  & 2.5  & $\ \ \ 5.0 \sigma$ & - \\
	\hline\hline
	\bottomrule
	\end{tabular}
	}
	% \label{table10-1}
\end{table*}

To increase signal yields,
a partial reconstruction method is employed for the $\ee\go\DDpipi$ process.
A $D^+$ meson is reconstructed via its high branching fraction (9.38\%) decay mode,
$D^+\go K^-\pi^+\pi^+$, and an additional $\pp$ pair is selected
from the remaining charged tracks. The recoil mass of the $\pp D^+$ system
is used to identify the $D^-$ meson. Unless explicitly mentioned,
the inclusion of charge conjugate modes is implied throughout the context.

Simulated data samples produced with a {\sc geant4}-based~\cite{geant4}
Monte Carlo (MC) package, which includes the geometric description of
the BESIII detector and the detector response, are used to determine
detection efficiencies and to estimate backgrounds.
The simulation models the beam energy spread and ISR in
the $e^+e^-$ annihilations with the generator {\sc kkmc}~\cite{kkmc}.

In order to estimate the potential background contributions,
inclusive MC samples generated at $\sqrt{s}=$ 4.230, 4.360, 4.420,
and 4.600 GeV are used.
The inclusive MC sample includes the production of open charm processes,
the ISR production of vector charmonium(-like) states,
and the continuum processes incorporated in {\sc kkmc}~\cite{kkmc}.
The known decay modes are modelled with {\sc evtgen}~\cite{evtgen}
using branching fractions taken from the Particle Data Group (PDG)~\cite{pdg},
and the remaining unknown charmonium decays are modelled
with {\sc lundcharm}~\cite{lundcharm}. Final state radiation~(FSR)
from charged final state particles is incorporated using
the {\sc photos} package~\cite{photos}.

For the optimization of the selection criteria and signal extraction, the following MC samples
are produced at each $\sqrt{s}$: $\ee\go\DststD$,
with $D_1(2420)^+\go D^+\pi^+\pi^-$, $\ee\go\pipipsipp$,
with $\psi(3770)\go D^+D^-$, where
$D_{1}(2420)^{+}D^{-}$ and $\pi^{+}\pi^{-}\psi(3770)$ are uniformly distributed in the phase space, and $\ee\go\PHSP$ where the
$\DDpipi$ events are uniformly distributed in the phase space
to represent the processes with unknown intermediate states.
For the $D_{1}(2420)^{+}\rightarrow D^{+}\pi^{+}\pi^{-}$ process, the $D^{+}\pi^{+}\pi^{-}$ events are also uniformly distributed in the phase space.

\section{\boldmath Event Selection}
\label{sec:evtsel}
Charged tracks detected in the MDC are required to be within a polar angle ($\theta$) range of $|\cos\theta|<0.93$, where $\theta$ is defined with respect to the z-axis, which is the symmetry axis of the MDC.
For charged tracks not originating from $K_S^0$ or $\Lambda$ decays, the distance of closest approach to the interaction point (IP)
must be less than 10\,cm
along the $z$-axis, $|V_{z}|$,
and less than 1\,cm
in the transverse plane, $|V_{xy}|$.
A charged track should have a good quality in the track fitting and be
within the angle coverage of the MDC, $|\cos\theta|<0.93$.  A good
charged track (excluding those from $K_S^0$ or $\Lambda$ decays)
is required to be within 1~cm of the $\ee$ annihilation interaction
point (IP) transverse to the beam line ($|V_{xy}|<1$~cm) and
within 10~cm of the IP along the beam axis ($|V_{z}|<10$~cm).

Particle identification~(PID) for charged tracks combines measurements of the energy deposited in the MDC~(d$E$/d$x$) and the flight time in the TOF to form likelihoods $\mathcal{L}(h)~(h=p,K,\pi)$ for each hadron $h$ hypothesis.
Tracks are identified as protons when the proton hypothesis has the greatest likelihood ($\mathcal{L}(p)>\mathcal{L}(K)$ and $\mathcal{L}(p)>\mathcal{L}(\pi)$), while charged kaons and pions are identified by comparing the likelihoods for the kaon and pion hypotheses, $\mathcal{L}(K)>\mathcal{L}(\pi)$ and $\mathcal{L}(\pi)>\mathcal{L}(K)$, respectively.

Particle identification~(PID) for charged tracks combines
measurements of the energy loss in the MDC~($dE/dx$) and the flight time
in the TOF. Likelihoods $\mathcal{L}(h)$~($h=K$, $\pi$, $p$) for each
hadron $h$ hypothesis are formed and each track is assigned to the particle
type corresponding to the hypothesis with the greatest likelihood.
A proton (or an anti-proton) is identified
if $\mathcal{L}(p)>\mathcal{L}(\pi)$ and $\mathcal{L}(p)>\mathcal{L}(K)$.
In order to suppress background from $\ee\go \LcLc$ and other possible charmed baryons,
events with proton or anti-proton tracks are rejected.
Charged kaons and pions are identified by comparing the likelihoods
for the kaon and pion hypotheses,  $\mathcal{L}(K)>\mathcal{L}(\pi)$ and $\mathcal{L}(\pi)>\mathcal{L}(K)$, respectively.
	
To reconstruct the $D^+$ meson, one $K^-$ and two $\pi^+$ candidate tracks
are selected. They are required to originate from a common vertex and
the quality of the vertex fit is required to satisfy $\chi^2_{VF}<100$.
All possible $\Kpipi$ combinations in the event which satisfy these criteria
are kept as $D^+$ candidates for further analysis. There are 1.1 $D^{+}$
candidates per event on average after $M(\Kpipi)$ and $RM(\Dpipi)$ requirements mentioned in the following paragraph.
For each $D^+$ candidate, a $\pp$ pair is selected from the charged tracks
not used in $D^+$ reconstruction (referred to as $\pi_d^+$ and $\pi_d^-$)
and the recoil mass of $\Dpipi$ ($RM(\Dpipi)$) is calculated to identify
the $D^-$ candidate.

\begin{figure*}[htbp]
	\centering
	\includegraphics[width=2.1in]{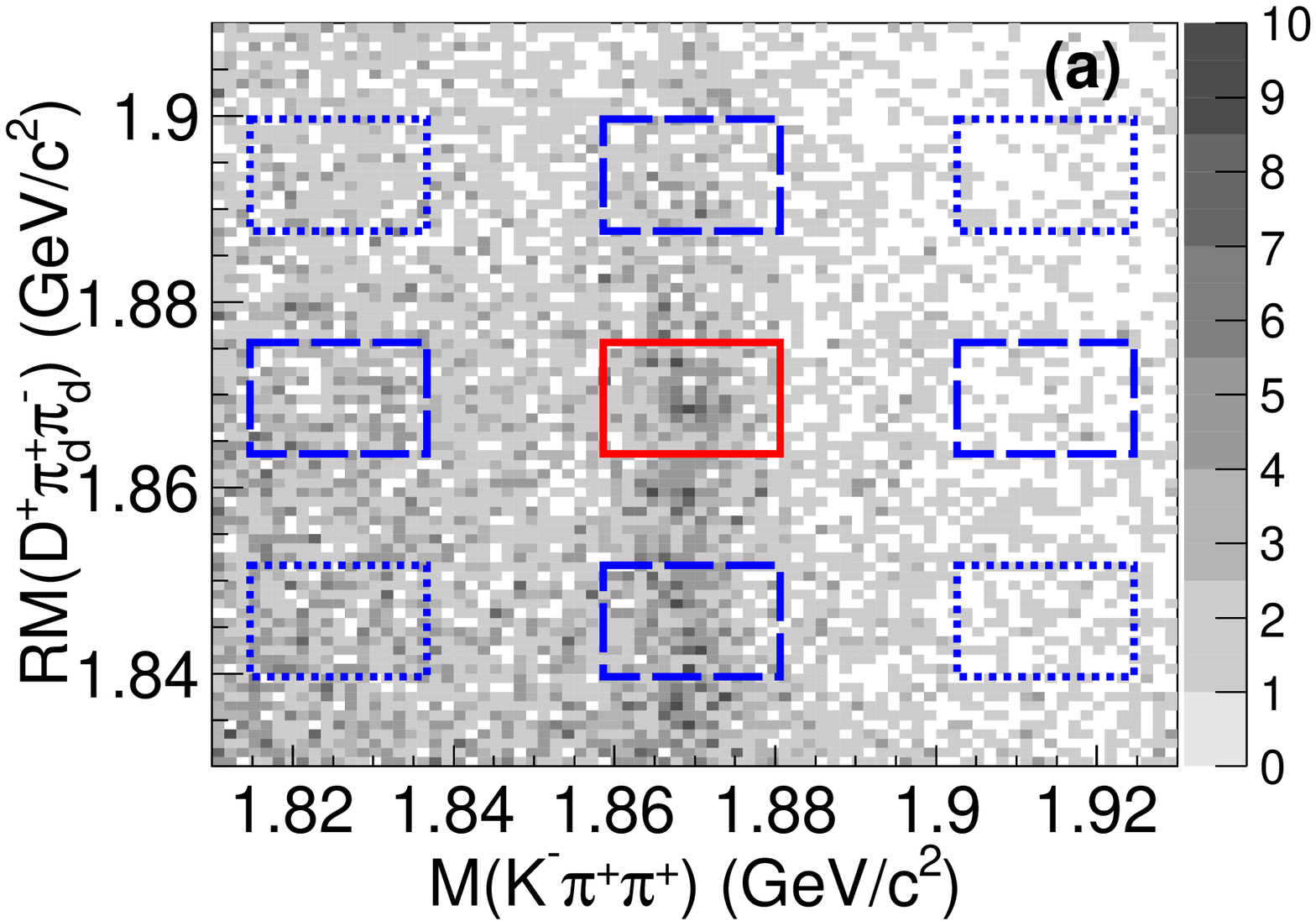}
	\includegraphics[width=2.1in]{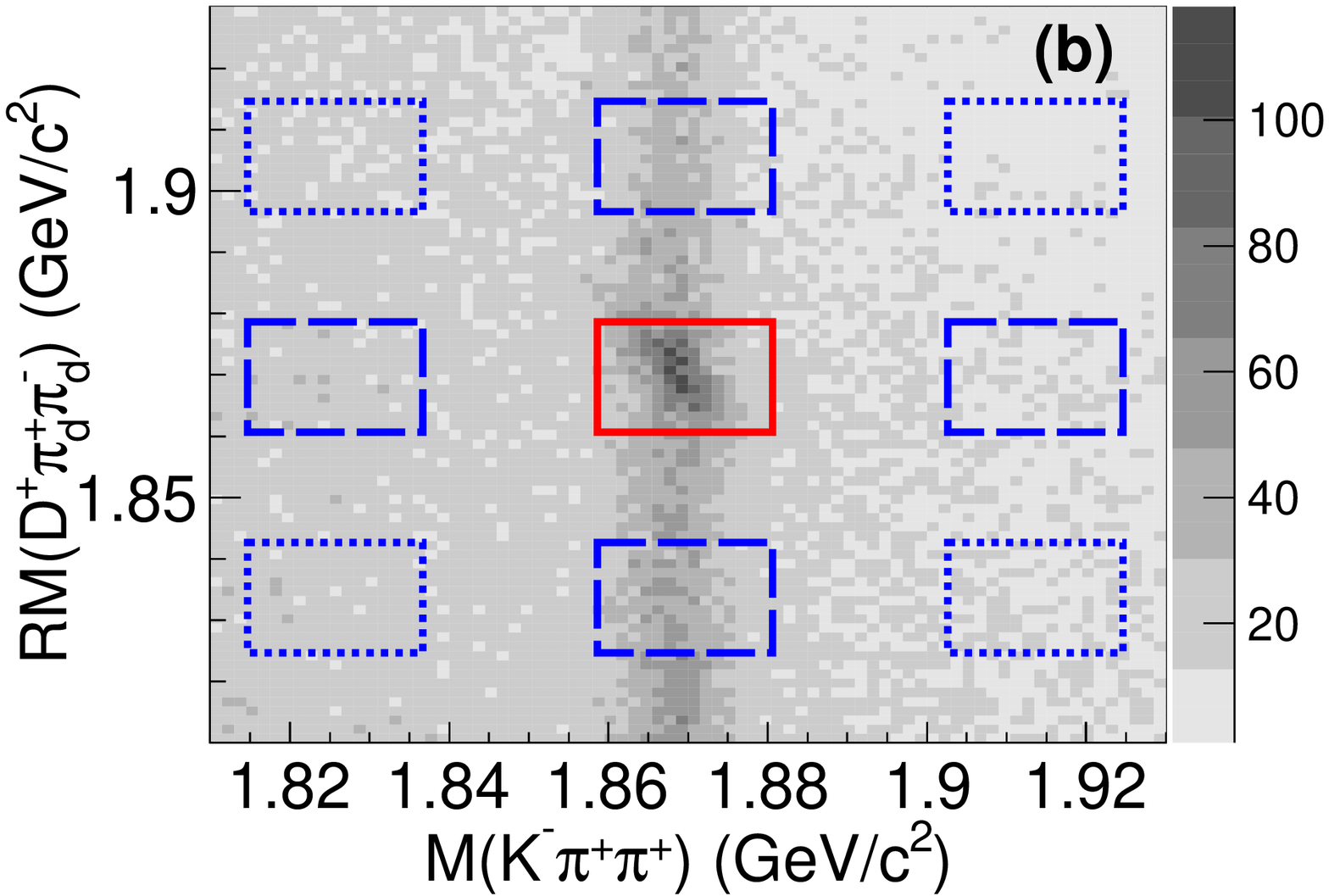}
	\includegraphics[width=2.1in]{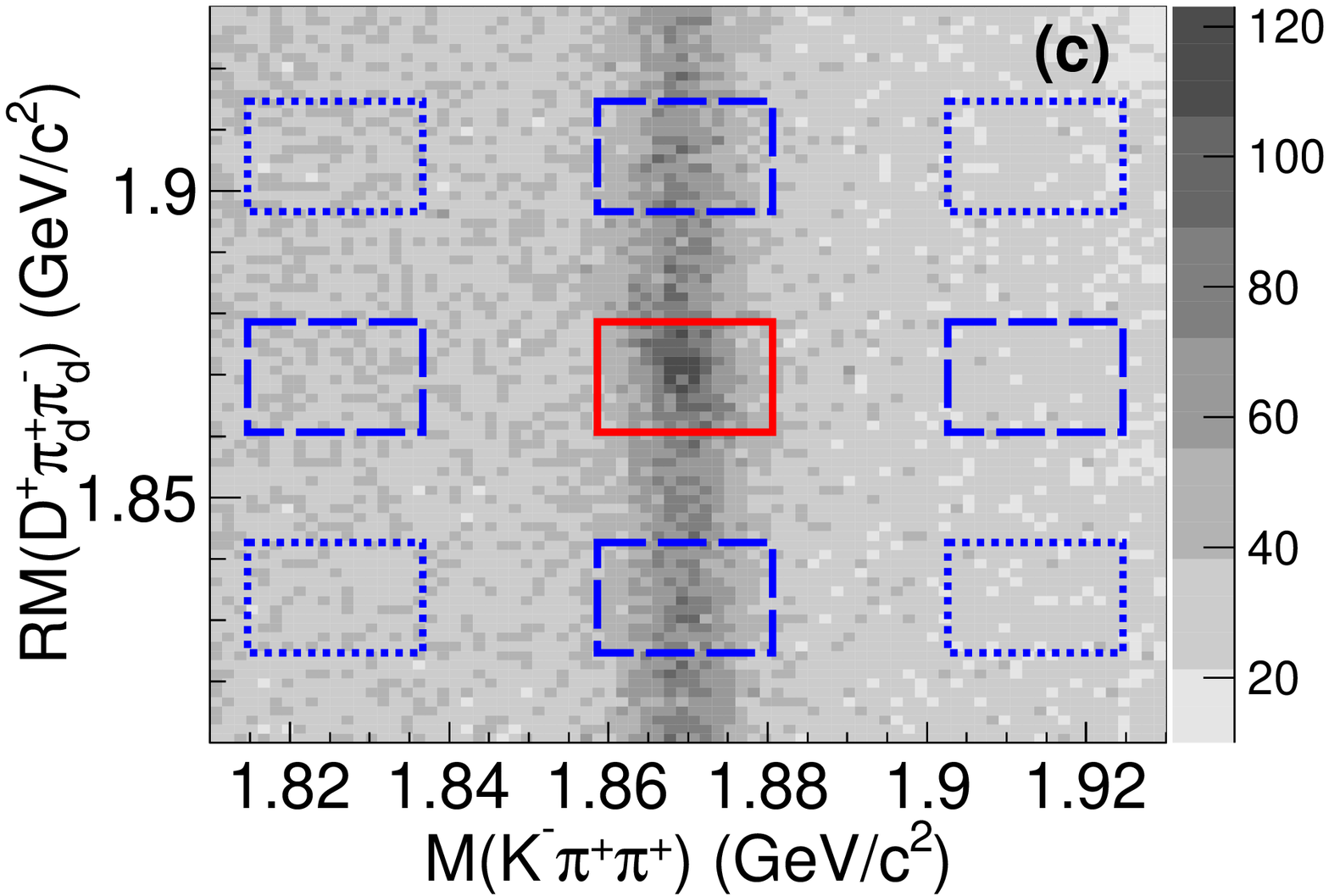}
\caption{Distributions of $RM(\Dpipi)$ versus $M(\Kpipi)$ for data samples at $\sqrt{s}$ =
4.230~(a), 4.420~(b), and 4.680~(c)~GeV.
The red solid box shows the signal region, the blue dashed boxes
the sideband regions with one real $D^+$ or $D^-$ candidate, and the blue dotted boxes
the sideband regions with fake $D^+$ and $D^-$ candidates.
The indices of the boxes from top to bottom and left to right are ($-$1, 1), (0, 1), (1, 1), ($-$1, 0), (0, 0), (1, 0), ($-$1, $-$1), (0, $-$1), and (1, $-$1), respectively.
The region with index (0, 0) is the signal region, while the others are the sideband regions (color version online).
}
	\label{fig:2D}
\end{figure*}

Figure~\ref{fig:2D} shows $RM(\Dpipi)$ versus the invariant mass of the $\Kpipi$
($M(\Kpipi)$) for data samples at $\sqrt{s}=$ 4.230, 4.420, and 4.680 GeV. Clear $D^-$ and $D^+$
signal peaks can be seen in the $RM(\Dpipi)$ and $M(\Kpipi)$ distributions, respectively.
The $\DDpipi$ signal region is defined as
$|M(\Kpipi)-m_{D^+}|<d_M$ and
$|RM(\Dpipi)-m_{D^-}|<d_{RM}$, and
the sideband regions as
$-5d_M<M(\Kpipi)-m_{D^+}<-3d_M$ or
$3d_M<M(\Kpipi)-m_{D^+}<5d_M$
 , and $-5d_{RM}<RM(\Dpipi)-m_{D^{-}}<-3d_{RM}$ or
 $3d_{RM}<RM(\Dpipi)-m_{D^{-}}<5d_{RM}$
, where $m_{D^{\pm}}=1.86966$ GeV/$c^{2}$ is the known $D^{\pm}$ mass \cite{pdg}.
 The signal and sideband regions are indicated in Fig.~\ref{fig:2D}.
 A linear mass or recoil mass dependence is assumed in estimating
 the background level in the signal region.
The widths of the window are $d_M=11$~MeV/$c^2$ for all the data samples,
and $d_{RM}=6$~MeV/$c^2$ for data samples with $\sqrt{s}$ smaller than 4.310~GeV, and
$d_{RM}=9$~MeV/$c^2$ for data samples with $\sqrt{s}$ greater than or equal to 4.310~GeV. Each
sideband has the same width as that of the signal region.

\begin{figure*}[htbp]
	\centering
	\includegraphics[width=2.1in]{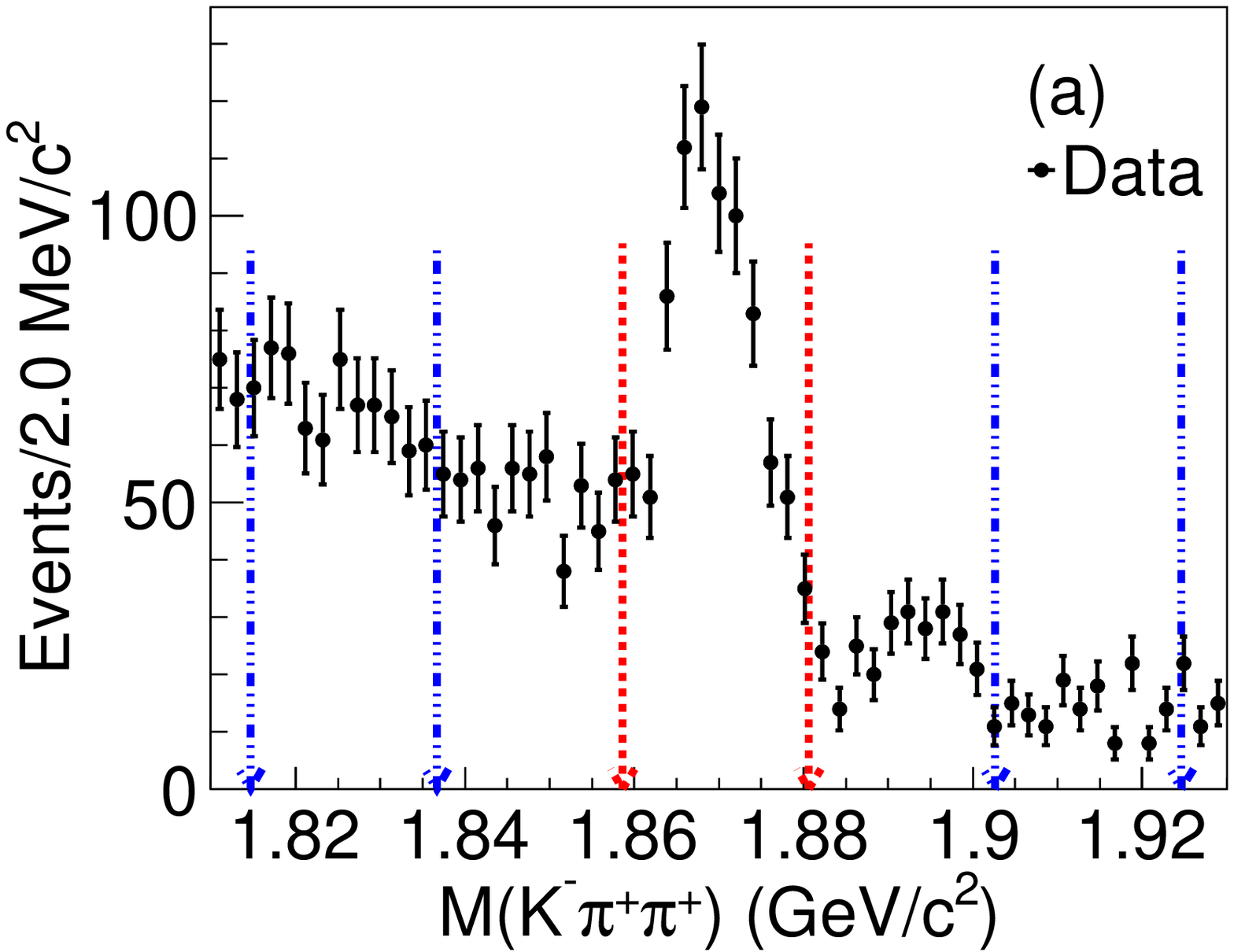}
	\includegraphics[width=2.1in]{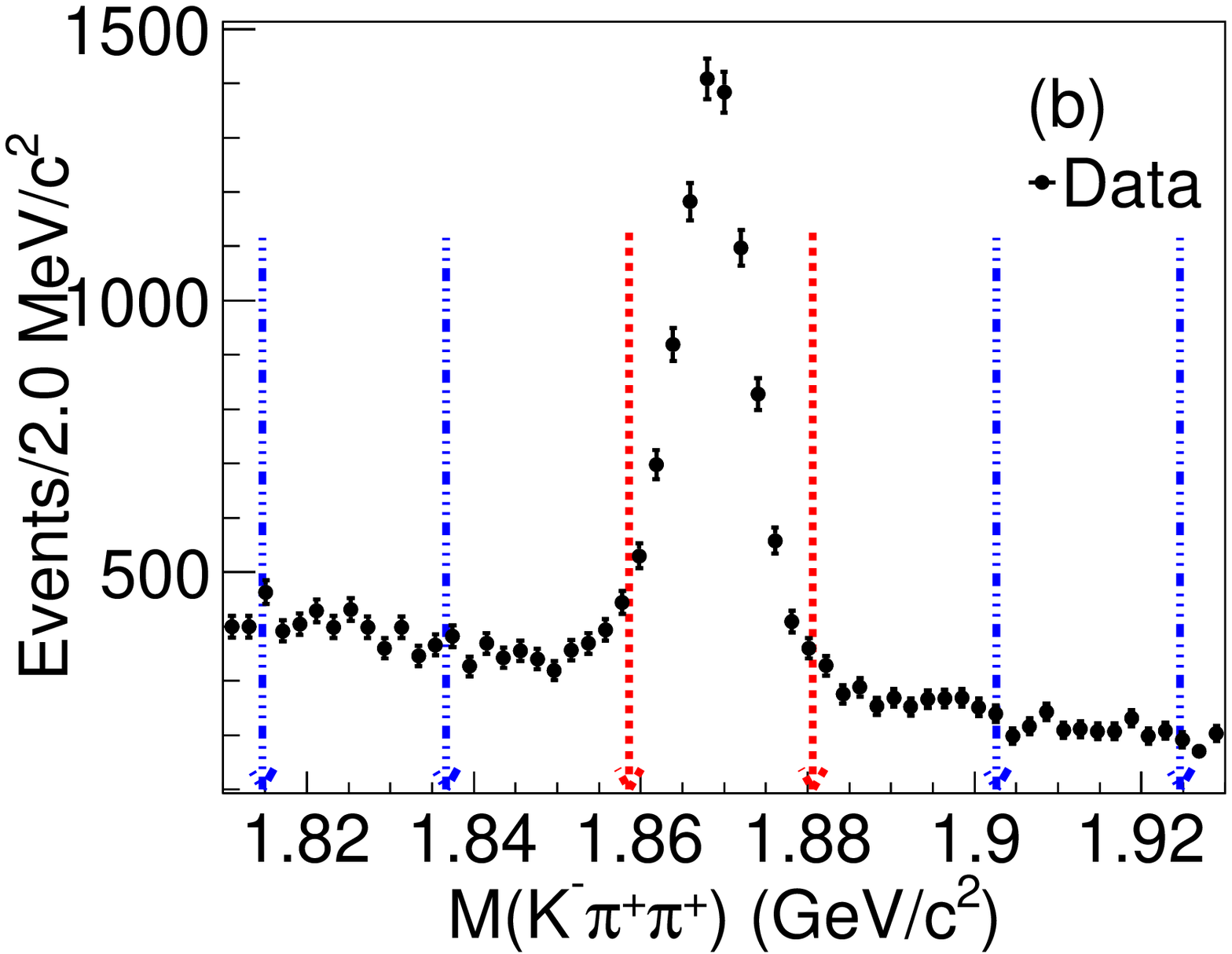}
	\includegraphics[width=2.1in]{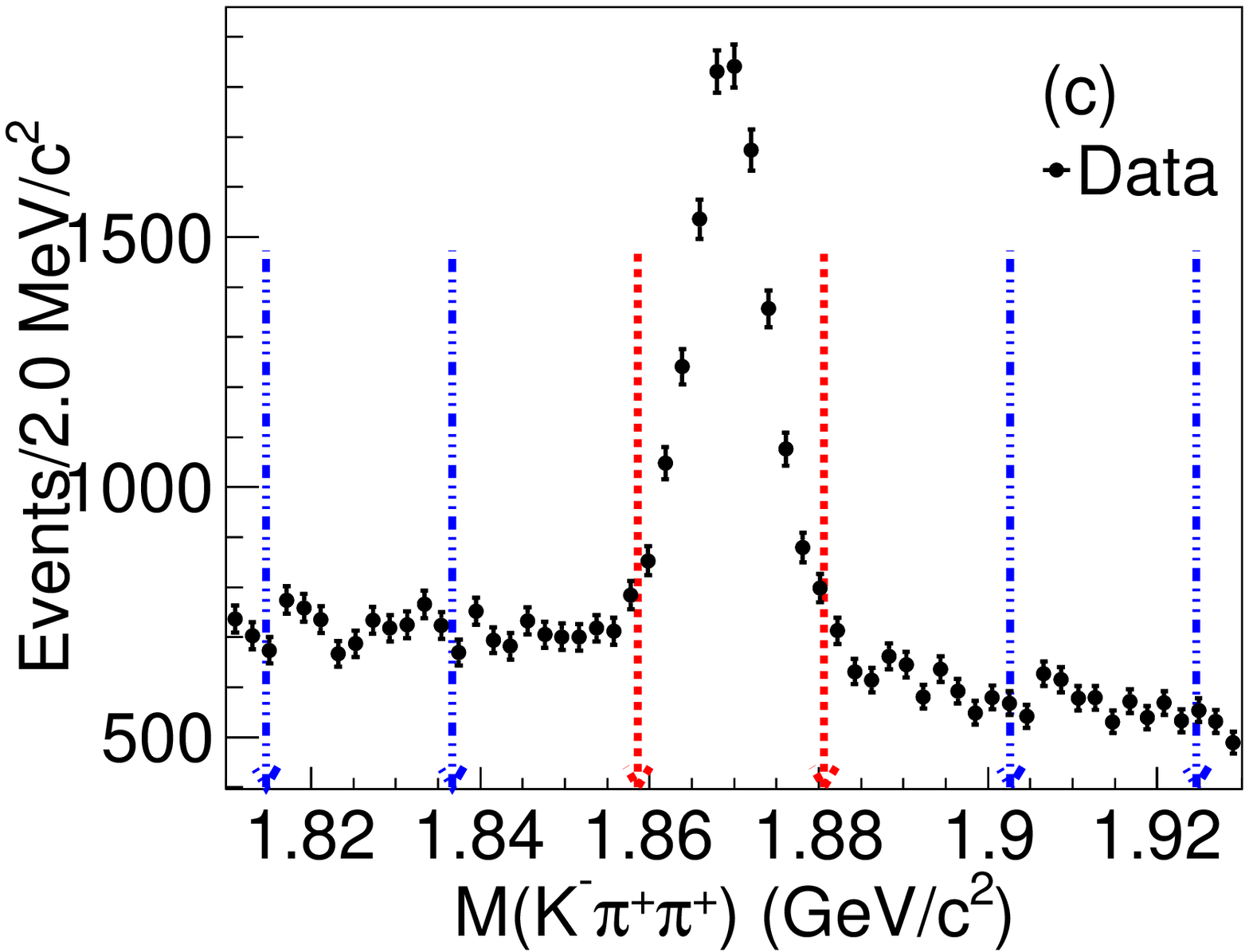}
\caption{The $\Kpipi$ invariant mass distributions for data samples at $\sqrt{s}=$ 4.230~(a),
4.420~(b), and 4.680~(c)~GeV.
The black dots with error bars are data, the regions between the two red
dashed arrows are $D^+$ signal regions and those between blue dash-dotted arrows
are sideband regions (color version online).}
	\label{fig:Mkpipi}
\end{figure*}

After requiring $|RM(\Dpipi)-m_{D^-}|<d_{RM}$, the $M(\Kpipi)$ distributions are shown in
Fig.~\ref{fig:Mkpipi} for  data samples at $\sqrt{s}=$ 4.230, 4.420, and 4.680 GeV
as examples. In the following analysis, the $\Kpipi$ combination in the signal region is constrained
to the known $D^+$ mass, $m_{D^+}$, with a kinematic fit to improve
its momentum resolution, and those in the sideband regions are constrained
to the central value of the corresponding sideband region.

\begin{figure*}[htbp]
    \centering
    \includegraphics[width=2.1in]{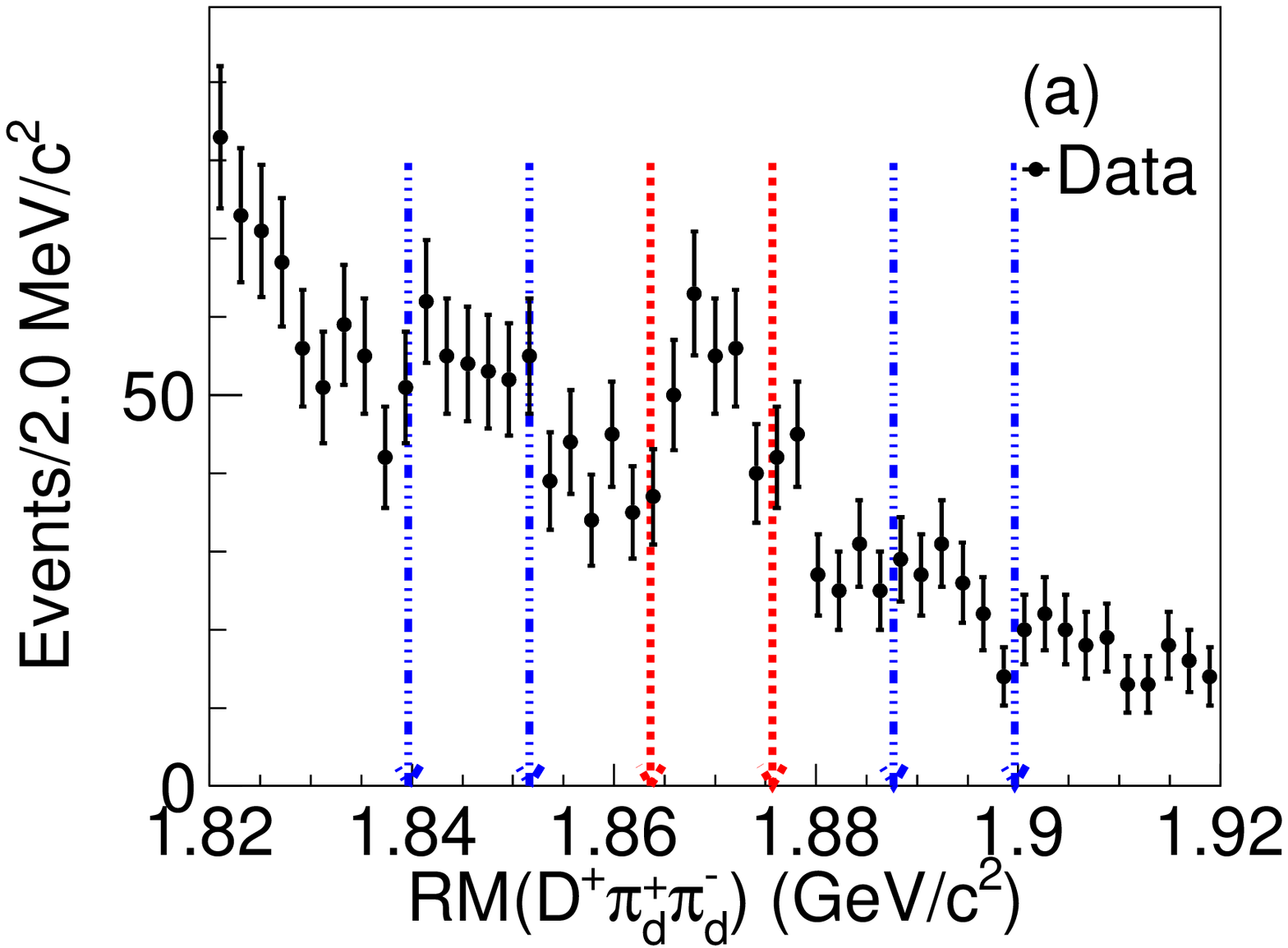}
    \includegraphics[width=2.1in]{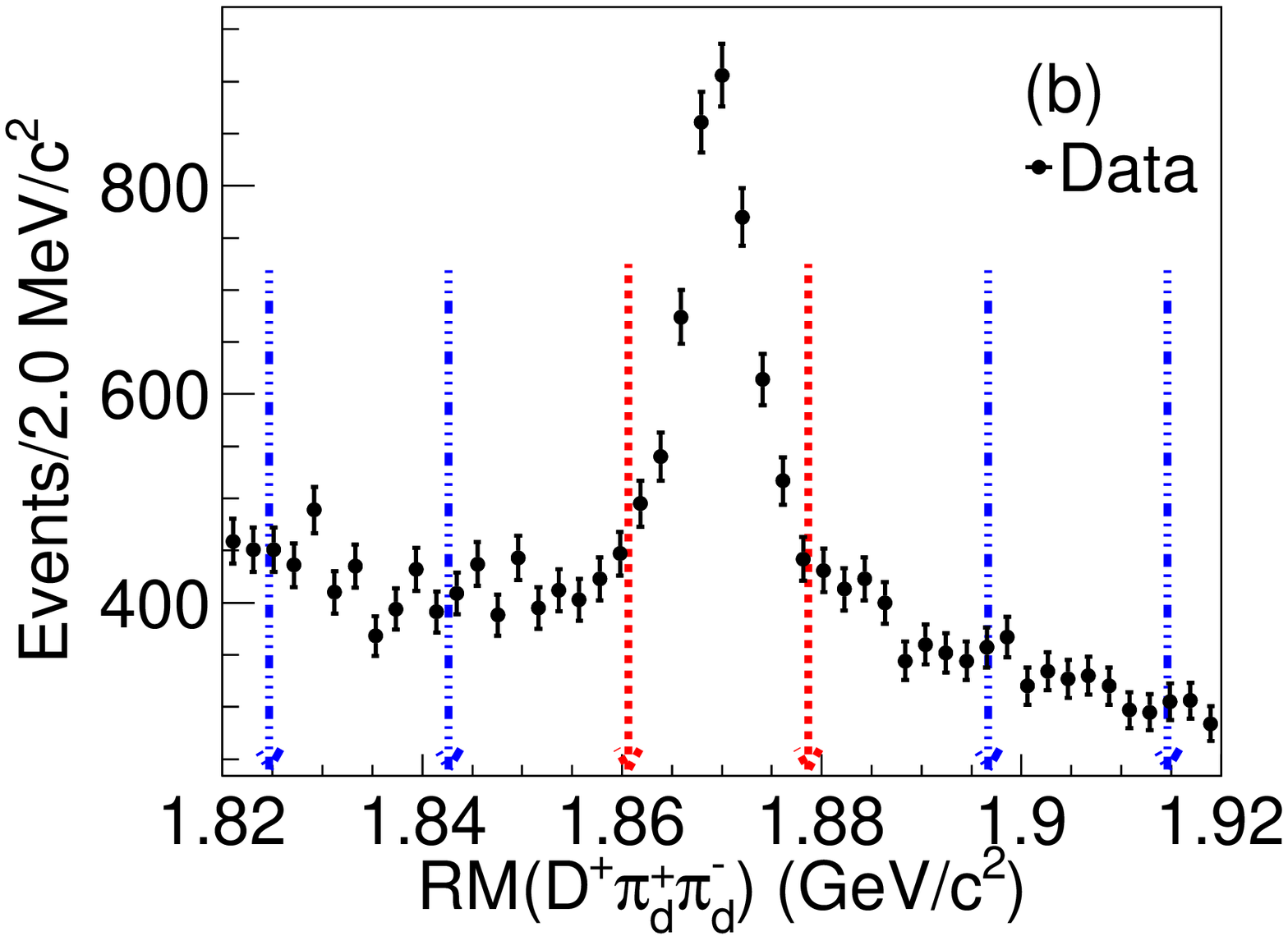}
    \includegraphics[width=2.1in]{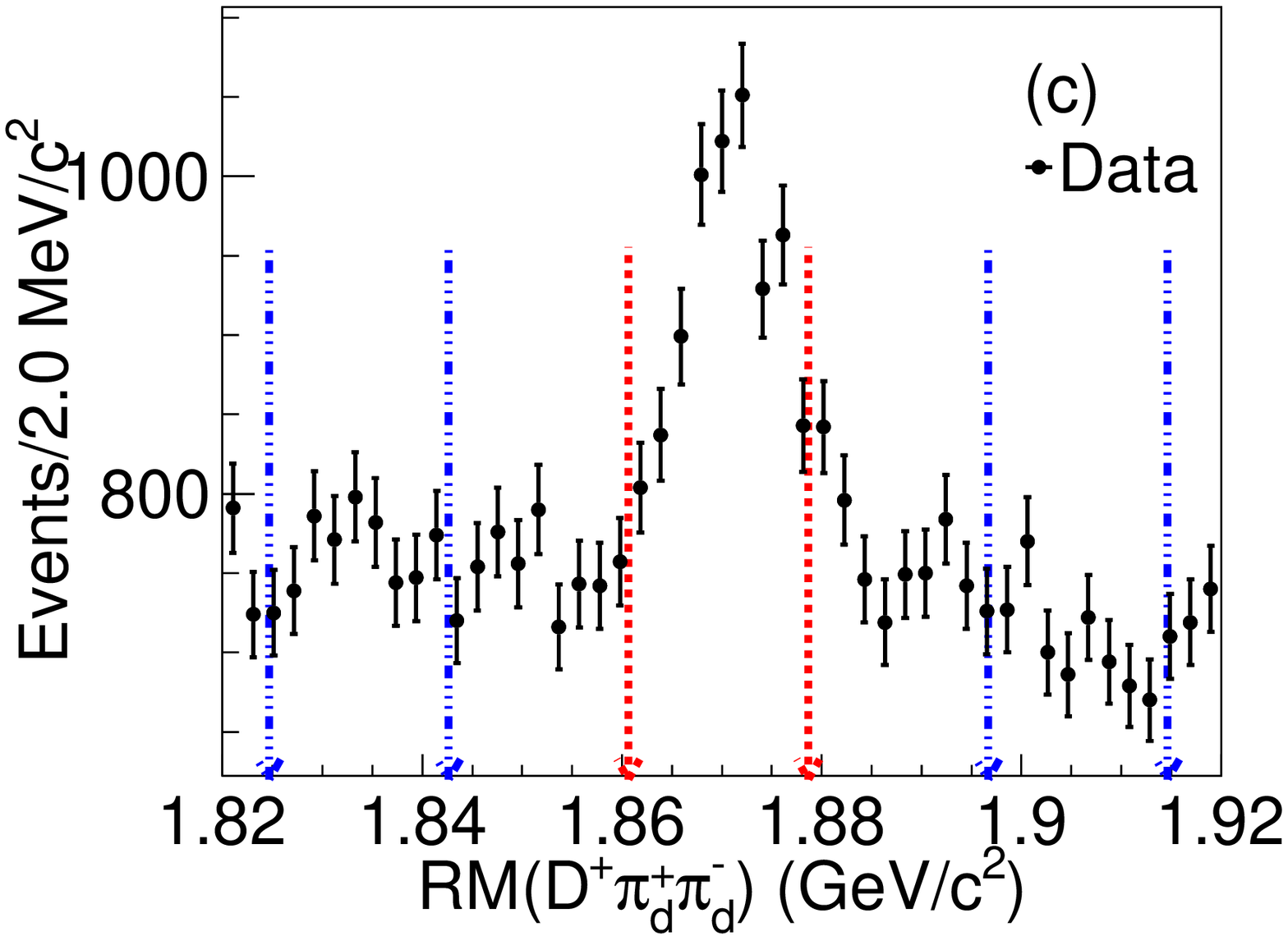}
\caption{Distributions of $RM(\Dpipi)$ in the $M(\Kpipi)$ signal
region for data samples at $\sqrt{s}=$ 4.230~(a), 4.420~(b), and 4.680~(c)~GeV.
The black dots with error bars are data, the regions between the two red dashed arrows
are $D^-$ signal regions and those between blue dash-dotted arrows are sideband regions (color version online).}
\label{fig:RMDpipi_before}
\end{figure*}

Figure~\ref{fig:RMDpipi_before} shows the $RM(\Dpipi)$ distributions
after requiring $|M(\Kpipi)-m_{D^+}|<d_M$  for data samples at $\sqrt{s}=$
4.230, 4.420, and 4.680~GeV.
Clear $D^-$ signal peaks are visible in all data samples.
 The $D^-$ signal and sideband regions are indicated by
 the arrows.

\begin{figure*}[htbp]
	\centering
	\includegraphics[width=2.1in]{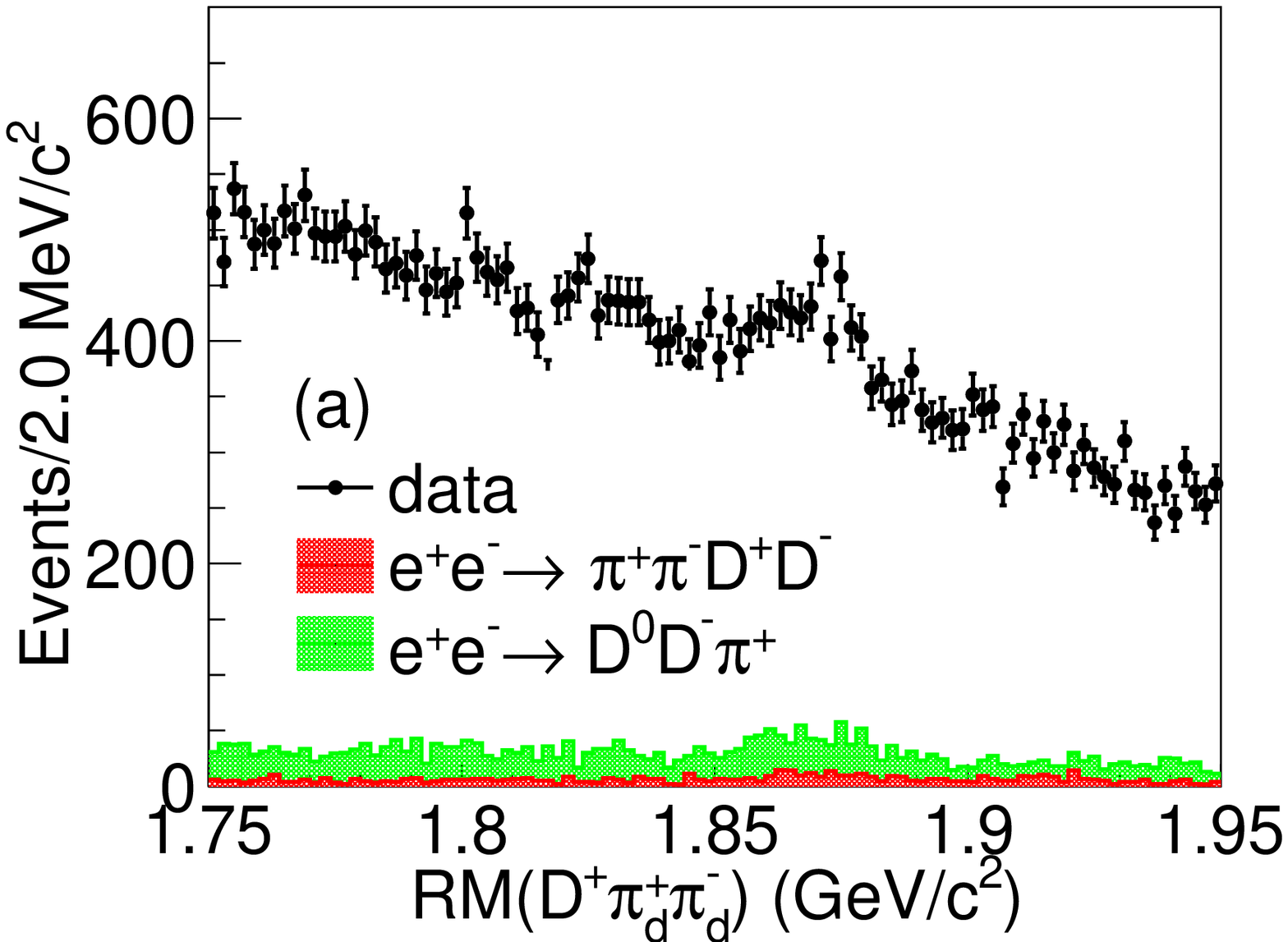}
	\includegraphics[width=2.1in]{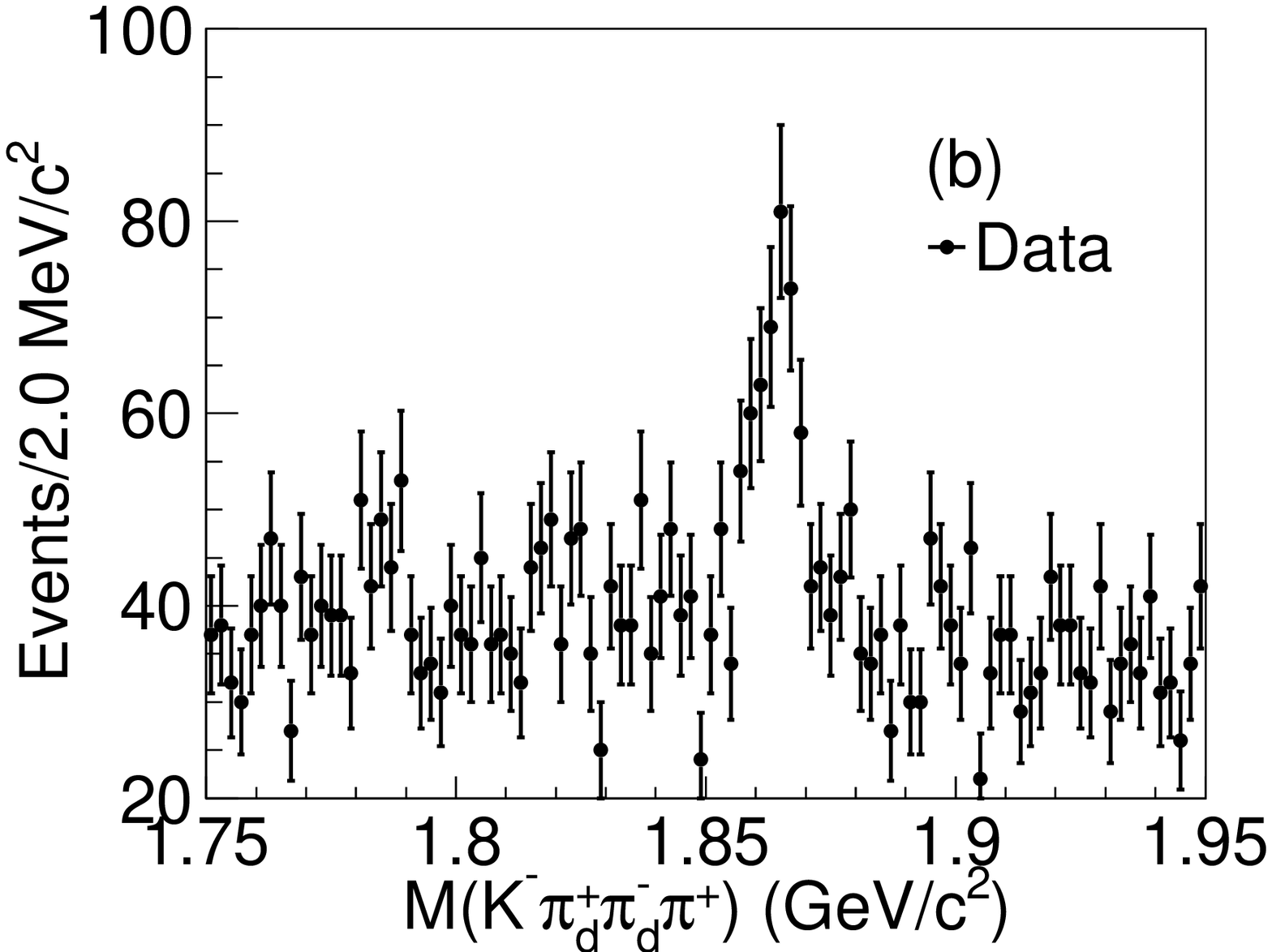}
	\includegraphics[width=2.1in]{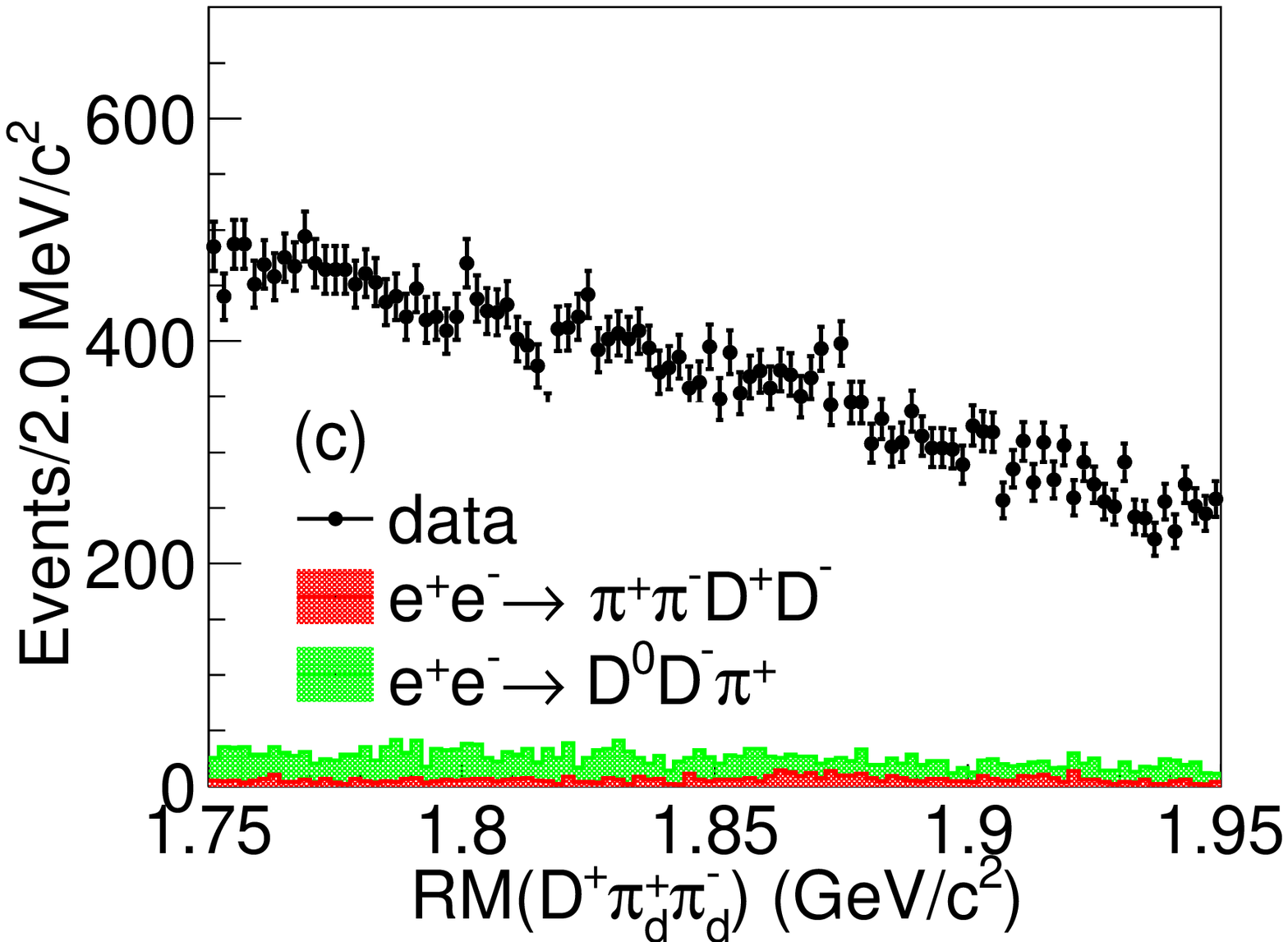}
\caption{The $RM(\Dpipi)$ distribution for combinations in the $M(\Kpipi)$ sideband
before (a) and after (c) the $D^0$ veto at $\sqrt{s}=$ 4.420~GeV.
The $M(K^-\pi_d^+\pi_d^-\pi^+)$ distribution
in the $M(\Kpipi)$ signal region is shown in the middle plot (b). The black dots with error bars are data and
the red and green histograms are MC simulations of $\ee\go \PHSP$ and
$\ee\go \DDpi$ processes, respectively, with inclusive decays of both $D$ mesons (color version online). The normalizations of $\ee\go \PHSP$ and $\ee\go \DDpi$ processes are accroding to cross sections of total $\ee\go\DDpipi$ process and $\ee\go\DDpi$ process measured from data samples, respectively.}
	\label{fig:DDpi}
\end{figure*}

The $\ee\go \DDpi$ process produces a peaking background in the $RM(\Dpipi)$ distribution
as shown in Fig.~\ref{fig:DDpi}(a). The peaking background may come from $\ee\go D^0D^-\pi^+$,
with $D^0\go K^-\pi_d^+\pi_d^-\pi^+$, where a directly produced $\pi^{+}$, together with $\pi^{+}$ and $K^{-}$
from $D^{0}$, forms the tagged $D^{+}$. Figure~\ref{fig:DDpi}(b) shows the
$M(K^-\pi_d^+\pi_d^-\pi^+)$\footnote{Here, $\pi^{+}$ could be either of the charged pions in the decay $D^{+}\go\Kpipi$.} distribution,
where a clear $D^0$ peak is seen.
We require $|M(K^-\pi_d^+\pi_d^-\pi^+)-m_{D^0}|>0.01$~GeV/$c^2$ to veto these
$D^0$ background contributions, where $m_{D^{0}}=$ 1.86484 GeV/$c^{2}$ \cite{pdg}. The value of $0.01$~GeV/$c^2$ corresponds to twice the resolution of $M(K^-\pi_d^+\pi_d^-\pi^+)$, which is 0.0045 GeV/$c^2$. The effectiveness of this veto can be seen in Fig.~\ref{fig:DDpi}(c). The number of $\ee\go\DDpipi$ events from fits to the $RM(\Dpipi)$ distributions in the $M(\Kpipi)$ sideband region before and after the veto are 614 $\pm$ 92 and 216 $\pm$ 72, respectively.

 After applying all the above selection
criteria, we compare distributions for events in the $D^+$ and $D^-$ signal region
($S$ sample) and sideband regions ($B$ sample) to further suppress
non-$\DDpipi$ background. The $B$ sample is defined as
\begin{equation}
	\begin{split}
		B = f_1\cdot (B_{-1,0}+B_{1,0})+f_2\cdot (B_{0,-1}+B_{0,1}) - \\f_3\cdot (B_{-1,-1}+B_{1,-1}+B_{-1,1}+B_{1,1}),
	\end{split}
\end{equation}
where $B_{i,j}$ is the sideband region defined in
Fig.~\ref{fig:2D}, $f_1=0.5$, $f_2=0.5$, and $f_3=f_1 f_2=0.25$ are the
normalization factors assuming a linear mass dependence in the background
distributions. In order to improve the momentum resolutions
of the final state particles, $D^+$ and $D^-$ mass constraints
and a total four-momentum conservation constraint to that of the
initial $\ee$ system are applied. For events in the $D^+$ or $D^-$
sidebands, the masses of the $D^+$ and $D^-$ combinations are
constrained to the central values of the corresponding sideband region.
	
\begin{figure*}[htbp]
	\centering
	\includegraphics[width=2.1in]{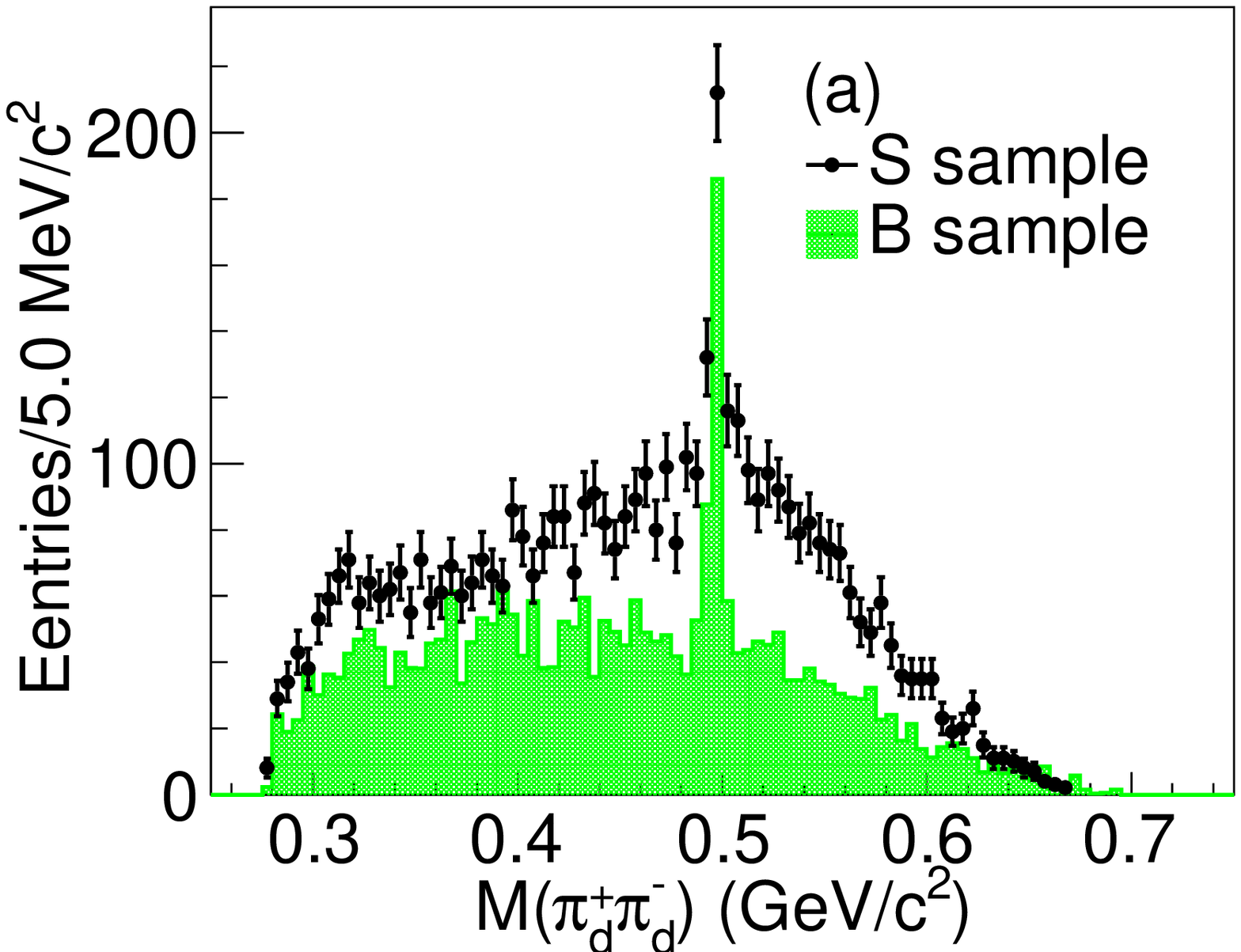}
	\includegraphics[width=2.1in]{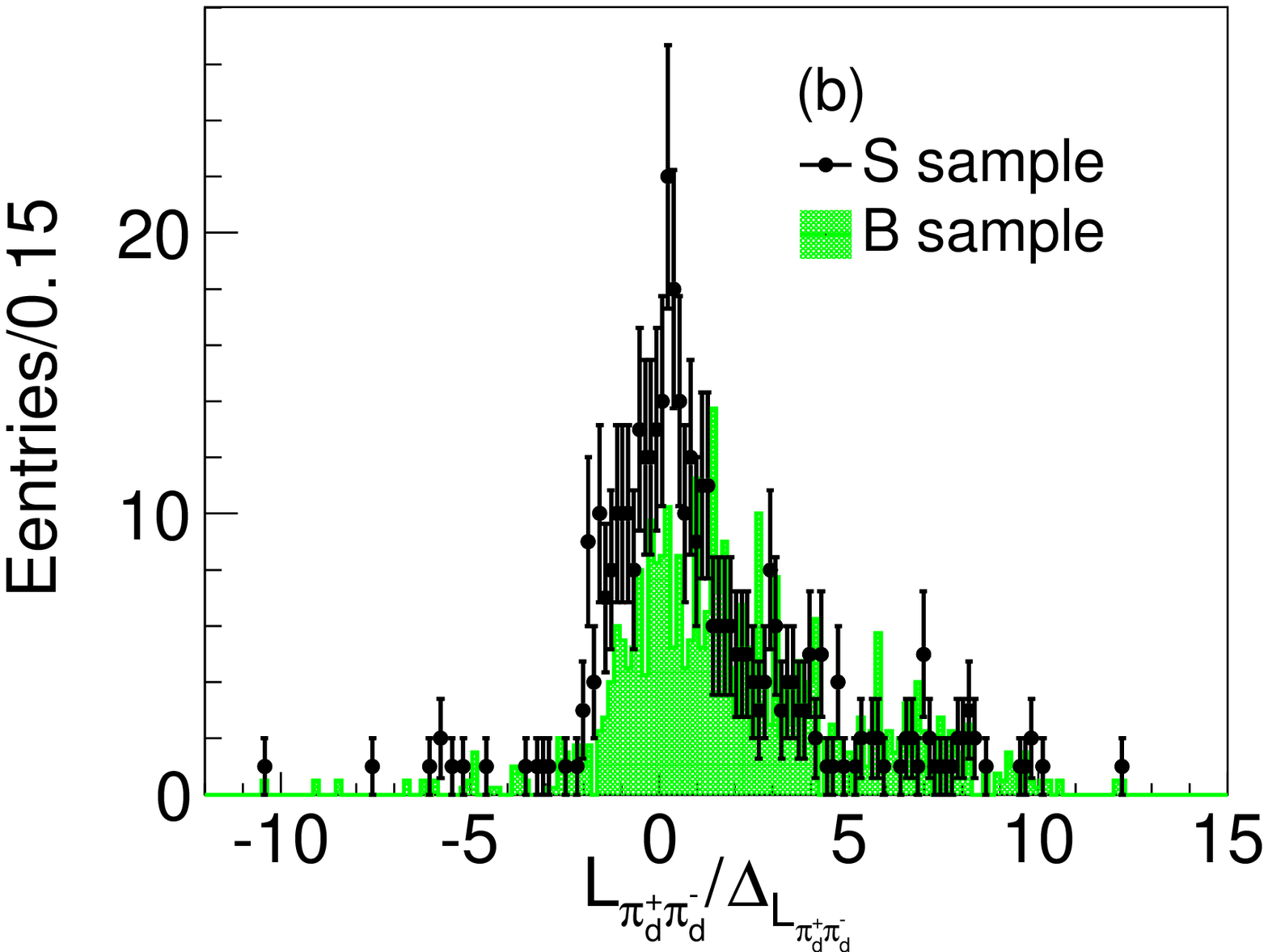}
	\includegraphics[width=2.1in]{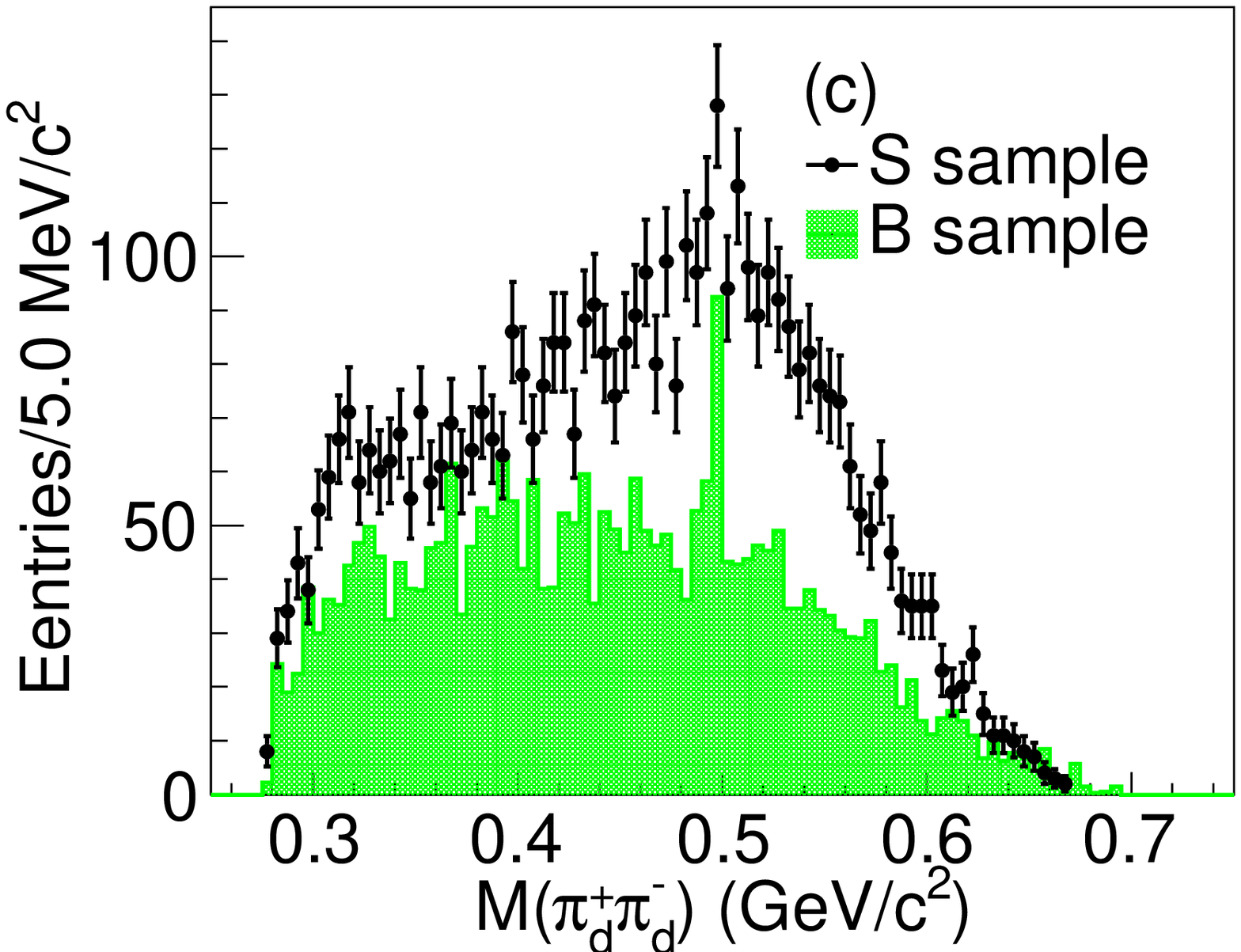}
\caption{Distributions of $M(\pipi)$ before (a) and after (c)
the $|L_{\pipi}/\Delta_{L_{\pipi}}|<2$ requirement,
and the $L_{\pipi}/\Delta_{L_{\pipi}}$ distribution (b) at $\sqrt{s}=$ 4.420~GeV.
The black dots with error bars stand for the $S$ sample and the green shaded histograms
for the $B$ sample (color version online).}
	\label{fig:KS}
\end{figure*}

The invariant mass distribution of the $\pipi$ pair
is shown in Fig.~\ref{fig:KS}(a), where clear $K_{S}^0$ peaks can
be seen in both the $S$ and $B$ samples. In order to veto the $K_{S}^0\go \pipi$
background, a secondary vertex fit is performed on the $\pipi$ pair.
The decay length $L_{\pipi}$ divided by its uncertainty $\Delta_{L_{\pipi}}$
of the combinations with $\pipi$ invariant mass between 491.0 and
503.5~MeV/$c^2$ is shown in Fig.~\ref{fig:KS}(b).
By requiring $|L_{\pipi}/\Delta_{L_{\pipi}}|<2$ the $K_{S}^0$
background is suppressed significantly as shown in Fig.~\ref{fig:KS}(c).
	
\begin{figure*}[htbp]
	\centering
	\includegraphics[width=2.1in]{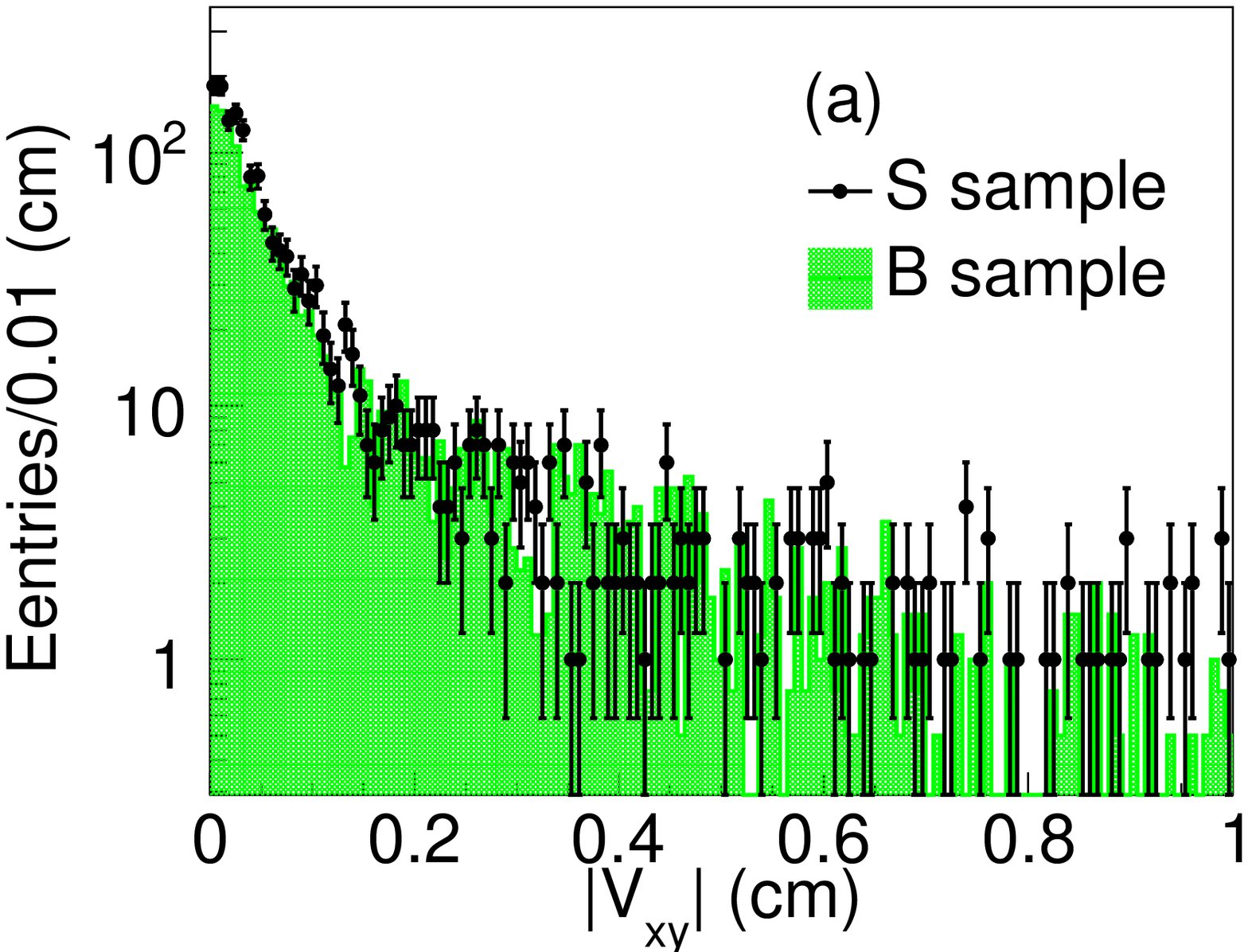}
	\includegraphics[width=2.1in]{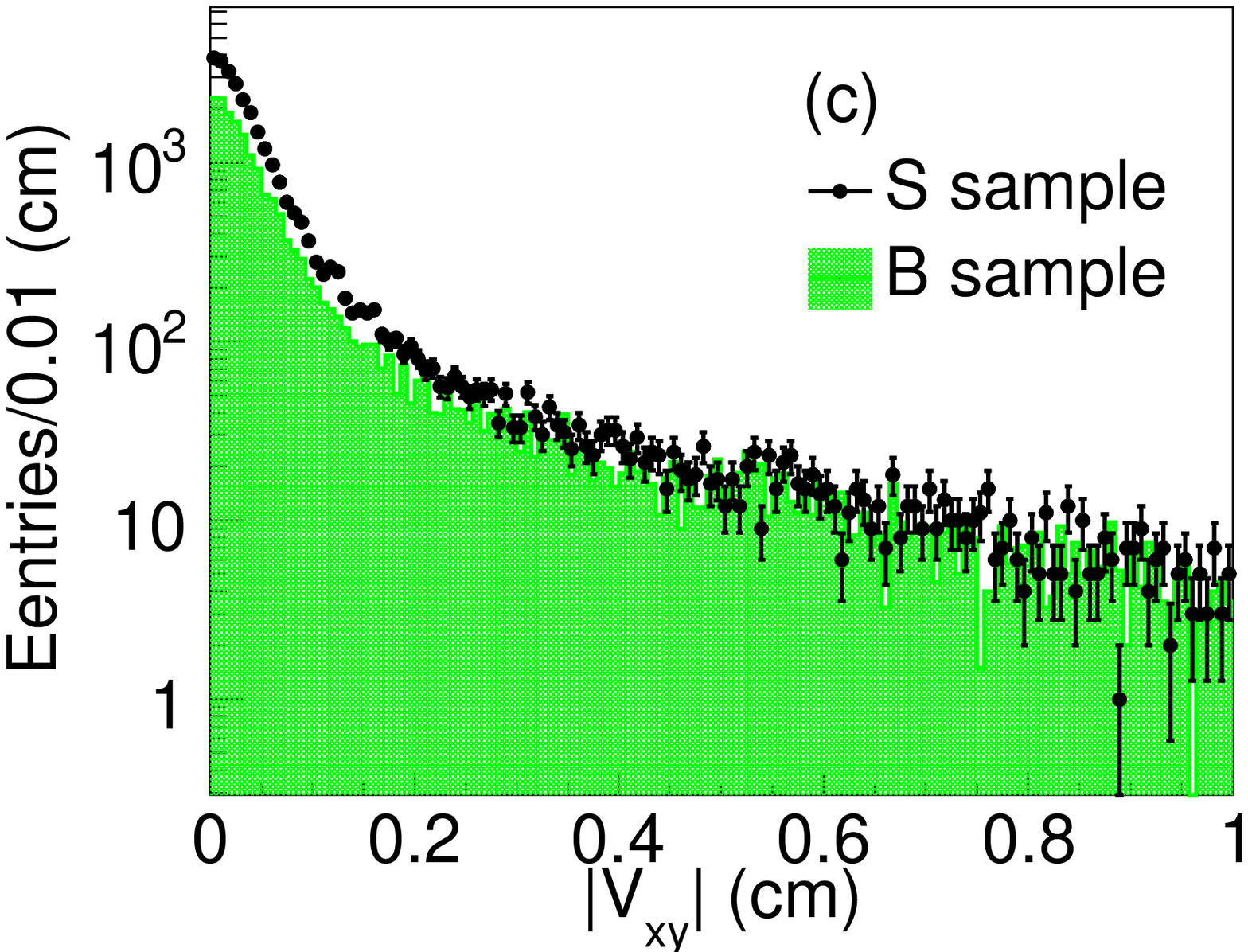}
	\includegraphics[width=2.1in]{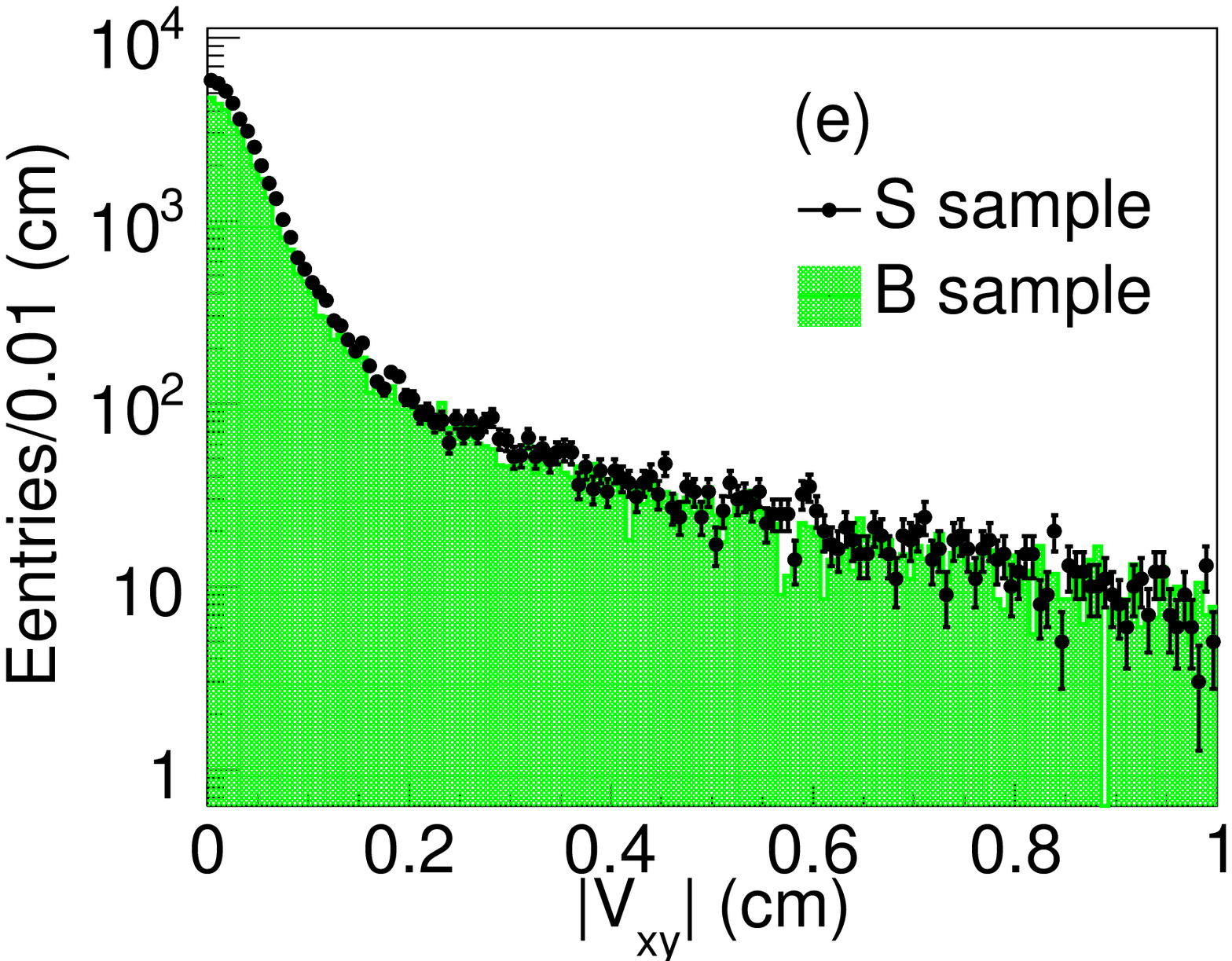}
	\includegraphics[width=2.1in]{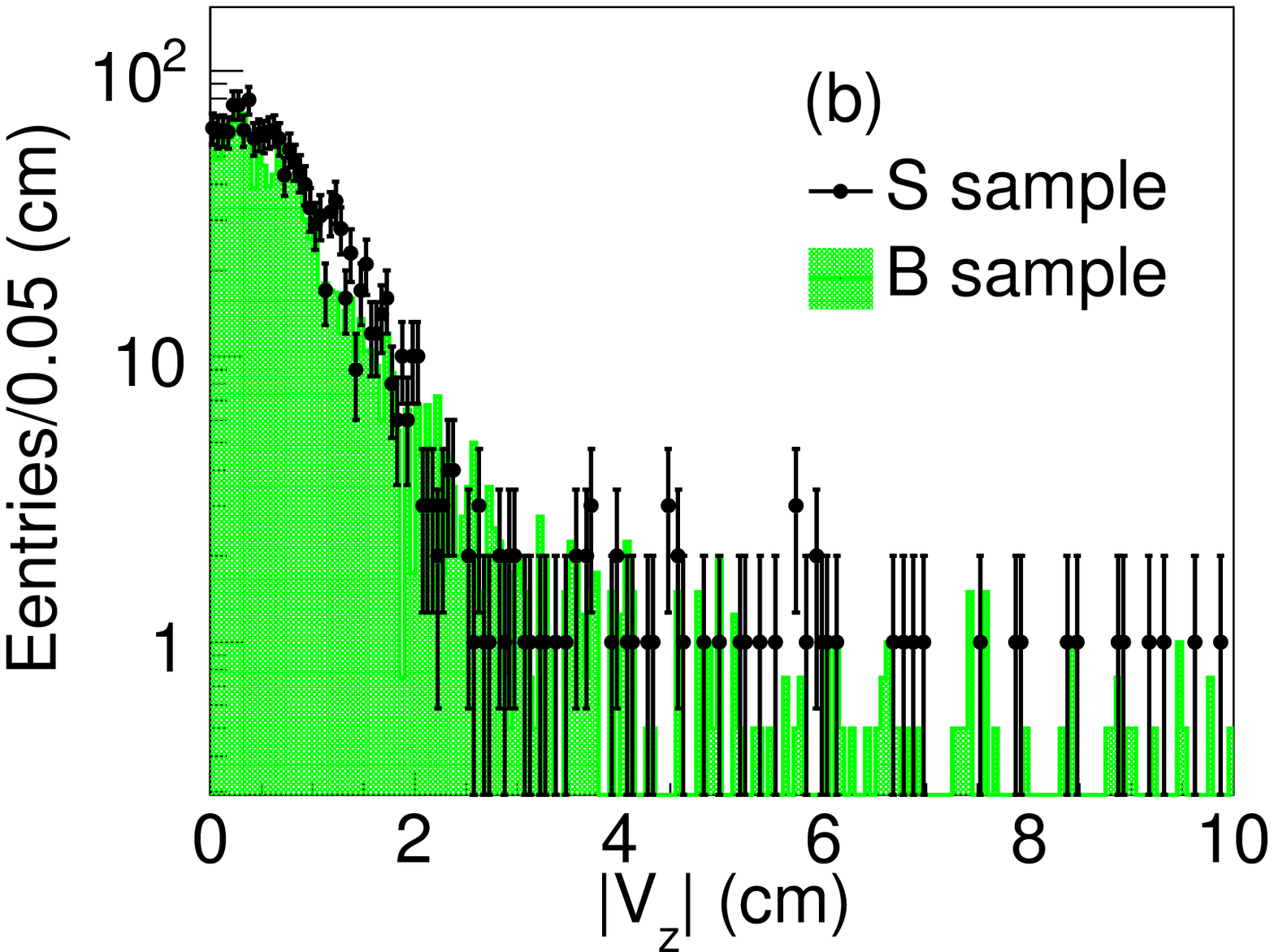}
	\includegraphics[width=2.1in]{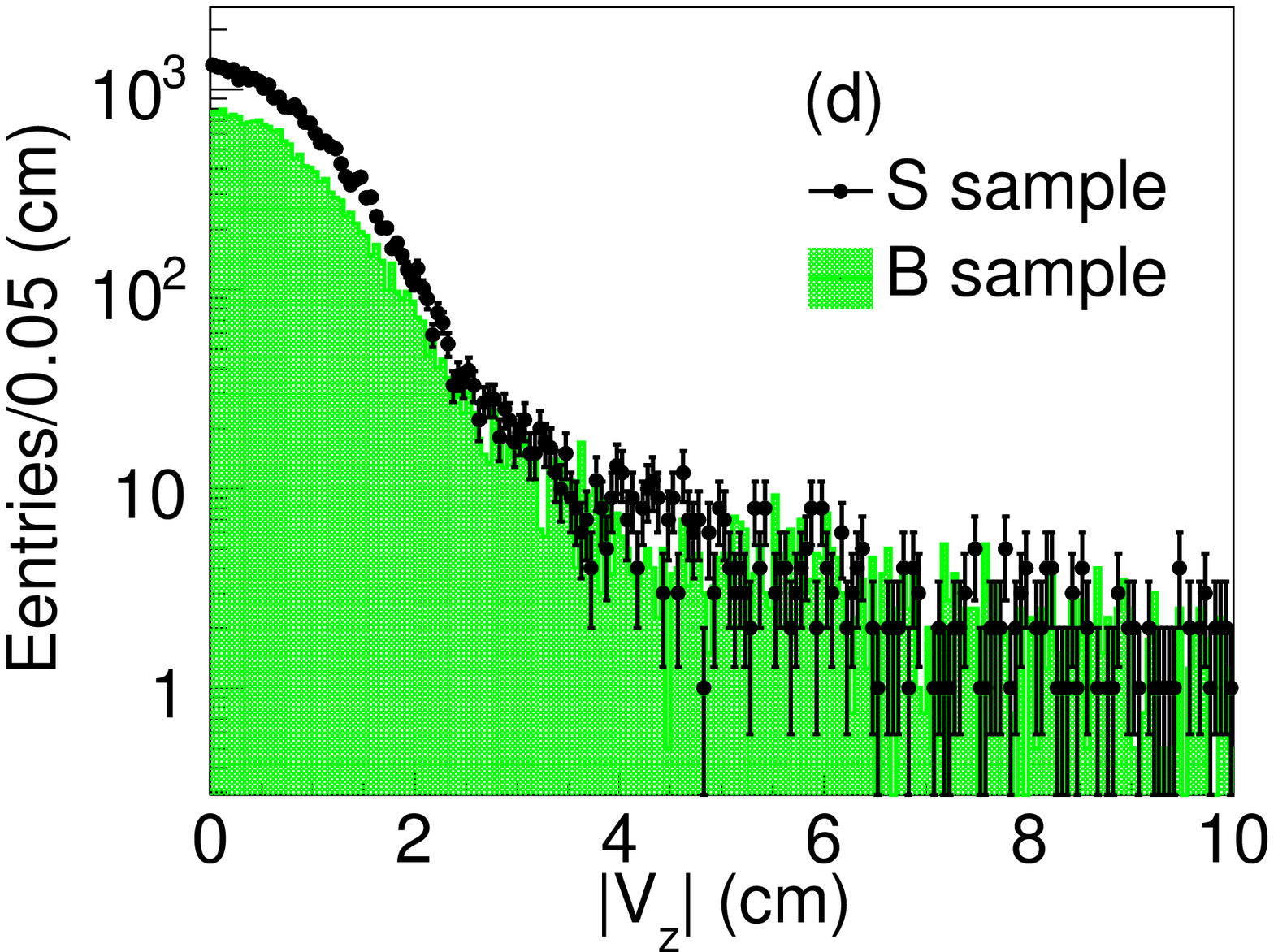}
	\includegraphics[width=2.1in]{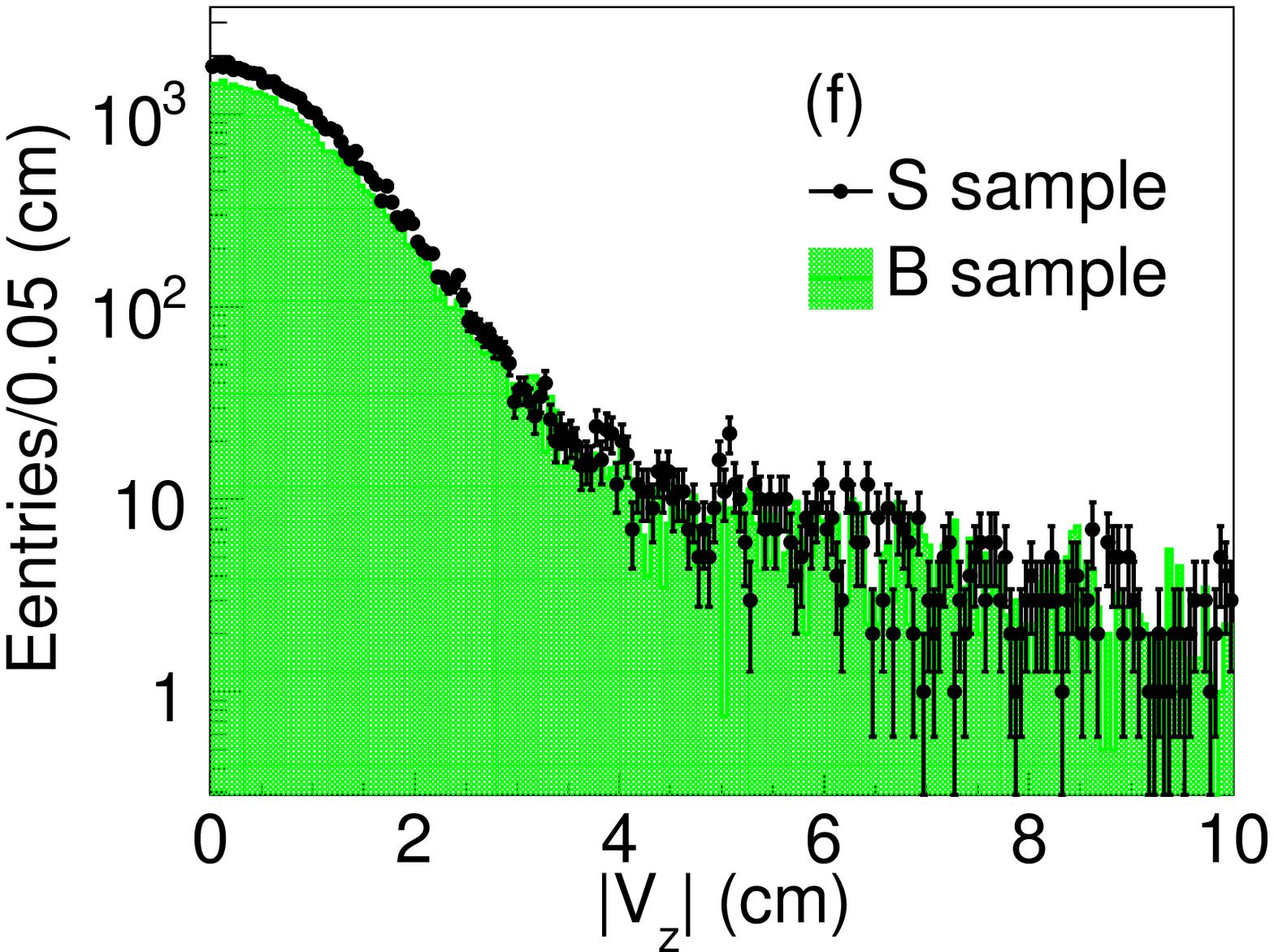}
\caption{Distributions of $|V_{xy}|$ for data samples at $\sqrt{s}=$ 4.230~(a),
4.420~(c), and 4.680~(e)~GeV, and those of $|V_{z}|$ for data samples at $\sqrt{s}=$ 4.230~(b),
4.420~(d), and 4.680~(f)~GeV.
The black dots with error bars correspond to the $S$ sample and the shaded histograms
to the $B$ sample (color version online).}
	\label{fig3-5}
\end{figure*}

The $|V_{xy}|$ and $|V_{z}|$ distributions of the $K^-$ and $\pi^+$ tracks
used in $D^+$ tag, and the $\pi_d^+$ and $\pi_d^-$ tracks from direct
$\ee$ annihilation are shown in Fig.~\ref{fig3-5}. Compared with the
typical requirements of less than 1~cm and 10~cm for $|V_{xy}|$ and
$|V_{z}|$, respectively, a set of tighter selection criteria $|V_{xy}|<0.55$~cm and $|V_{z}|<3$~cm is identified by optimizing the $\pi^{+}\pi^{-}D^+D^-$ signal significance.
	
\begin{figure*}[htbp]
    \centering
    \includegraphics[width=2.1in]{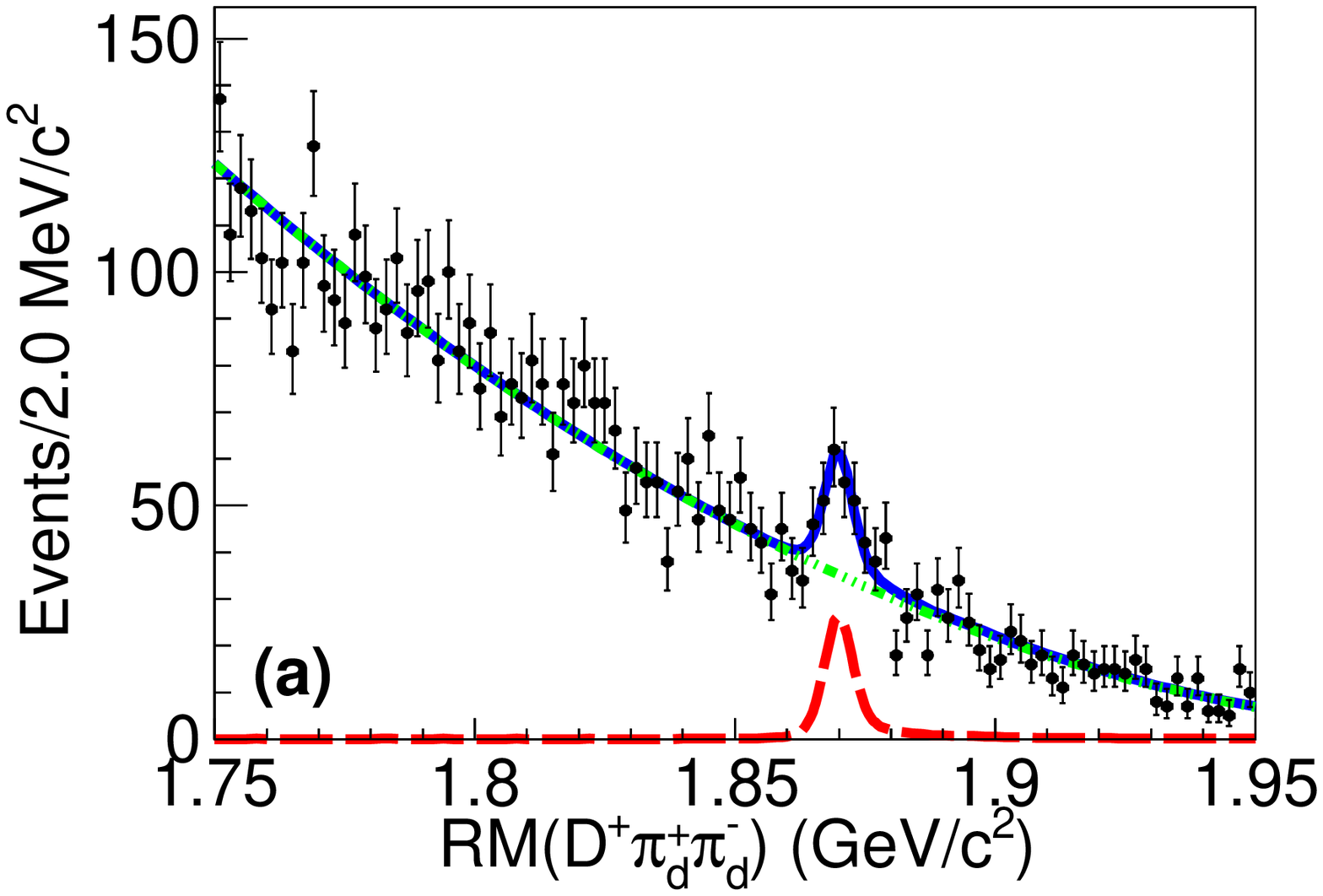}
    \includegraphics[width=2.1in]{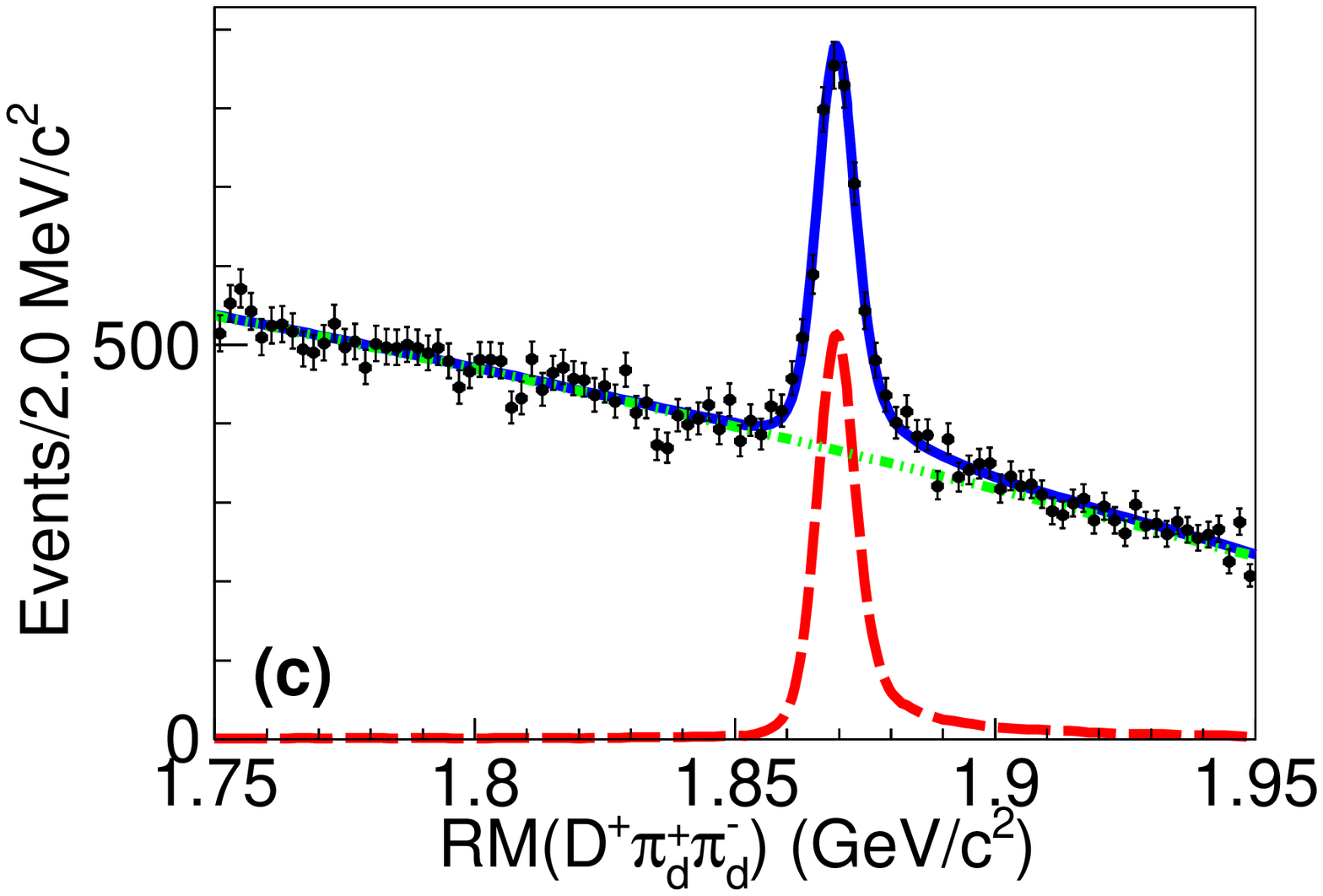}
    \includegraphics[width=2.1in]{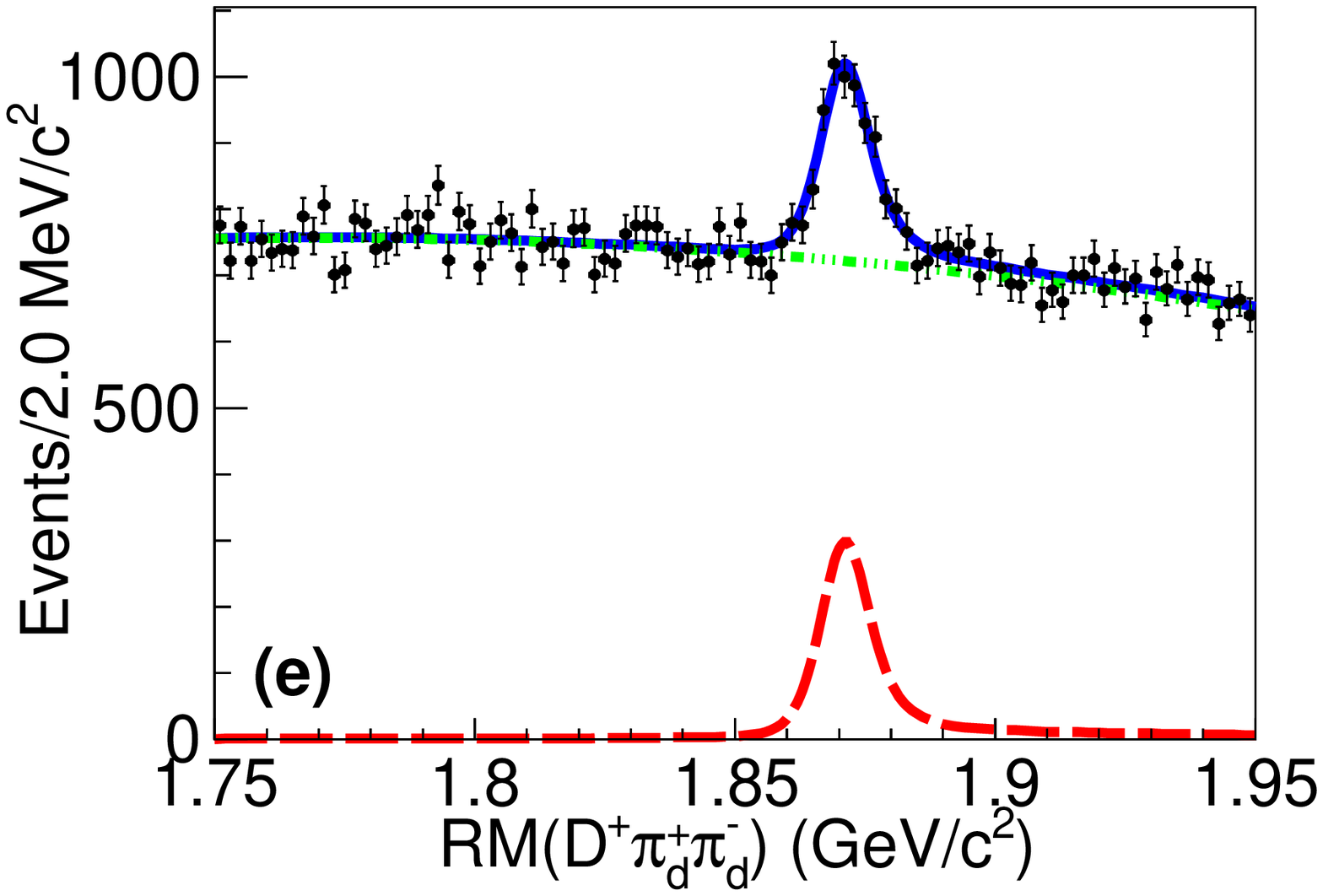}
    \includegraphics[width=2.1in]{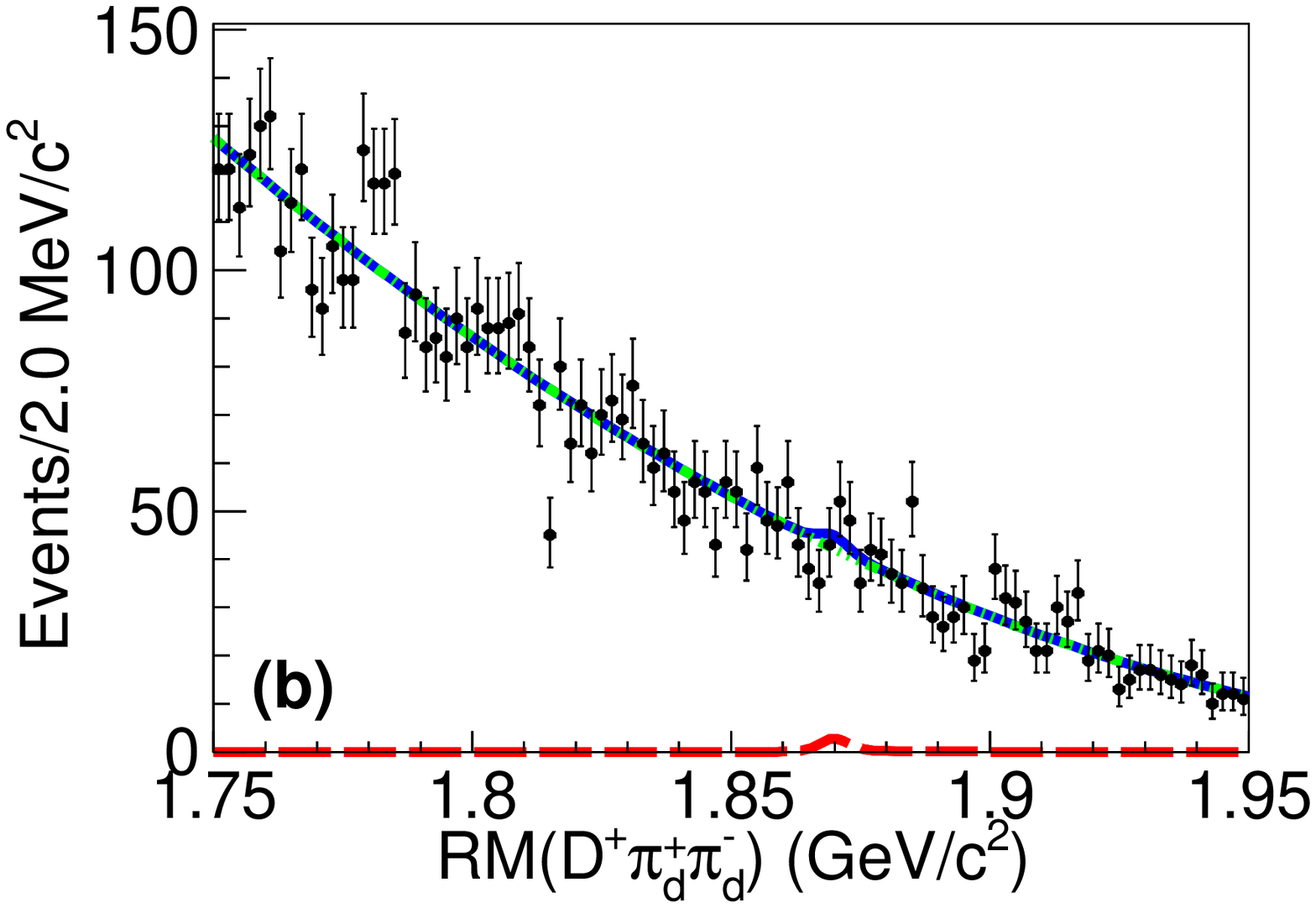}
    \includegraphics[width=2.1in]{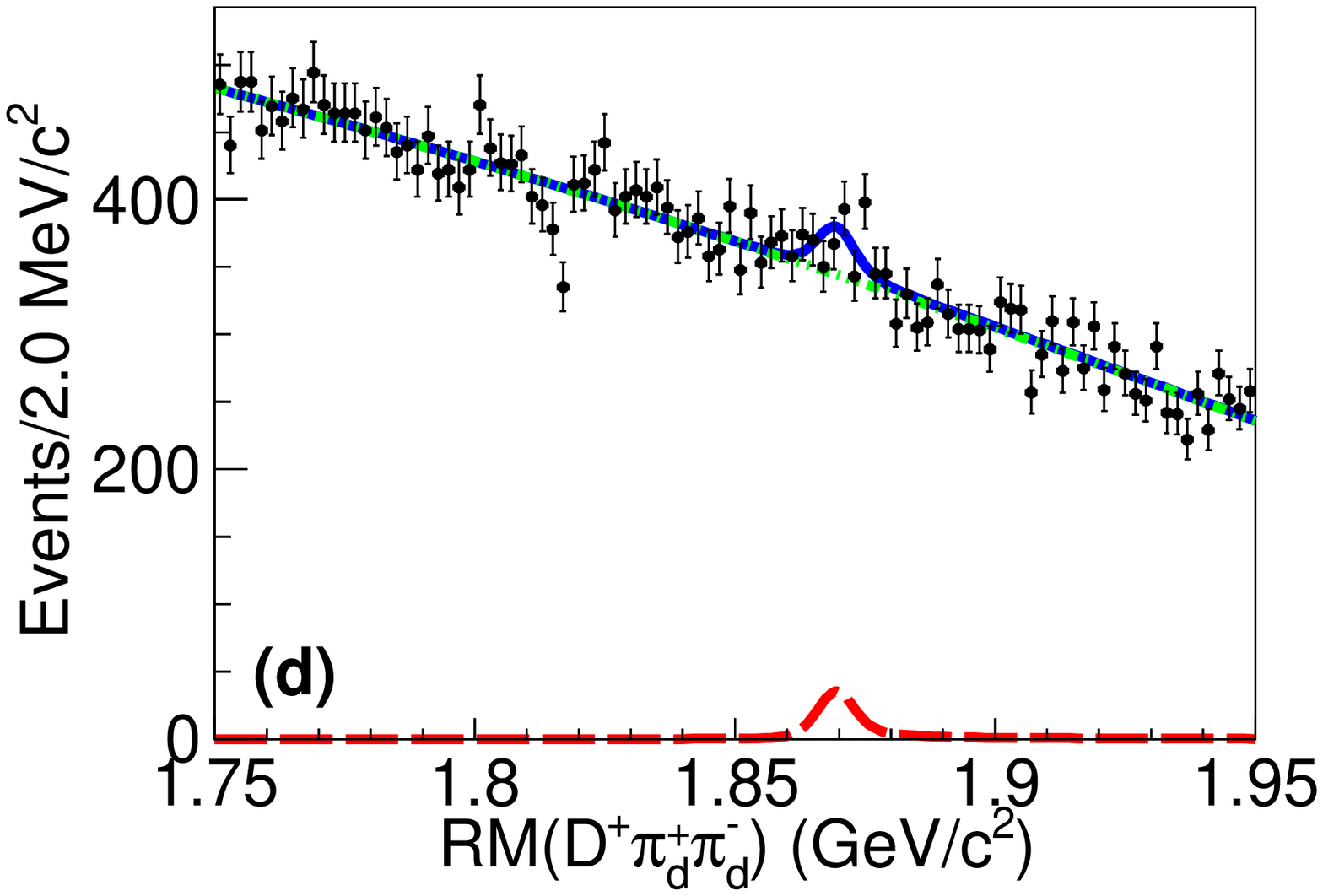}
    \includegraphics[width=2.1in]{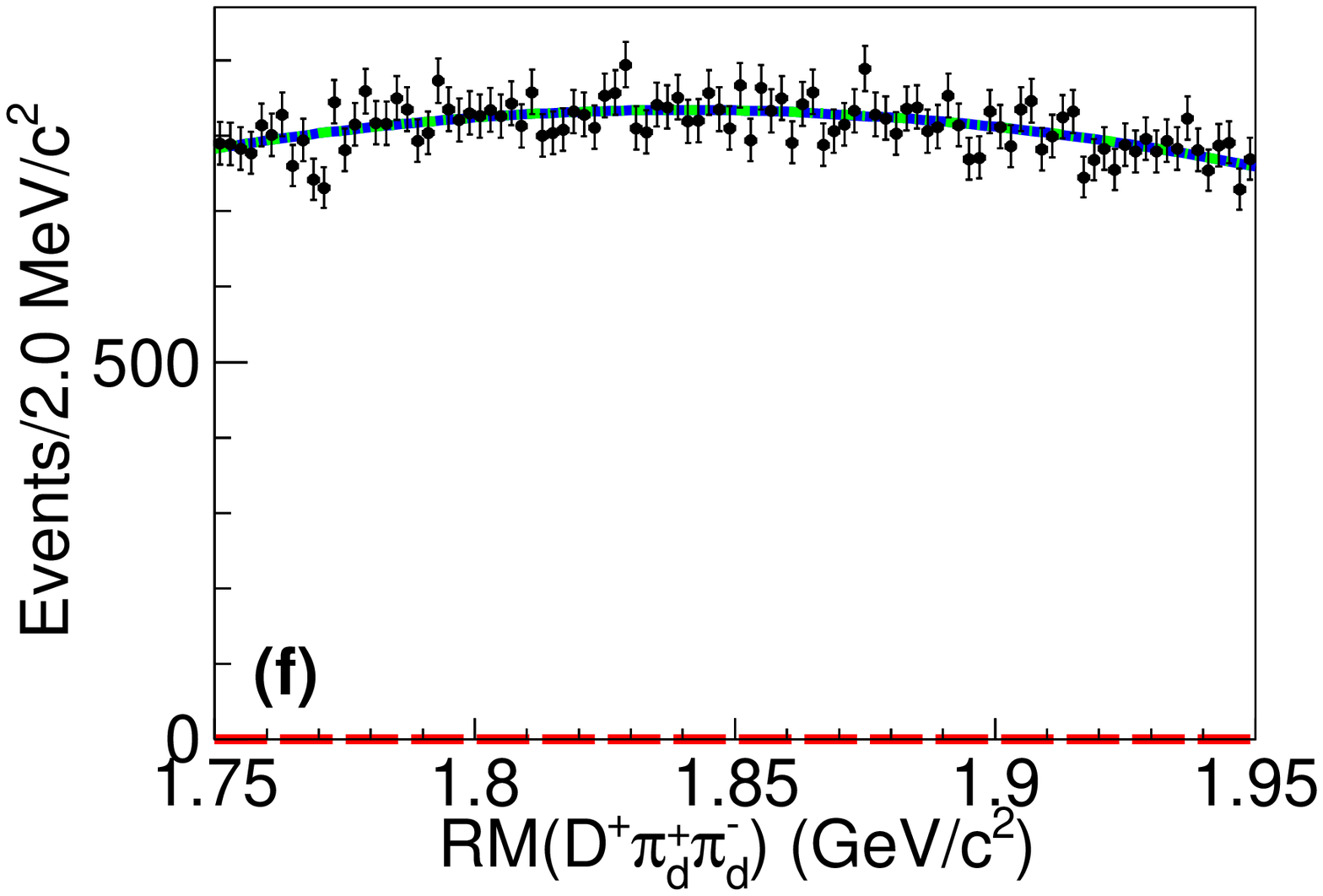}
\caption{Distributions of $RM(\Dpipi)$ in $M(\Kpipi)$ signal (a,~c,~e) and sideband
(b,~d,~f) regions for data samples at $\sqrt{s}=$ 4.230~(a,~b), 4.420~(c,~d),
and 4.680~(e,~f)~GeV, and the best fits to the distributions.
The black dots with error bars are data,
the red dashed, green dash-dotted, and blue solid lines are the signal, background, and total fit,
respectively (color version online). The fit qualities are tested using a $\chi^{2}$-test method, with $\chi^{2}/\rm {n.\ d.\ f.}=$93.72/91, 96.39/95, 83.74/93, 104.25/95, 98.69/93, and 85.12/95 for (a), (b), (c), (d), (e), and (f), respectively.}
\label{fig:RMDpipi_after}
\end{figure*}

After applying all the above selection criteria, requiring $|M(\Kpipi)-m_{D^+}|<d_M$, and constraining
$M(\Kpipi)$ to the $D^+$ mass,
we obtain
the $RM(\Dpipi)$ distributions (Figs.~\ref{fig:RMDpipi_after}(a,~c,~e))
for data samples at $\sqrt{s}=$ 4.230, 4.420, and 4.680~GeV.
Clear $D^-$ signal peaks are observed in all these data samples.
The non-$\pi^{+}\pi^{-}D^+D^-$ background is studied
by examining the $RM(\Dpipi)$ distributions (Figs.~\ref{fig:RMDpipi_after}(b,~d,~f))
for $\Kpipi$ combinations in the $D^+$ mass sideband regions,
defined as $-5d_M<M(\Kpipi)-m_{D^+}<-3d_M$ or
$3d_M<M(\Kpipi)-m_{D^+}<5d_M$. No significant $D^-$ signal peaks are
observed in the sideband samples. In calculating $RM(\Dpipi)$ for the
sideband events, the $M(\Kpipi)$ is constrained to the central values
of the corresponding sideband region rather than to the known mass $m_{D^+}$. The number of
$\ee\to \DDpipi$ signal events is obtained by subtracting
the number of $D^-$ signal candidates in the $D^+$ sideband regions
from that in the $D^+$ signal region, as discussed
in Sec.~\ref{xs_total}.

\section{\boldmath Cross sections of the $\ee\go\DDpipi$ process}
\label{xs_total}

The cross section of the $\ee\go\DDpipi$ process is calculated with
\begin{equation}
\sigma = \frac{N_{\rm signal}-N_{\rm sideband}/2}
 {2f(\sum_i\omega_i\epsilon_i(1+\delta)_i)\frac{1}{|1-\Pi|^2}\mathscr{B}\mathscr{L}},
\end{equation}
where $N_{\rm signal}$ and $N_{\rm sideband}$ are the number of
$\ee\go\DDpipi$ events from fits to $RM(\Dpipi)$ distributions
(Fig.~\ref{fig:RMDpipi_after}) in the $M(\Kpipi)$ signal and
sideband regions, respectively,
$\frac{1}{|1-\Pi|^2}$ is the vacuum polarization factor,
$\mathscr{B}$ is the branching fraction of the decay $D^+\go\Kpipi$ \cite{pdg},
and $\mathscr{L}$ is the integrated luminosity of the data sample.
$f$ denotes an efficiency correction factor
\begin{equation}
	\begin{split}
		f=f^{M(\Kpipi)}f^{K\nrightarrow p}
		f^{V_{xy,\ z}}f^{L/\Delta_{L}}\\f^{RM(\Dpipi)},
	\end{split}
\end{equation}
with $f^{v}$ referring to the efficiency correction factor caused by
selection criterion $v$, which includes $M(\Kpipi)$ and $RM(\Dpipi)$
mass window requirements, $p/\bar{p}$ veto ($K\nrightarrow p$),
$V_{xy,\ z}$ requirements, and $L_{\pipi}/\Delta_{L_{\pipi}}$
requirement for $K_{S}^0$ background suppression.
Details on the evaluation of $f^{v}$ can be found in Sec.~\ref{sec:sys_mes}.
$(1+\delta)_i$ is the ISR correction factor, and
$\omega_i$ and $\epsilon_i$ are the fraction and the detection efficiency
of subprocess $i$, respectively,
here, $i=0$, $1$, and $2$ correspond to $\ee\go\DststD\go\DDpipi$,
$\ee\go\pipipsipp\go\DDpipi$, and $\ee\go\PHSP$ subprocess, respectively.
$\omega_{i}$ is estimated from the $S$ sample by a one dimentional simultaneous extended-unbinned-likelihood fit
to $RM(D^{+})$, $RM(D_{\rm miss}^{-})$, and $RM(\pipi)$ distributions,
and the background is estimated by the $B$ sample.
For data samples with $\sqrt{s}$ larger than $4.315$~GeV, $i=0$, $1$, and $2$,
while for data samples with $\sqrt{s}$ smaller than or equal to $4.315$~GeV, $i=1$ and $2$, since the threshold of $D_{1}(2420)\bar{D}$ is 4291.75 MeV/$c^{2}$, and no significant $D_{1}(2420)\bar{D}$ events  are observed at $\sqrt{s}=$ 4.310 and 4.315 GeV.
%and the other parameters are the same as  those in
%Eq.~(\ref{eq_sigmai}).

\begin{figure*}[htbp]
    \centering
 	\includegraphics[width=3.in]{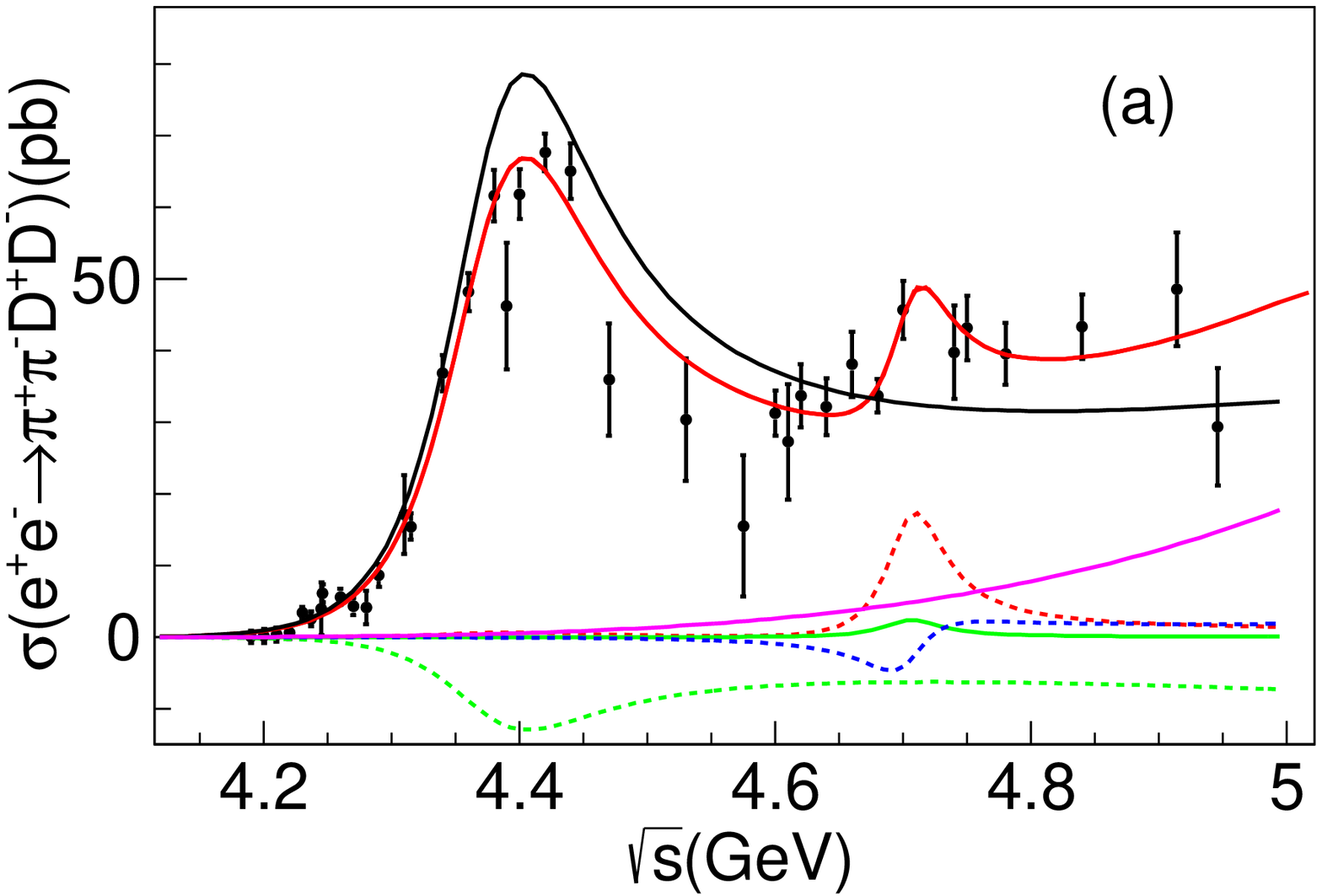}
 	\includegraphics[width=3.in]{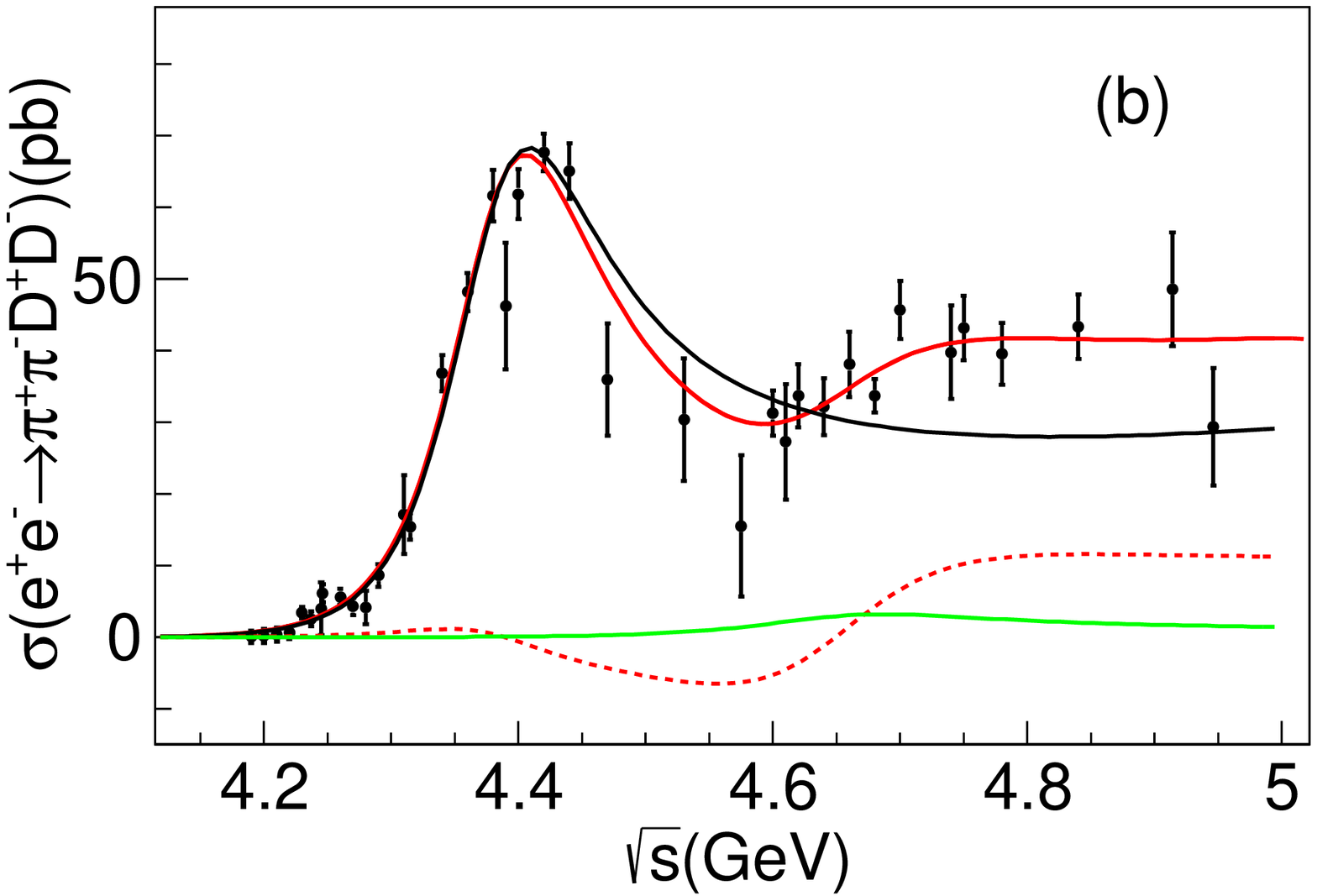}
\caption{Cross section of the process $\ee\to \DDpipi$ and fit with the coherent sum
of two BW functions and a phase space term (Solution \uppercase\expandafter{\romannumeral4})
(a), and with the coherent sum
of two BW functions only (Solution \uppercase\expandafter{\romannumeral2}) (b).
Other solutions of (a) and (b) could be found in Table \ref{table_solution1} and \ref{table_solution2}, respectively.
Dots with error bars are data with
the statistical uncertainties and the red lines show the best fit results. For (a), the black, green, and pink solid lines describe $BW_{0}$, $BW_{1}$, and $\Phi_4$ components, respectively,
and the red, green, and blue dashed lines describe interferences between $BW_{0}$ and $BW_{1}$, $BW_{0}$ and $\Phi_4$, and $BW_{1}$ and $\Phi_4$, respectively.
For (b), the black and green solid lines describe $BW_{0}$ and $BW_{1}$ components, respectively, and the red dashed line describes the inteference between $BW_{0}$ and $BW_{1}$. (color version online).
% and the curves show the best fit and the contribution from each component.
The fit qualities are tested using a $\chi^{2}$-test method, with $\chi^{2}/\rm {n.\ d.\ f.}=$47.1/28 and 46.1/30 for (a) and (b), respectively. The $\rm{n.\ d.\ f.}$ denotes the number of degrees of freedom. The $\chi^{2}$ function is constructed as $\chi^{2}=\Sigma\frac{(\sigma_{i}-\sigma^{fit}_{i})^{2}}{\delta_{i}^{2}}$, here, $\sigma_{i}$ and $\sigma^{fit}_{i}$ are the measured and fitted cross section of the $i^{th}$ data sample, respectively, and $\delta_{i}$ is the standard deviation of the measured cross section, which includes the statistical uncertainties only.}
 	\label{figxs_total}
\end{figure*}

Figure~\ref{fig:RMDpipi_after} shows the fit results
of data samples at $\sqrt{s}=$ 4.230, 4.420, and 4.680~GeV.
The signal shape is modelled by the $RM(\Dpipi)$ distributions
in MC simulation of each subprocess weighted according to $\omega_i$ and
convolved with a Gaussian function to take the resolution
difference between data and MC simulation into account.
The background shape is described by a second-order Chebychev
polynomial function.
At each $\sqrt{s}$, the signal shape for the fit in the $M(\Kpipi)$
sideband regions is the same as that for the fit in the $M(\Kpipi)$ signal region.
The results for $N_{\rm signal}$ and $N_{\rm sideband}$
obtained from the fits are listed in Table~\ref{tab:CS}, together with the
fit results for all other data samples. The calculated cross section of the
$\ee\go\DDpipi$ process is shown in Fig.~\ref{figxs_total}.

For data samples where no significant $\ee\go\DDpipi$ signal peaks are observed
(statistical significance smaller than $5\sigma$), the upper limits on
the cross section are calculated
using a Bayesian method~\cite{2003UPL}. By fitting the $RM(\Dpipi)$ distribution
for the events in the $D^+$ signal region with fixed values for the signal yield,
a scan of the likelihood distribution as a function of the cross section is obtained.
To take the total systematic uncertainty (listed in Table \ref{table4-1})
into consideration, the likelihood distribution is convolved with a Gaussian
function with a width corresponding to the overall systematic uncertainty.
The upper limit on the cross section at 90\% C.L. is obtained from
\(  \int_0^{\sigma} L(x) d x / \int_0^{\infty} L(x) d x=0.9. \)
The upper limits on the cross sections are listed in Table~\ref{tab:CS}.

\section{\boldmath Resonances in the $\ee\go\DDpipi$ Cross Section line shape}

Clear resonant structures around $\sqrt{s}=$ 4.390 and 4.700~GeV can be seen in Fig.~\ref{figxs_total},
and there is no significant signal at other energies, including at the expected
masses of the $Y(4230)$, $Y(4360)$, and $Y(4660)$ states.

A fit to the measured $\ee\go\DDpipi$ cross section line shape is performed
with a coherent sum of two Breit-Wigner (BW) functions and a phase space term
\begin{equation}
	\begin{split}
    \sigma(\sqrt{s})=|B W_0(\sqrt{s})+
         BW_1(\sqrt{s})e^{i \phi_1}\\+ce^{i\phi_2}\Phi_4(\sqrt{s})|^2,
    \end{split}
\end{equation}
with the BW function defined as
\begin{equation}
	BW_j(\sqrt{s})=\frac{\sqrt{12 \pi {\Gamma^{\ee}_j} \Gamma^{\rm tot}_j  \mathscr{B}_j}}
	{s-m_j^2+i m_j\Gamma^{\rm tot}_j} \sqrt{\frac{\Phi_4(\sqrt{s})}{\Phi_4(m_j)}},
\end{equation}
where $m_j$, $\Gamma^{\rm tot}_j$, and ${\Gamma^{\ee}_j}$ are the mass, width,
and electronic partial width of the $j^{th}$ resonance ($R_j$), respectively;
$\mathscr{B}_j$ is the branching fraction of the decay $R_j\go\DDpipi$,
$\phi_j$ is the relative phase between the $j^{th}$ resonance, as well as the phase space term,
$\Phi_4(\sqrt{s})$ is the phase space factor of the
four-body decay $R\go\DDpipi$, and $c$ is a constant describing
the magnitude of $\Phi_4(\sqrt{s})$.

There are four solutions with the same fit quality and identical resonance parameters for $R_{0}$ and $R_{1}$ as well as $c$, but different $\Gamma^{\ee}_{j}\mathscr{B}_{j}$ and $\phi_{j}$, as listed in Table \ref{table_solution1}. 	
The fitted parameters for $R_0$ are
in agreement with those of the $Y(4390)$ resonance observed by
the BESIII Collaboration in the $\ee\go\pi^+\pi^-h_c$
process~\cite{BESIII:2016adj}.
The statistical significance of $R_1$ is determined to be $4.1\sigma$
by comparing the likelihood of the baseline fit and that of the fit
without $R_1$.

\begin{table*}[htp]
	\centering
	\caption{The fitted parameters of the cross sections of $e^{+}e^{-}\rightarrow\pi^{+}\pi^{-}D^{+}D^{-}$ with the coherent sum of two Breit-Wigner functions and a phase space term. The first uncertainties are statistical and the second systematic.}
	\begin{tabular}{crrrrr}
	\hline\hline
	Parameters & Solution \uppercase\expandafter{\romannumeral1} & Solution \uppercase\expandafter{\romannumeral2} & Solution \uppercase\expandafter{\romannumeral3} & Solution \uppercase\expandafter{\romannumeral4} &  \\
	\hline
	 $c\ (\rm{MeV}^{-3/2})$ & \multicolumn{4}{c}{$\ \ \ \ \ \ \ \ \ \ \ \ \ \ \ \ \ (1.6\pm0.9\pm0.1)\times10^{3}$} & \\
	 $m_{0}\ (\rm{MeV}/c^{2})$ & \multicolumn{4}{c}{$4373.1\pm4.0\pm1.0$} & \\
	 $\Gamma_{0}\ (\rm{MeV})$ & \multicolumn{4}{c}{$\ \ 146.5\pm7.4\pm1.1$} & \\
	 $m_{1}\ (\rm{MeV}/c^{2})$ & \multicolumn{4}{c}{$\ \ 4706\ \pm11\ \pm4\ \ $} & \\
	 $\Gamma_{1}\ (\rm{MeV})$ & \multicolumn{4}{c}{$\ \ \ \ \ \ 45\ \pm28\ \pm9\ \ $} & \\
	 $\Gamma^{\ee}_{0}\mathscr{B}_{0}\ (\rm eV)$ & 9.3$\ \pm\ $0.8$\ \pm\ $1.8 & 9.3$\ \pm\ $0.8$\ \pm\ $1.3 & 13.0$\ \pm\ $1.5$\ \pm\ $1.7 & 9.9$\ \pm\ $1.1$\ \pm\ $1.4 & \\
	 $\Gamma^{\ee}_{1}\mathscr{B}_{1}\ (\rm eV)$ & 11.9$\ \pm\ $6.5$\ \pm\ $3.2 & 0.2$\ \pm\ $0.1$\ \pm\ $0.1 & 0.2$\ \pm\ $0.1$\ \pm\ $0.1 & 10.8$\ \pm\ $5.3$\ \pm\ $2.8 & \\
	 $\phi_{1}\ (\rm{rad})$ & 4.9$\ \pm\ $0.2$\ \pm\ $0.0 & 0.3$\ \pm\ $0.4$\ \pm\ $0.0 & 1.1$\ \pm\ $0.7$\ \pm\ $0.0 & 4.1$\ \pm\ $0.3$\ \pm\ $0.0 & \\
	 $\phi_{2}\ (\rm{rad})$ & 4.6$\ \pm\ $0.3$\ \pm\ $1.0 & 1.6$\ \pm\ $0.3$\ \pm\ $1.0 & 1.6$\ \pm\ $0.3$\ \pm\ $1.0 & 4.5$\ \pm\ $0.3$\ \pm\ $1.0 & \\
	\hline\hline
	\end{tabular}
	\label{table_solution1}
\end{table*}

If we omit the phase space term from the baseline fit, the fit quality becomes slightly worse, indicating the statistical
significance of this amplitude is only $1.4\sigma$ and the solutions of this amplitude could be found in Table \ref{table_solution2}. However, the fit in the
high energy region becomes very different, as shown in Fig.~\ref{figxs_total}(b).
For the fit with the coherent sum of two BW functions only, there are two solutions with the same fit quality and identical resonance parameters for $R_{0}$ and $R_{1}$, but different $\Gamma^{\ee}_{j}$ and $\phi_{j}$, as listed in Table \ref{table_solution2}.
The statistical significance of $R_1$ is $7.0\sigma$.

\begin{table*}[htp]
	\centering
	\caption{The fitted parameters of the cross sections of $e^{+}e^{-}\rightarrow\pi^{+}\pi^{-}D^{+}D^{-}$ with the coherent sum of two Breit-Wigner functions. The uncertainties are statistical.}
	\begin{tabular}{ccrr}
	\hline\hline
	Parameters & Solution \uppercase\expandafter{\romannumeral1} & Solution \uppercase\expandafter{\romannumeral2} &  \\
	\hline
	 $m_{0}\ (\rm{MeV}/c^{2})$ & \multicolumn{2}{c}{$4378.0\pm8.5$} & \\
	 $\Gamma_{0}\ (\rm{MeV})$ & \multicolumn{2}{c}{$\ \ \ \ 152\pm14$} & \\
	 $m_{1}\ (\rm{MeV}/c^{2})$ & \multicolumn{2}{c}{$\ \ 4605\pm90$} & \\
	 $\Gamma_{1}\ (\rm{MeV})$ & \multicolumn{2}{c}{$\ \ \ \ 245\pm67$} & \\
	 $\Gamma^{\ee}_{0}\mathscr{B}_{0}\ (\rm{eV})$ & 21$\ \pm\ $12 & 12.2$\ \pm\ $5.8 & \\
	 $\Gamma^{\ee}_{1}\mathscr{B}_{1}\ (\rm{eV})$ & 54$\ \pm\ $15 & 1.3$\ \pm\ $2.7 & \\
	 $\phi_{1}\ (\rm{rad})$ & 4.1$\ \pm\ $0.3 & 5.6$\ \pm\ $2.6 & \\
	\hline\hline
	\end{tabular}
	\label{table_solution2}
\end{table*}

Other than the $R_0$ and $R_1$ contributions, we also tested the statistical significances
of the possible structures around $\sqrt{s}=$ 4.245 and 4.914~GeV.
By adding the $Y(4230)$ amplitude to the fit, with the mass and width
fixed according to the world averaged values~\cite{pdg}, its significance
is found to be only $0.3\sigma$. By adding a new resonance at high energy with free
mass and width, the statistical significance is found to be $1.3\sigma$.
Therefore, such additional structures are not considered at
the upper and lower mass regions.

Note that there are four points ($\sqrt{s}$ from 4.400 to 4.600 GeV) systematically below the fitted
line. Since the integrated luminosities of these data samples are very low,
larger data samples are needed to draw a conclusion.
 	
\section{\boldmath Evidence for $\ee\go \pi^+\pi^-X(3842)$}

To search for the $X(3842)$ state, the $S$ sample defined in Sec.~\ref{sec:evtsel}
with the additional $K^0_S$ veto and stringent $|V_{xy}|$ and $|V_z|$
requirements ($|V_{xy}|<0.55$ cm and $|V_{z}|<3$ cm) is used. In order to suppress the $\ee\go\DststD$ background,
the $D_1(2420)$ signal is suppressed by requiring
$|RM(D^+)-m_{D_1(2420)^-}|>0.01$~GeV/$c^2$
and $|RM(D_{\rm{miss}}^-)-m_{D_1(2420)^+}|>0.01$~GeV/$c^2$,
where $m_{D_{1}(2420)^{\pm}}=2.4221$ GeV/$c^{2}$ is the known $D_{1}(2420)^{\pm}$ mass~\cite{pdg} \footnote{Here, the selection criteria correspond to the resolutions of $RM(D^{+})$ and $RM(D_{\rm miss}^{-})$, both equal to 0.01 GeV/$c^{2}$.}.

\begin{figure*}[htbp]
    \centering
    \includegraphics[width=2.1in]{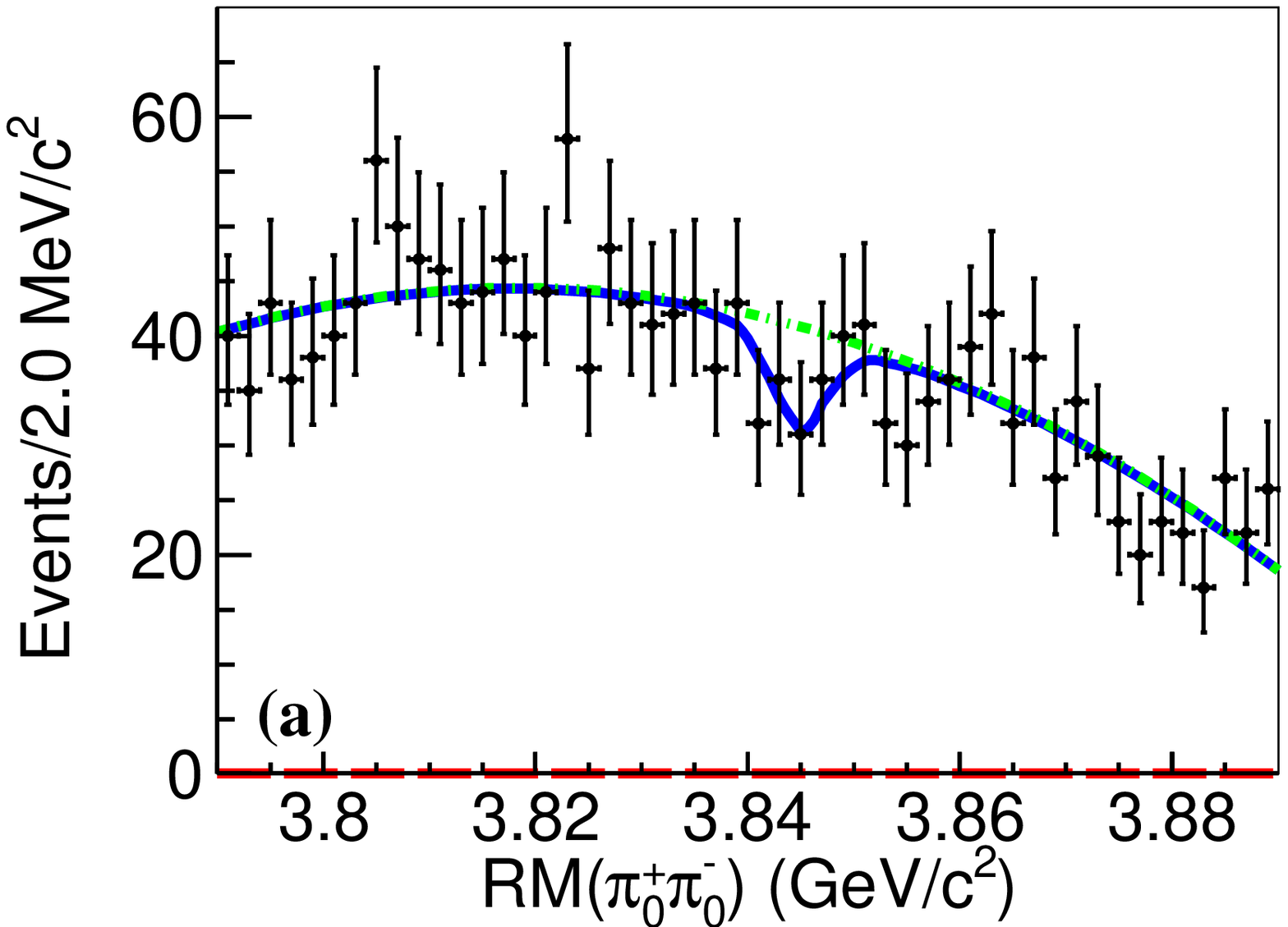}
    \includegraphics[width=2.1in]{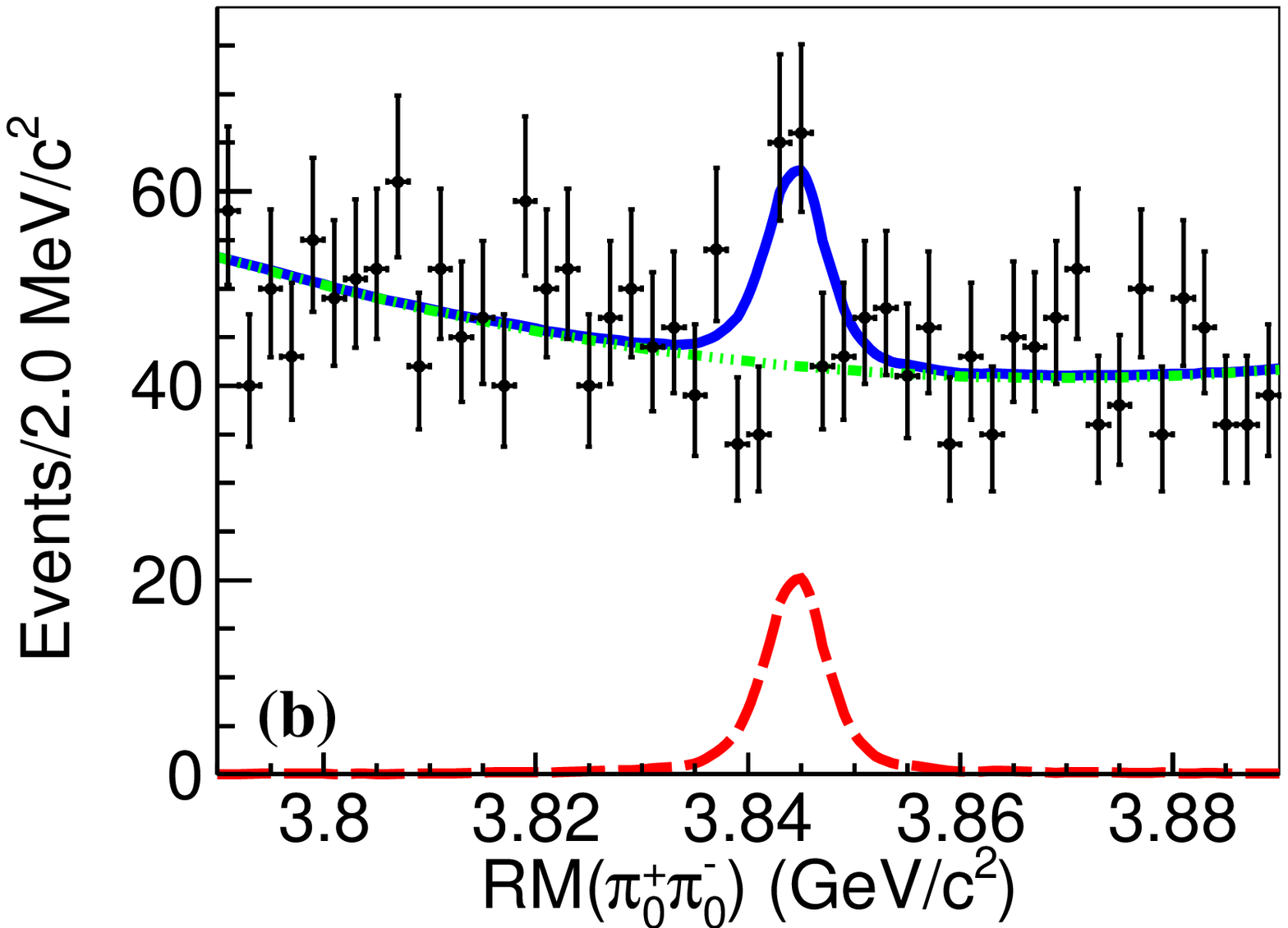}
    \includegraphics[width=2.1in]{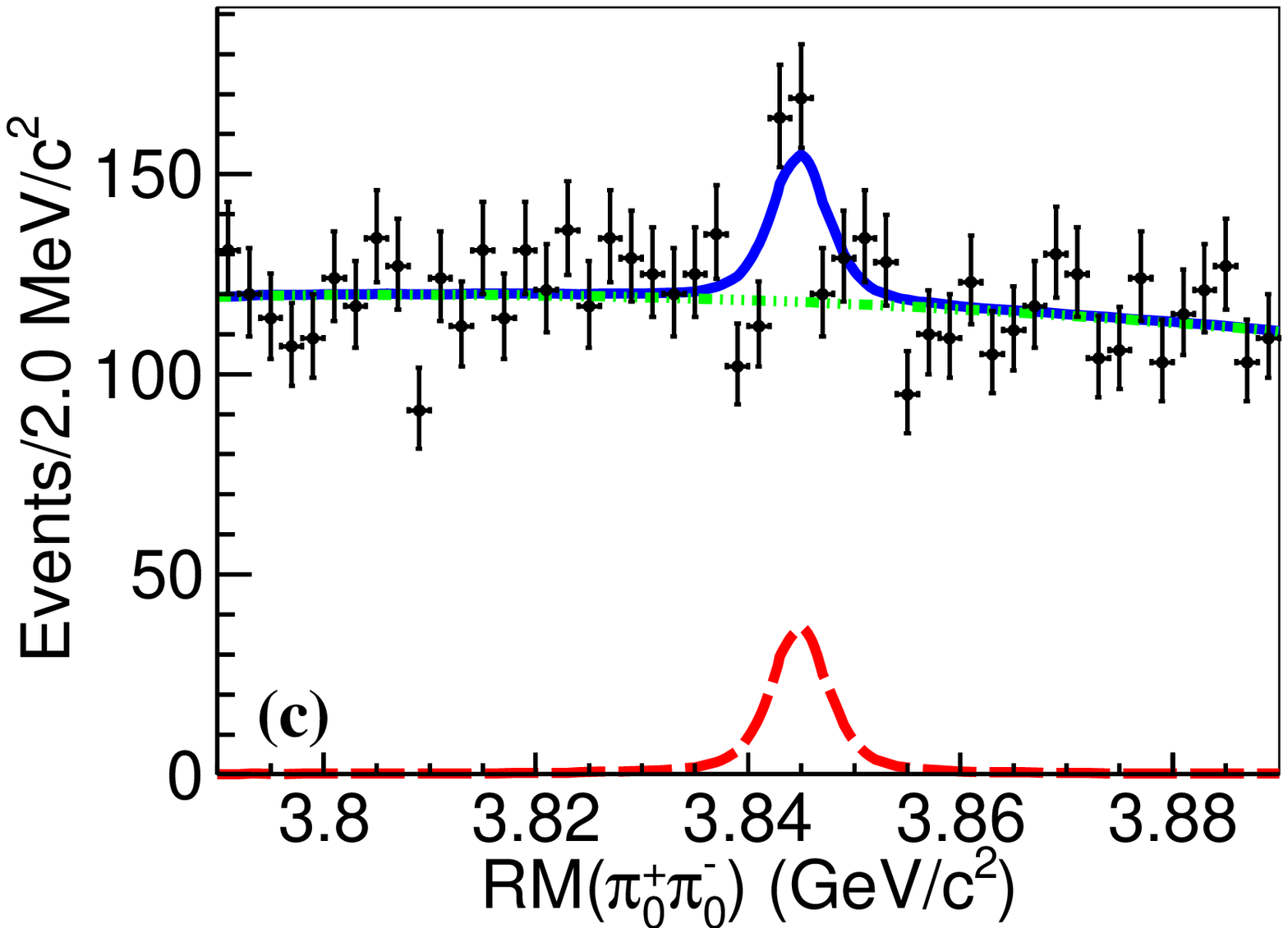}
\caption{The $RM(\pipi)$ distributions and the fits at $\sqrt{s}=$
4.420~(a), 4.680~(b) GeV, and data samples
with $\sqrt{s}$ from 4.600 to 4.700~GeV~(c).
The black dots with error bars are the $S$ sample, and the red dashed, green dash-dotted, and blue
solid curves are the signal shape, background shape, and total fit,
respectively (color version online). The fit qualities are tested using a $\chi^{2}$-test method, with $\chi^{2}/\rm {n.\ d.\ f.}=$26.3/45, 41.4/43, and 57.07/45 for (a), (b), and (c), respectively.}
    \label{fig:X_3842}
\end{figure*}

The $RM(\pipi)$ (equivalent to the invariant mass of $D^+D_{\rm miss}^-$) distributions
in all data samples are examined. While no significant signal is observed
at any single $\sqrt{s}$, there is evidence for an $X(3842)$ resonance
for $\sqrt{s}$ from 4.600 to 4.700~GeV.
Figure~\ref{fig:X_3842} shows the $RM(\pipi)$ distributions at $\sqrt{s}=$
4.420, 4.680~GeV, and data samples with $\sqrt{s}$ from 4.600 to 4.700~GeV.

To fit the $RM(\pipi)$ distributions, the $X(3842)$ signal shape
is obtained from MC simulation of the $\ee\go f_{0}(500) X(3842)\go\DDpipi$  process\footnote{Two body
decay is assumed since it has the largest phase space and
$f_{0}(500)$ is the meson with $J^{P}=0^{+}$.},
and convolved with a Gaussian function to take the resolution difference
between data and MC simulation into account.
The mean and sigma values of the Gaussian function for other fits are fixed to the fit values obtained at $\sqrt{s}=$ 4.680 GeV as this sample contains the largest number of signal events.
The mass and width of the $f_{0}(500)$ are taken from Ref.~\cite{2001sigma}
when generating MC events. The background is described with a second-order Chebychev polynomial function.

Fit results of the $RM(\pipi)$ distributions are shown in
Figs.~\ref{fig:X_3842}(a,~b,~c).
The signal yields (statistical significances) are $-$39 $\pm$ 18 ($-2.0\sigma$),
58 $\pm$ 24 ($1.8\sigma$), and 155 $\pm$ 38 ($4.2\sigma$) at $\sqrt{s}=$
4.420, 4.680~GeV and data samples with $\sqrt{s}$
from 4.600 to 4.700~GeV, respectively.
%Besides the nominal fit
Furthermore, for data samples with $\sqrt{s}$
from 4.600 to 4.700~GeV, the fits are also performed by changing the fit range,
the signal shape, or the background shape. In all cases,
the minimum value of the $X(3842)$ resonance significance is $4.2\sigma$.
The fit results at other energies are listed in Table~\ref{table3-2}.
The cross sections of the $\ee\go\pi^+\pi^- X(3842)\go\DDpipi$ process
are calculated in a similar way as for other processes,
and the upper limits of the cross sections are determined using a similar strategy to
that described in Sec.~\ref{xs_total}. The
results are also listed in Table~\ref{table3-2}.

\begin{table*}[htp]
	\centering
	\caption{Results for the $e^{+}e^{-}\rightarrow \pi^{+}\pi^{-}X(3842)\rightarrow\pi^{+}\pi^{-}D^{+}D^{-}$ process. Here, $\sigma$ is the cross section of the $e^{+}e^{-}\rightarrow \pi^{+}\pi^{-}X(3842)\rightarrow\pi^{+}\pi^{-}D^{+}D^{-}$ process, where the first uncertainty is statistical and the second systematic; $S$ is the statistical significance; $\epsilon$, $(1+\delta)$, $N$, and $\sigma_{\rm ul}$ are the detection efficieny, ISR correction factor, signal yields, and the upper limit of cross section at 90\% confidence level.}
	% \resizebox{\textwidth}{55mm}{
	\begin{tabular}{c c c c r@{$\pm$}l r@{$\pm$}c@{$\pm$}l c c}
	\toprule
	\hline\hline
$\sqrt{s}^{\rm nominal}$ & $\epsilon (\%)$ & $(1+\delta)$ & $\mathscr{L}$ ($\rm{pb}^{-1}$) & \multicolumn{2}{c}{$N$} & \multicolumn{3}{c}{$\sigma\ (\rm{pb})$} & $S$ & $\sigma_{\rm{ul}}$ (\rm{pb}) \\
	\hline
	\midrule
	 4.190& 3.3 & 0.80  & 570.0   &  1   &  2   & 0.5  & 0.8  & 0.1  & $\ \ \ \  0.6\sigma$ & 2.5  \\
	 4.200& 4.8 & 0.81 & 526.0   &  3   &  3   & 0.8  & 0.8  & 0.1  & $\ \ \ \  1.1 \sigma$ & 2.6  \\
	 4.210& 5.8 & 0.82 & 572.1   &  $-$1  &  1   & $-$0.2 & 0.4  & 0.0  & - & 0.9  \\
	 4.220& 7.0 & 0.83 & 569.1   &  $-$2  &  2   & $-$0.4 & 0.4  & 0.1  & - & 0.7  \\
	 4.230& 9.1 & 0.83 & 1100.9  &  0   &  4   & 0.0  & 0.3  & 0.0  & $\ \ \ \  0.1 \sigma$ & 0.6  \\
	 4.237& 9.7 & 0.84 & 530.0   &  0   &  3   & 0.1  & 0.4  & 0.0  & $\ \ \ \  0.3 \sigma$ & 1.0  \\
	 4.246& 10.4 & 0.84 & 538.1   &  $-$3  &  2   & $-$0.4 & 0.3  & 0.1  & - & 0.5  \\
	 4.269& 11.8 & 0.85 & 825.7   &  $-$8  &  4   & $-$0.6 & 0.3  & 0.1  & - & 0.3  \\
	 4.270& 12.2 & 0.85 & 531.1   &  6   &  4   & 0.6  & 0.4  & 0.1  & $\ \ \ \  1.5 \sigma$ & 1.5  \\
	 4.290& 12.2 & 0.86 & 502.4   &  0   &  4   & $-$0.0 & 0.4  & 0.0  & $\ \ \ \  0.1 \sigma$ & 0.9  \\
	 4.315& 14.8 & 0.87 & 501.2   &  2   &  6   & 0.2  & 0.5  & 0.0  & $\ \ \ \  0.4 \sigma$ & 0.9  \\
	 4.340& 15.6 & 0.88 & 505.0   &  $-$8  &  7   & $-$0.7 & 0.6  & 0.1  & - & 0.6  \\
	 4.360& 17.1 & 0.88 & 544.0   &  $-$7  &  9   & $-$0.5 & 0.6  & 0.1  & - & 1.0  \\
	 4.380& 16.2 & 0.89 & 522.7   & $-$19  &  8   & $-$1.3 & 0.6  & 0.2  &- & 1.1  \\
	 4.400& 16.4 & 0.89 & 507.8   &  11  &  12  & 0.8  & 0.9  & 0.1  & $\ \ \ \  1.0 \sigma$ & 4.6  \\
	 4.420& 18.3 & 0.90  & 1090.7  & $-$39  &  18  & $-$1.2 & 0.6  & 0.2  &- & 0.6  \\
	 4.440& 16.7 & 0.90 & 569.9   &  7   &  15  & 0.5  & 0.9  & 0.1  & $\ \ \ \  0.5 \sigma$ & 3.1  \\
	 4.600& 19.6 & 0.92 & 586.9   &  31  &  13  & 1.6  & 0.7  & 0.2  & $\ \ \ \  2.5 \sigma$ & 3.3  \\
	 4.620& 18.5 & 0.93 & 521.5   &  27  &  13  & 1.6  & 0.8  & 0.2  & $\ \ \ \  2.2 \sigma$ & 3.4  \\
	 4.640& 18.8 & 0.93 & 552.4   &  17  &  13  & 1.0  & 0.7  & 0.1  & $\ \ \ \  1.4 \sigma$ & 2.3  \\
	 4.660& 19.1 & 0.93 & 529.6   &  13  &  13  & 0.8  & 0.7  & 0.1  & $\ \ \ \  1.1 \sigma$ & 2.0  \\
	 4.680& 19.1 & 0.93 & 1669.3  &  58  &  24  & 1.0  & 0.4  & 0.1  & $\ \ \ \  1.8 \sigma$ & 1.5  \\
	 4.700& 19.1 & 0.93 & 536.5   &  1   &  13  & 0.1  & 0.7  & 0.0  & $\ \ \ \  0.1 \sigma$ & 1.4  \\
	 4.750& 20.1 & 0.93 & 367.2   &  0   &  10  & $-$0.1 & 0.8  & 0.0  & $\ \ \ \  0.1 \sigma$ & 1.4  \\
	 4.780& 20.0 & 0.94 & 512.8   &  15  &  12  & 0.9  & 0.7  & 0.1  & $\ \ \ \  1.4 \sigma$ & 2.0  \\
	 4.840& 20.5 & 0.94 & 527.3   & $-$11  &  10  & $-$0.6 & 0.5  & 0.1  & - & 0.7  \\
	 4.916& 20.4 & 0.95 & 208.1   &  $-$6  &  5   & $-$0.9 & 0.8  & 0.1  & - & 1.1  \\
	 4.946& 20.0 & 0.95 & 160.4   & $-$10  &  3   & $-$1.7 & 0.6  & 0.2  & - & 0.8  \\
	\hline\hline
	\bottomrule
	\end{tabular}
	% }
	\label{table3-2}
\end{table*}

\section{\boldmath Systematic Uncertainties}

\subsection{Systematic uncertainties for the $\ee\go\DDpipi$ cross sections}
\label{sec:sys_mes}

The systematic uncertainties in the cross section measurement of the $\ee\go\DDpipi$
process stem from many sources.
The systematic uncertainties associated with the detection efficiencies,
including tracking~\cite{TRACKPID} and PID~\cite{TRACKPID},
are estimated as 1\% for each.
The systematic uncertainty associated with the integrated luminosity
measurement using Bhabha ($\ee\go\ee$) events is estimated as 1\%~\cite{Lum}. For the vacuum polarization factor calculation, the systematic uncertainty originates mainly from hadronic contributions, and is estimated as 0.1\% according to~\cite{VP}.
The systematic uncertainty coming from the input branching fraction of $D^+\go\Kpipi$ is estimated as 1.7\%~\cite{pdg}.
Details of further systematic uncertainties are given below.

The selection efficiency is obtained from MC simulation and corrected
according to the measurements with control samples selected from
data directly. The efficiency correction factor $f^{v}$ is defined as
\begin{equation}
	f^{v}  = \epsilon^{v}_{\rm{data}}/\epsilon^{v}_{\rm{MC}},
\end{equation}
with
\begin{equation}
	\epsilon^{v}_{\rm{data} (\rm{MC})} = N^{v}_{\rm{signal}_{\rm{data} (\rm{MC})}}/N^{v}_{\rm{all}_{\rm{data} (\rm{MC})}},
\end{equation}
where the subscript ``MC" represents MC simulation and the subscript ``data"
represents the data sample,
$N^{v}_{\rm{signal}_{\rm{data} (\rm{MC})}}$ is the number of events
in the signal region of a selection criterion $v$,
and $N^{v}_{\rm{all}_{\rm{data} (\rm{MC})}}$ is the number of events
in the full range of $v$.

The uncertainty of $\epsilon^{v}_{\rm{data} (\rm{MC})}$
\begin{equation}
	\sigma_{\epsilon^{v}_{\rm{data} (\rm{MC})}} = \sqrt{\frac{\epsilon^{v}_{\rm{data} (\rm{MC})}(1-\epsilon^{v}_{\rm{data} (\rm{MC})})}{N^{v}_{\rm{all}_{\rm{data} (\rm{MC})}}}},
\end{equation}
and
the uncertainty of $f^{v}$ is
\begin{equation}
	\frac{\sigma^2_{f^{v}}}{{f^{v}}^2} = \frac{\sigma_{\epsilon^{v}_{\rm{data}}}^2}{{\epsilon_{\rm{data}}^{v}}^2} + \frac{\sigma_{\epsilon^{v}_{\rm{MC}}}^2}{{\epsilon^{v}_{\rm{MC}}}^2},
\end{equation}
since data and MC simulation are independent.
For $f^{v}=(1\pm \Delta f^{v})\pm\sigma_{f^{v}}$,
if $|\frac{\Delta f^{v}}{\sigma_{f^{v}}}|<1.00$,
no correction will be applied and $|\Delta f^{v}|+\sigma_{f^{v}}$
will be taken as the systematic uncertainty, where $\Delta f^{v}$ is the deviation of $f^{v}$ from 1;
while if $|\frac{\Delta f^{v}}{\sigma_{f^{v}}}|>1.00$,
the MC efficiency will be corrected as
$\epsilon = \epsilon_{MC}\times f^{v}$, and
$\sigma_{f^{v}}$ will be taken as the systematic uncertainty.

In order to avoid effects from statistical uncertainty,
only data samples at $\sqrt{s}=$ 4.340, 4.360, 4.400, 4.420, 4.440, 4.600, and 4.680 GeV are used
to estimate the systematic uncertainty originating from
the $M(\Kpipi)$ ($RM(\Dpipi)$) mass window requirement.
A constant parameter
is used to fit the distributions of $f^{M(\Kpipi)}$ ($f^{RM(\Dpipi)}$)
among the data samples mentioned above, and the fitted
$f^{M(\Kpipi)}$ ($f^{RM(\Dpipi)}$) and
$\sigma_{f^{M(\Kpipi)}}$ ($\sigma_{f^{RM(\Dpipi)}}$) values
are 0.986 $\pm$ 0.003 (0.984 $\pm$ 0.005),
the value of $\Big|\frac{\Delta f^{M(\Kpipi)}}{\sigma_{f^{M(\Kpipi)}}}\Big|$
\Big($\Big|\frac{\Delta f^{RM(\Dpipi)}}{\sigma_{f^{RM(\Dpipi)}}}\Big|$\Big) is 5.6 (2.8),
therefore, the systematic uncertainty is taken as 0.3\% (0.5\%)
and $f^{M(\Kpipi)}$ ($f^{RM(\Dpipi)}$) is set to be 0.986 (0.984).

In order to avoid effects from the statistical uncertainty,
the same set of data samples as mentioned in the previous paragraph
is used to estimate the systematic uncertainties originating from the fit range
and background shape of $RM(\Dpipi)$.
The systematic uncertainty coming from the choice of the fit range is estimated
by varying the limits of the fit range from (1.75,~1.96)~GeV/$c^2$
to (1.77,~1.97)~GeV/$c^2$.
The background shape is varied from the first-order Chebychev
polynomial function to a second-order one
at $\sqrt{s}=$ 4.340 and 4.360~GeV,
and the second-order Chebychev polynomial function to the first-order one
at $\sqrt{s}=$ 4.400, 4.420, 4.440, 4.600, and 4.680 GeV.
The largest difference of the cross section compared with the
baseline value among the data samples mentioned above is taken as a
systematic uncertainty of 1.6\% (1.7\%) for the fit range (background shape).
	
The systematic uncertainty stemming from the $p/\bar{p}$ veto,
which is caused by the difference in mis-identification probability
of $K$ to $p/\bar{p}$ between data and MC simulation,
is estimated by the control sample
of $J/\psi\go K_{S}^0K^-\pi^++c.c.$ with the BESIII $J/\psi$ sample~\cite{KsKpi}.
The values of $f^{K\nrightarrow p}$ and
$\sigma_{f^{K\nrightarrow p}}$ are 0.996 $\pm$ 0.003, and
the value of $|\frac{\Delta f^{K\nrightarrow p}}{\sigma_{f^{K\nrightarrow p}}}|$
is 1.4, therefore, the systematic uncertainty is taken
as 0.3\% and $f^{K\nrightarrow p}$ is set to 0.996.
Similarly, using the control sample of $\ee\go\pi^+\pi^-J/\psi$ at
$\sqrt{s}=4.260$~GeV~\cite{2013Zc_3900}, $f^{L/\Delta_{L}}$ and $f^{V_{xy,\ z}}$ are estimated by performing a secondary vertex fit on $\pi^+$ and $\pi^-$ pair and comparing $V_{xy,\ z}$ of $\pi^+$, $\pi^-$, and lepton pair from $J/\psi$ in data and MC simulation, respectively. The values of $f^{L/\Delta_{L}}$ ($f^{V_{xy,\ z}}$) and $\sigma_{f^{L/\Delta_{L}}}$ ($\sigma_{f^{V_{xy,\ z}}}$) are 0.992 $\pm$ 0.010 (0.997 $\pm$ 0.001), $f^{L/\Delta_{L}}$ ($f^{V_{xy,\ z}}$) is set as 0.992 (0.997), and the systematic uncertainty associated with the $L_{\pipi}/\Delta_{L_{\pipi}}$ requirement for the $K_{S}^{0}$ veto ($V_{xy,\ z}$ requirements) is 1.0\% (0.1\%).

\begin{table*}[htp]
	\centering
	\caption{Systematic uncertainties (\%) from the scale factors $f_{1}$ and $f_{2}$ ($f_{1}$ and $f_{2}$), $\psi(3770)$ and $D_{1}(2420)^{+}$ shapes, including a new Breit-Wigner shape in the high energy region when parameterizing each subprocess cross section line shape, uncertainty of $\omega_{i}$ ($\omega_{i}$), and angular distribution modeling of $e^{+}e^{-}\rightarrow D_{1}(2420)^{+}D^{-}$ decay ({\sc helamp}). The last column shows the total systematic uncertainty obtained by summing up all sources of systematic uncertainties in quadrature assuming they are uncorrelated.}
	% \resizebox{\textwidth}{110mm}{
	\begin{tabular}{ccccccccc}
	\hline\hline
    $\sqrt{s}^{\rm nominal}$ & $f_{1}$ and $f_{2}$ & $\psi(3770)$ shape & $D_{1}(2420)^{+}$ shape & New Breit-Wigner & $\omega_{i}$ & {\sc helamp} & Total &\\
	\hline
	 4.190 & 0.0 & 0.0 & -   & 0.0 & 14.4 & -   & 16.6 &\\
	 4.200 & 0.0 & 0.0 & -   & 0.0 & 19.3 & -   & 21.0 &\\
	 4.210 & 0.0 & 0.0 & -   & 0.0 & 16.8 & -   & 18.7 &\\
	 4.220 & 0.0 & 0.0 & -   & 0.0 & 0.8  & -   & 8.3 &\\
	 4.230 & 0.6 & 1.7 & -   & 0.9 & 0.8  & -   & 8.6 &\\
	 4.237 & 0.0 & 1.6 & -   & 2.3 & 0.6  & -   & 8.8 &\\
	 4.245 & 0.5 & 0.3 & -   & 0.3 & 1.8  & -   & 8.5 &\\
	 4.246 & 0.5 & 0.7 & -   & 3.0 & 7.8  & -   & 11.8 &\\
	 4.260 & 0.2 & 0.9 & -   & 0.7 & 0.4  & -   & 8.4 &\\
	 4.270 & 0.0 & 0.5 & -   & 1.2 & 0.7  & -   & 8.4 &\\
	 4.280 & 0.2 & 1.4 & -   & 1.4 & 0.5  & -   & 8.5 &\\
	 4.290 & 0.1 & 0.1 & -   & 1.8 & 0.5  & -   & 8.5 &\\
	 4.310 & 1.5 & 0.3 & -   & 1.4 & 0.4  & -   & 8.5 &\\
	 4.315 & 0.1 & 0.4 & -   & 0.3 & 0.2  & -   & 8.3 &\\
	 4.340 & 0.1 & 0.4 & 0.7 & 1.2 & 0.3  & 0.5 & 8.4 &\\
	 4.360 & 0.2 & 0.0 & 0.2 & 2.0 & 0.1  & 0.2 & 8.5 &\\
	 4.380 & 0.2 & 0.6 & 0.1 & 0.4 & 0.3  & 0.8 & 8.4 &\\
	 4.390 & 0.1 & 0.3 & 1.2 & 1.0 & 0.8  & 0.3 & 8.5 &\\
	 4.400 & 0.1 & 0.5 & 1.0 & 0.1 & 0.7  & 0.6 & 8.4 &\\
	 4.420 & 0.0 & 0.1 & 0.6 & 0.1 & 1.8  & 0.8 & 8.5 &\\
	 4.440 & 0.2 & 0.9 & 3.3 & 0.2 & 1.5  & 0.4 & 9.1 &\\
	 4.470 & 0.2 & 1.4 & 0.9 & 0.5 & 6.6  & 0.2 & 10.7 &\\
	 4.530 & 0.2 & 0.2 & 0.4 & 0.3 & 6.0  & 0.7 & 10.3 &\\
	 4.575 & 0.5 & 0.3 & 1.8 & 0.0 & 2.1  & 1.3 & 8.8 &\\
	 4.600 & 0.5 & 0.4 & 1.7 & 2.8 & 0.8  & 0.8 & 9.0 &\\
	 4.612 & 0.1 & 0.2 & 0.0 & 0.3 & 1.2  & 0.5 & 8.4 &\\
	 4.620 & 0.1 & 1.5 & 0.9 & 0.3 & 1.1  & 0.4 & 8.5 &\\
	 4.640 & 0.5 & 0.2 & 0.8 & 1.6 & 1.4  & 0.2 & 8.6 &\\
	 4.660 & 0.2 & 0.9 & 0.3 & 2.6 & 1.0  & 0.2 & 8.8 &\\
	 4.680 & 0.2 & 0.4 & 0.9 & 0.0 & 0.4  & 0.1 & 8.3 &\\
	 4.700 & 0.0 & 0.8 & 0.7 & 2.3 & 0.5  & 0.0 & 8.7 &\\
	 4.740 & 0.1 & 0.1 & 2.1 & 0.5 & 0.8  & 0.3 & 8.6 &\\
	 4.750 & 0.5 & 0.8 & 0.6 & 1.9 & 1.2  & 0.2 & 8.7 &\\
	 4.780 & 0.1 & 0.3 & 0.2 & 0.4 & 0.8  & 0.3 & 8.3 &\\
	 4.840 & 0.3 & 0.9 & 0.7 & 0.5 & 0.6  & 1.9 & 8.6 &\\
	 4.914 & 0.0 & 0.6 & 1.1 & 1.1 & 1.4  & 1.7 & 8.7 &\\
	 4.946 & 0.8 & 0.6 & 0.4 & 0.6 & 1.6  & 0.9 & 8.6 &\\
	\hline\hline
	\end{tabular}
	%}
	\label{table4-1}
\end{table*}
	
$\ee\go\pipipsipp\go\DDpipi$ and $\ee\go\DststD\go\DDpipi$ processes
are simulated when estimating $\omega_{i}$,
for the estimation of the systematic uncertainty stemming from the
$\psi(3770)$ ($D_1(2420)^+$) shape,
alternative MC samples are produced by varying the width of
$\psi(3770)$ ($D_1(2420)^+$) by
one standard deviation of its world average value~\cite{pdg}.
The difference of the cross section of $\ee\go\DDpipi$ process
compared with the baseline value is taken as the systematic uncertainty
as listed in Table~\ref{table4-1}.

In Sec.~\ref{sec:data_sets}, $\ee$ is assumed to annihilate into $\DststD$ directly,
and the systematic uncertainty stemming from modeling the angular distribution of the
$\ee\go\DststD$ process is estimated by repeating the analysis procedure with the new model.
For the $\ee\go\DststD$ process, two extreme cases of the angular distribution following $1+\cos^2\theta_{D_1}$ and
$1-\cos^2\theta_{D_1}$ are assumed, where $\theta_{D_1}$ is the helicity angle of
the $D_1(2420)^+$ in the rest frame of the initial $\ee$ system.
The fractions of these two cases are estimated by fitting
to the $\rm{cos}\ \theta_{D^+}$ distribution,
where $\theta_{D^+}$ is the polar angle of $D^+$ in the rest frame
of the initial $\ee$ system, the detection efficiency of
$\ee\go\DststD\go\DDpipi$ process is recalculated according to
the detection efficiencies and fractions of these two cases,
and the cross section of $\ee\go\DDpipi$ process is recalculated as well.
The difference of the cross section of the $\ee\go\DDpipi$ process
compared with the baseline value is taken as the systematic uncertainty
as listed in Table~\ref{table4-1}.

In Sec.~\ref{sec:evtsel}, the normalization factor $f_1$ ($f_2$)
in the $B$ sample is estimated
by assuming a linear background distribution
in $M(\Kpipi)$ \big($RM(\Dpipi)$\big).
A second-order Chebychev polynomial function is used as the
background shape to fit the $M(\Kpipi)$ \big($RM(\Dpipi)$\big) distribution to estimate $f_1$ ($f_2$).
The signal shape is modelled by the $M(\Kpipi)$ \big($RM(\Dpipi)$\big) distributions
in MC simulation of each subprocess weighted according to fractions of each subprocess, $\omega_i$,
and convolved with a Gaussian function to take the resolution
difference between data and MC simulation into consideration.
$\omega_{i}$ is re-estimated according to the new $f_{1}$
and $f_{2}$, and the cross section of the $\ee\go\DDpipi$ process is recalculated.
The difference from the baseline value is taken as the systematic uncertainty
originating from this source as listed in Table~\ref{table4-1}.
	
The systematic uncertainty due to the uncertainty of the fraction of each subprocess, $\omega_i$, is estimated
by varying $\omega_i$ 500 times according to the convariant matrix
in the simultaneous fit of $RM(D^+)$, $RM(D_{\rm{miss}}^-)$,
and $RM(\pipi)$ distributions for each $\sqrt{s}$.
In each iteration, the difference between the cross section
of the $\ee\go\DDpipi$ process and the baseline value is calculated,
and the distribution of the differences is sampled at each $\sqrt{s}$,
the width of the distribution is taken as the systematic uncertainty
as listed in Table~\ref{table4-1}.
	
The systematic uncertainty of the radiative correction is calculated by
using the {\sc kkmc} package.
Initially, the observed signal events are assumed to originate from
the $Y(4260)$ resonance to obtain the efficiency and ISR correction factor.
Then, the measured line shape is used as input to calculate the efficiency
and ISR correction again.
This procedure is repeated until the difference between the subsequent
iteration is comparable with the statistical uncertainty.
The systematic uncertainty due to the input line shapes of
subprocesses is estimated as described below.

The input line shape of each subprocess is varied 500 times
according to the convariant matrix when parametrizing,
and the $\sum\omega_i\epsilon_i(1+\delta)_i$ distribution is sampled
at $\sqrt{s}=$ 4.380, 4.390, 4.400, 4.420, and 4.440 GeV.
The maximum fraction of width and mean values of the distributions,
2.8\%, is taken as the systematic uncertainty due to the
input line shapes in the ISR correction.
Moreover, new resonances around $\sqrt{s}=$ 4.700~GeV are added when
parameterizing the line shape of each subprocess since there is an evidence
around $\sqrt{s}=4.700$ GeV in the $\ee\go\DDpipi$ cross section line shape,
and the difference is taken as the systematic uncertainty associated with the new BW resonance in the high energy regions as listed
in Table~\ref{table4-1}.

Table~\ref{table4-1} summarizes the total systematic uncertainties.
The total systematic uncertainty at each $\sqrt{s}$ is obtained
by summing up all sources of systematic uncertainties in quadrature,
assuming that they are uncorrelated.
	
\subsection{Systematic uncertainties in resonance parameters}
\label{res_sys}
	
The systematic uncertainties when parameterizing the resonances in
the \\ $\ee\go\DDpipi$ cross section line shape mainly stem from the
absolute $\sqrt{s}$ measurement, the $\sqrt{s}$ spread, global shift of the
$\sqrt{s}$ for data samples taken in the same period, and
the systematic uncertainty of the cross section measurement.

The absolute $\sqrt{s}$ of data samples with $\sqrt{s}$ smaller than 4.610 GeV are measured with
dimuon events, with an uncertainty of
\\ $\pm$ 0.8 MeV, while those with $\sqrt{s}$ larger
than or equal to 4.610 GeV are measured with $\Lambda_{c}^{+}\bar{\Lambda}_{c}^{-}$ events
with an uncertainty of $\pm$ 0.6 MeV.
\\ Thus, 0.8 MeV is taken as the systematic uncertainty, and propagates to the
masses of the resonances by the same amount.

The systematic uncertainty from the $\sqrt{s}$ spread is estimated
by convolving the fit formula with a Gaussian function with a width of 1.6 MeV,
which is the beam spread, determined from measurement results of the Beam Energy Measurement System~\cite{2011BEMS} at other $\sqrt{s}$.

The systematic uncertainty from global shift of the $\sqrt{s}$ for data samples
taken in the same period is estimated by shifting the $\sqrt{s}$ of corresponding
data samples by 3 MeV and deviations of parameters is taken as the systematic uncertainties.

The systematic uncertainty from the cross section measurement is divided into two parts.
The first part covers uncorrelated systematic uncertainties among the different
data samples (those in Table~\ref{table4-1}).
The corresponding systematic uncertainty is estimated by including
the uncertainty in the fit to the cross section,
and taking the differences on the parameters as the systematic uncertainties.
The second part includes all the other systematic uncertainties (8.3\%),
which is common for all data samples, and only affects
the $\Gamma^{\ee}_{j}\mathscr{B}_{j}$ measurement.

Table~\ref{tableres} summarizes the systematic uncertainties
in the parameters of resonances for the four solutions. The total systematic uncertainty
is obtained by summing up all sources of systematic uncertainties
in quadrature, assuming they are uncorrelated.

\begin{table*}[htp]
	\centering
	\caption{Systematic uncertainties in the measurement of the resonances parameters. $\sqrt{s}$ represents the systematic uncertainty from the center-of-mass measurement. $\sqrt{s}$ shift represents the systematic uncertainty from the global shift of $\sqrt{s}$ for data samples taken in the same period. Cross $\rm{section}_{a(b)}$ represents the systematic uncertainty from the cross section measurements which are uncorrelated (common) in each data sample. The units of $m_{i}$, $\Gamma^{\rm{tot}_{i}}$, $c$, $\Gamma^{\ee}_{j}\mathscr{B}_{j}$, and $\phi_{j}$ are MeV/$c^{2}$, MeV, MeV$^{-3/2}$, eV and rad, respectively.}
	\begin{tabular}{ccccccccccccccc}
	\hline\hline
	& Sources & $m_{0}$ & $\Gamma_{0}$ & $m_{1}$ & $\Gamma_{1}$ & $c$ & $\Gamma^{\ee}_{0}\mathscr{B}_{0}$ & $\Gamma^{\ee}_{1}\mathscr{B}_{1}$ & $\phi_{0}$ & $\phi_{1}$ &\\
	\hline
	\multirow{5}*{Solution \uppercase\expandafter{\romannumeral1}} & $\sqrt{s}$        & 0.8 & - & 0.8 & - & - & - & - & - & - &\\
	 ~                                                              & $\sqrt{s}$ shift  & 2.0 & 0.7 & 0.6 & 0.0 & 70 & 0.1 & 0.0 & 0.0 & 1.0 &\\
	 ~                                                              & $\sqrt{s}$ spread & 0.0 & 0.1 & 0.1 & 0.3 & 92 & 1.6 & 1.3 & 0.0 & 1.0 &\\
	 ~                                                      & Cross $\rm{Section}_{a}$ & 0.6 & 1.1 & 3.8 & 8.8 & 58 & 0.1 & 2.9 &                                   - &                                   - &\\
	 ~                                                      & Cross $\rm{Section}_{b}$ &  - &  - &  - &  - & - & 0.8 & 0.0 &  - &  - &\\
	 ~                                                              & Overall & 2.2 & 1.3 & 3.9 & 8.8 & 13$\times$10 & 1.8 & 3.2 & 0.0 & 1.4 &\\
	\hline
	\multirow{5}*{Solution \uppercase\expandafter{\romannumeral2}} & $\sqrt{s}$        & 0.8 & - & 0.8 & - & - & - & - & - & - &\\
	 ~                                                              & $\sqrt{s}$ shift  & 2.0 & 0.7 & 0.6 & 0.0 & 70 &0.1 & 0.0 & 0.0 & 3.2 &\\
	 ~                                                              & $\sqrt{s}$ spread & 0.0 & 0.1 & 0.1 & 0.3 & 92 & 1.0 & 0.0 & 0.0 & 0.0 &\\
	 ~                                                      & Cross $\rm{Section}_{a}$ & 0.6 & 1.1 & 3.8 & 8.8 & 58 & 0.1 & 0.1 &                                   - &                                   - &\\
	 ~                                                      & Cross $\rm{Section}_{b}$ &  - &  - &  - &  - & - & 0.8 & 0.0 &  - &  - &\\
	 ~                                                              & Overall & 2.2 & 1.3 & 3.9 & 8.8 & 13$\times$10 & 1.3 & 0.1 & 0.0 & 3.2 &\\
	\hline
	\multirow{5}*{Solution \uppercase\expandafter{\romannumeral3}} & $\sqrt{s}$        & 0.8 & - & 0.8 & - & - & - & - & - & - &\\
	 ~                                                              & $\sqrt{s}$ shift  & 2.0 & 0.7 & 0.6 & 0.0 & 70 & 0.1 & 0.0 & 0.1 & 3.1 &\\
	 ~                                                              & $\sqrt{s}$ spread & 0.0 & 0.1 & 0.1 & 0.3 & 92 & 1.5 & 0.0 & 0.0 & 3.1 &\\
	 ~                                                      & Cross $\rm{Section}_{a}$ & 0.6 & 1.1 & 3.8 & 8.8 & 58 & 0.3 & 0.1 &                                   - &                                   - &\\
	 ~                                                      & Cross $\rm{Section}_{b}$ &  - &  - &  - &  - & - & 0.8 & 0.0 &  - &  - &\\
	 ~                                                              & Overall & 2.2 & 1.3 & 3.9 & 8.8 & 13$\times$10 & 1.7 & 0.1 & 0.1 & 4.4 &\\
	\hline
	\multirow{5}*{Solution \uppercase\expandafter{\romannumeral4}} & $\sqrt{s}$        & 0.8 & - & 0.8 & - & - & - & - & - & - &\\
	 ~                                                              & $\sqrt{s}$ shift  & 2.0 & 0.7 & 0.6 & 0.0 & 70 & 0.1 & 0.0 & 0.0 & 1.0 &\\
	 ~                                                              & $\sqrt{s}$ spread & 0.0 & 0.1 & 0.1 & 0.3 & 92 & 1.1 & 1.2 & 0.0 & 0.0 &\\
	 ~                                                      & Cross $\rm{Section}_{a}$ & 0.6 & 1.1 & 3.8 & 8.8 & 58 & 0.0 & 2.5 &                                   - &                                   - &\\
	 ~                                                      & Cross $\rm{Section}_{b}$ &  - &  - &  - &  - & - & 0.8 & 0.0 &  - &  - &\\
	 ~                                                              & Overall & 2.2 & 1.3 & 3.9 & 8.8 & 13$\times$10 & 1.4 & 2.8 & 0.0 & 1.0 &\\
	\hline\hline
	\end{tabular}
	\label{tableres}
\end{table*}

\subsection{Systematic uncertainties in $X(3842)$ measurement}

Except for the fit range and the background shape of the $RM(\pipi)$, $RM(D^+)$
and $RM(D_{\rm{miss}}^-)$ mass window requirements,
other sources of systematic uncertainties associated with this measurement are the same as
those in Sec.~\ref{sec:sys_mes}, but with the fit range and background shape of $RM(\Dpipi)$ excluded.

The systematic uncertainty originating from the fit
range of $RM(\pipi)$ is estimated by varying the limits of
the fit range from (3.79, 3.89) GeV/$c^2$ to (3.81, 3.91) GeV/$c^2$.
The difference of the cross section from the baseline value
in the data sample at $\sqrt{s}=4.680$ GeV is taken as the systematic uncertainty, and is 10.4\%.
The background shape is varied from a second-order Chebychev polynomial function to
a first order one in the data sample taken at $\sqrt{s}=4.680$ GeV, the difference of the cross section compared
with the baseline value is taken as the systematic uncertainty, and is 1.9\%.

The systematic uncertainty stemming from the $RM(D^+)$ and $RM(D_{\rm{miss}}^-)$
mass window requirements,
which is mainly caused by the difference between distributions of
data and MC simulation in the corresponding selection criterion ranges,
is estimated by producing alternative MC samples where the mass and
width of $f_{0}(500)$ are varied by one standard deviation in the data sample at $\sqrt{s}=4.680$ GeV.
The difference of the cross section compared with the baseline value
is taken as the systematic uncertainty, and is 1.9\%.

The total systematic uncertainty for data samples with $\sqrt{s}$
smaller than or equal to 4.315 GeV are equal to 12.9\% ,
and for those with $\sqrt{s}$ larger than 4.315 GeV are equal to 13.1\%
by summing up all sources of systematic uncertainties in quadrature,
assuming they are uncorrelated.

\section{\boldmath Summary}
	
Using data samples taken at $\sqrt{s}$ from 4.190 to 4.946~GeV,
the cross section of the $\ee\go\DDpipi$ process
is reported for the first time by a partial reconstruction method.

In the cross section of the $\ee\go\DDpipi$ process, a structure with 0.3$\sigma$ significance is visible around $\sqrt{s}=4.245$~GeV. This
might be the $Y(4230)$ resonance, however, due to its low significance, it is not possible to assign it to the $Y(4230)$ or the $Y(4260)$ state.

By fitting the $\ee\go\DDpipi$ cross section line shape, we observe a resonance with a mass of (4373.1 $\pm$ 4.0 $\pm$ 2.2) MeV/$c^2$
and a width of (146.5 $\pm$ 7.4 $\pm$ 1.3) MeV, which is in agreement with the $Y(4390)$.
There is evidence with a statistical significance of 4.1$\sigma$ for a second resonance
with a mass of (4706 $\pm$ 11 $\pm$ 4) MeV/$c^2$ and
a width of (45 $\pm$ 28 $\pm$ 9) MeV.
The first uncertainties are statistical and the second are systematic.

The $X(3842)$ resonance is searched for in the $RM(\pipi)$ distribution and evidence is found in the $M(\pipi)$ distribution in data samples with
$\sqrt{s}$ from 4.600 to 4.700~GeV, and its significance is $4.2\sigma$.
By comparing this study with previous studies, the cross section of the $\ee\go\pipipsipp\go\DDpipi$ process peaks around 4.390 GeV which indicates this process might be produced via the $Y(4390)$ state \cite{2019HY, 2020ZY}; the process
$\ee\go\pi^{+}\pi^{-}\psi_{2}(3823)(\go\gamma\chi_{c1})$ peaks around $\sqrt{s}=4.360$ and $4.420$ GeV, which means this process might be produced via the $Y(4360)$ and the $\psi(4415)$ \cite{2015X3823} resonances.
There is evidence that the cross section of the $\ee\go\pi^{+}\pi^{-}X(3842)$ process peaks at $\sqrt{s}$ from 4.600 to 4.700 GeV, but no significant signal is observed in samples collected at $\sqrt{s}$ around 4.400 GeV.
This indicates that the production mechanism of the $\ee\go\pi^{+}\pi^{-}\psi(1D)$ processes might be different and could proceed via different $Y$ or $\psi$ states. More data samples and more precise measurements are needed to reveal the mechanism \cite{2020BESWP}.
	
\acknowledgments

The BESIII collaboration thanks the staff of BEPCII and the IHEP computing center for their strong support. This work is supported in part by National Key R\&D Program of China under Contracts Nos. 2020YFA0406300, 2020YFA0406400; National Natural Science Foundation of China (NSFC) under Contracts Nos. 11635010, 11735014, 11835012, 11935015, 11935016, 11935018, 11961141012, 12022510, 12025502, 12035009, 12035013, 12192260, 12192261, 12192262, 12192263, 12192264, 12192265; the Chinese Academy of Sciences (CAS) Large-Scale Scientific Facility Program; Joint Large-Scale Scientific Facility Funds of the NSFC and CAS under Contract No. U1832207; CAS Key Research Program of Frontier Sciences under Contract No. QYZDJ-SSW-SLH040; 100 Talents Program of CAS; INPAC and Shanghai Key Laboratory for Particle Physics and Cosmology; ERC under Contract No. 758462; European Union's Horizon 2020 research and innovation programme under Marie Sklodowska-Curie grant agreement under Contract No. 894790; German Research Foundation DFG under Contracts Nos. 443159800, Collaborative Research Center CRC 1044, GRK 2149; Istituto Nazionale di Fisica Nucleare, Italy; Ministry of Development of Turkey under Contract No. DPT2006K-120470; National Science and Technology fund; National Science Research and Innovation Fund (NSRF) via the Program Management Unit for Human Resources \& Institutional Development, Research and Innovation under Contract No. B16F640076; STFC (United Kingdom); Suranaree University of Technology (SUT), Thailand Science Research and Innovation (TSRI), and National Science Research and Innovation Fund (NSRF) under Contract No. 160355; The Royal Society, UK under Contracts Nos. DH140054, DH160214; The Swedish Research Council; U. S. Department of Energy under Contract No. DE-FG02-05ER41374.

	% \nolinenumbers %% end of line numbers
\end{document}